\documentclass[lettersize,journal]{IEEEtran}
\usepackage{amsmath,amsfonts}
\usepackage{algorithmic}
\usepackage{algorithm}
\usepackage{array}
\usepackage[caption=false,font=normalsize,labelfont=sf,textfont=sf]{subfig}
\usepackage{textcomp}
\usepackage{stfloats}
\usepackage{url}
\usepackage{verbatim}
\usepackage{graphicx}
\usepackage{cite}
\usepackage{booktabs}
\usepackage{threeparttable}
\usepackage{caption}                    
\usepackage{subfig}
\usepackage{xcolor}

\newenvironment{SUBENVhighlight}[1]{\color{#1}}{\color{black}}

\begin{document}

\title{Dynamic Thermal Feedback in Highly Immersive VR Scenarios: a Multimodal Analysis of User Experience}

\author{Sophie Villenave{},
        Pierre Raimbaud{},
       and Guillaume Lavoué{}

\IEEEcompsocitemizethanks{\IEEEcompsocthanksitem S. Villenave, P. Raimbaud, G. Lavoué are affiliated to Ecole Centrale de Lyon, CNRS, INSA Lyon, Universite Claude Bernard Lyon 1, LIRIS, UMR5205, ENISE, 42023 Saint Etienne, France. E-mail: name.surname@liris.cnrs.fr.\protect}

\thanks{Manuscript received xxx; revised xxx}}

\markboth{Journal of \LaTeX\ Class Files,~Vol.~XX, No.~X, XXX}%
{Villenave \MakeLowercase{\textit{et al.}}: Dynamic Thermal Feedback in Highly Immersive VR Scenarios: a Multimodal Analysis of User Experience}

\IEEEpubid{0000--0000/00\$00.00~\copyright~2021 IEEE}

\maketitle

\begin{abstract}
Thermal feedback is critical to a range of Virtual Reality (VR) applications, such as firefighting training or thermal comfort simulation. Previous studies showed that adding congruent thermal feedback positively influences User eXperience (UX). However, existing work did not compare different levels of thermal feedback quality and mostly used less immersive virtual environments. To investigate these gaps in the scientific literature, we conducted a within-participant user study in two highly-immersive scenarios, Desert Island (n=25) and Snowy Mountains (n=24). Participants explored the scenarios in three conditions (\textit{Audio-Visual only}, \textit{Static-Thermal Feedback}, and \textit{Dynamic-Thermal Feedback}). To assess the complex and subtle effects of thermal feedback on UX, we performed a multimodal analysis by crossing data from questionnaires, semi-structured interviews, and behavioral indicators. Our results show that despite an already high level of presence in the \textit{Audio-Visual only} condition, adding thermal feedback increased presence further. Comparison between levels of thermal feedback quality showed no significant difference in UX questionnaires, however this result is nuanced according to participant profiles and interviews. Furthermore, we show that although the order of passage did not influence UX directly, it influenced user behavior. We propose guidelines for the use of thermal feedback in VR, and the design of studies in complex multisensory scenarios.

\end{abstract}


\section{Introduction}
\IEEEPARstart{T}hermal perception is a fundamental part of the human sensory experience, playing a crucial role in comfort~\cite{liuEffectsLocalHeating2022}, in emotion~\cite{yangThermotionExplorationFacilitating2019}, and most importantly in body thermal regulation~\cite{battistelInvestigationHumansSensitivity2023}. Although this sense is often sidelined in conventional VR experiences, some applications require thermal feedback to be relevant, e.g., firefighter training~\cite{monteiroDeliveringCriticalStimuli2021}, or thermal comfort assessment in human-built environment~\cite{lyuImmersiveMultisensoryVirtual2023}. To this end, various thermal feedback devices dedicated to VR have been developed and evaluated in the literature. These evaluations serve at least one of two main objectives: ensure the device is able to provide a plausible thermal feedback, and understand how thermal feedback influences the VR experience.

In previous VR studies that include thermal feedback, participants are often immersed in low-fidelity Virtual Environments (VEs) that offer limited interactions. As a result, the positive aspects of thermal feedback may be overestimated. Moreover, as only one quality level of thermal feedback is presented for a given virtual object or environment, comparisons are only made between the absence and the presence of a thermal modality. Yet, thermal feedback stems from heat transfer, a physical phenomenon which can be very diverse in type (radiation, conduction, and convection), direction (exothermic or endothermic), and surface area. When thermal feedback is included in VR, one must also take into account: its congruence with the VE, the level of adaptation to interactions (i.e, dynamism), and the time between an action and the corresponding stimulation (i.e., latency). While congruence with the VE has already been explored, showing that congruent feedback has a positive impact on UX~\cite{dasilveiraThermalWindDevices2023b}, as well as that incongruent feedback has a negative one~\cite{ragozin_thermoquest_2021}, the influence of other factors remains underexplored.

To address these gaps, we investigate here \textit{dynamism} as a quality factor for thermal feedback, and do so in a setting where audio-visual quality is high, and where users are engaged in task-oriented scenarios. We conducted a within-subject user study to evaluate the effects of dynamic ambient thermal feedback on UX where participants experienced one of two distinct VR scenarios, Desert Island for hot stimulation (n = 25) or Snowy Mountains for cold stimulation (n = 24). These task-oriented scenarios feature high-fidelity audio-visual environments that can be explored using teleportation, and our apparatus allowed users to move freely in a 2\textit{m} by 2\textit{m} physical space. Participants repeated the scenario three times in different thermal feedback conditions: \textbf{Control} (i.e., no thermal feedback), \textbf{Static} (i.e., congruent thermal feedback that does not adapt to the user's actions), \textbf{Dynamic} (i.e., congruent thermal feedback that adapts to the user's actions).

We measure thermal perception, general~\cite{schubertExperiencePresenceFactor2001} and physical presence~\cite{hartmannSpatialPresenceExperience2015}, sense of possible actions~\cite{hartmannSpatialPresenceExperience2015}, and haptic experience~\cite{anwarFactorsHapticExperience2023} using post-experience questionnaires. To gain more nuance in the results, as our experimental conditions provide a subtle difference in thermal feedback quality, we also gathered qualitative insights through semi-structured interviews after the last session (the third one) . While essential to understand UX, questionnaires and interviews are limited to a subjective and asynchronous report of the actual experience~\cite{putzeBreakingExperienceEffects2020}. Therefore, to gather synchronous and objective data, we also systematically record VR sessions using the PLUME toolbox~\cite{javerliatPLUMERecordReplay2024} and analyze several user behavioral indicators: the physical and virtual positions of participants, as well as the number of performed teleportations. Leveraging these three complementary data sources, we propose a multimodal analysis of the user experience that takes two main factors into account: quality of thermal feedback (none, static, dynamic) and order of passage induced by the within-subject design (first, second, third).

Our results extend earlier findings and show that even in highly immersive VEs, congruent ambient thermal feedback significantly enhances presence. However, we found no statistically significant difference in UX questionnaires between static and dynamic thermal feedback. In spite of this results, by taking into account differences in thermal sensitivity and reported experience through interviews, we found that the difference in UX between static and dynamic thermal feedback does exist, but that its perception highly depends on individuals. This suggests that while dynamic thermal feedback might not be essential in high-immersion settings, a user-driven approach considering individual thermal sensitivity could enhance its effectiveness. Regarding the influence of the order of passage on UX, while we found no main effect on questionnaires, it did induce a main effect on behavioral metrics. Actually, physical and virtual traveled distance tend to decrease from one session to the next in both scenarios, and so does teleportation count as well as roaming and movement entropy in the Desert Island scenario.

To summarize, our contributions are the following: 
(1)	Our work extends earlier findings by measuring the influence of ambient thermal feedback on UX in a highly immersive VE.
(2)	This novel study on thermal feedback quality factors on UX tackles the lack thereof by comparing the benefits of static versus dynamic thermal stimulation.
(3)	We propose an original multimodal analysis framework to reveal subtle effects of thermal feedback on UX within highly immersive VR scenarios.
4) Finally we provide methodological recommendations for investigating the impact of thermal parameters on user experience in complex VR settings.

\section{Related Work}
%% Restructure this section by : reducing the technical related work to ambient thermal feedback only
%% Keep the section on effects on realism, immersion and presence
%% Add a section on multimodal analysis (behavior, questionnaire, interviews, physiological) of user experience
\label{Related_Work}
With multisensory VR systems, user stimulation extends beyond audiovisual cues, meaning that at least one additional sensory modality is engaged in response to the virtual environment (e.g., haptic~\cite{adilkhanovHapticDevicesWearabilityBased2022}, olfactory~\cite{tewellReviewOlfactoryDisplay2024}, taste~\cite{ranasingheVirtualTasteDigital2023}). Haptic modalities, including tactile and kinesthetic feedback\cite{adilkhanovHapticDevicesWearabilityBased2022}, are the most extensively studied~\cite{meloMultisensoryStimuliBenefit2020b}. as they allow users to perceive and manipulate virtual objects with greater realism and precision. (Selection and manipulation are the main interaction with virtual object) A less commonly addressed facet of touch is thermal perception~\cite{meloMultisensoryStimuliBenefit2020b}, which holds potential for enhancing realism and immersion in VR~\cite{dasilveiraThermalWindDevices2023b}. In this section, we review VR studies that have used devices capable of generating ambient thermal stimuli. We highlight key findings, particularly regarding how thermal feedback interacts with other sensory modalities, which is an important consideration given the complexity of cross-modal perception in multisensory VR. We also examine how thermal feedback contributes to the overall UX and specifically presence. Finally, we conclude with an overview of multimodal methodologies that support comprehensive UX evaluation in immersive environments.

\subsection{Ambient thermal feedback displays for VR}
Thermal feedback systems for VR have evolved significantly, from early non-wearable setups that provided non-contact effects using infrared lamps and/or fans~\cite{dionisio_virtual_1997, lecuyer_homere_2003} to modern miniaturized wearable devices that leverage Peltier elements~\cite{peiris_thermovr_2017, zhuSenseIceFire2019b}. A comprehensive review by da Silveira et al.~\cite{dasilveiraThermalWindDevices2023b} outlines this evolution and highlights common challenges such as latency and limited responsiveness due to thermodynamic constraints. These challenges arise when producing either an ambient thermal feedback, aimed at simulating the overall environmental temperature of the VE, or a localized one, that simulates touching a hot or cold object within the VE. For the present work, we focused on the creation and evaluation of ambient thermal feedback, which has diverse use-cases of interest: environmental cues such as weather conditions~\cite{hanHapmosphereSimulatingWeathers2019a, gunther_therminator_2020}, fire evacuation drill~\cite{shaw_heat_2019}, thermal conditions in human-built environment~\cite{nytsch-geusenDevelopmentInteractiveVirtual2021, lyuImmersiveMultisensoryVirtual2023}.

Back in 1997, Dionisio’s~\cite{dionisio_virtual_1997} prototyped an advanced thermal feedback system for non-immersive VR. His system combined IR lamps and fans placed around the user, as well as Peltier elements integrated into the armrests of the chair in which the user had to sit to enjoy the VR experience. Later, the idea of combining IR lamps and fans was then taken up again, but adapted to immersive VR: in a CAVE by Hülsmann et al.~\cite{hulsmann_wind_2014}, as well as for HMD setups by Villenave et al.~\cite{villenaveDynamicModularThermal2025}. The latter is open-source, fully integrated with Unity, which is the most widely used game engine for VE development, and perceptually characterized. Further extending this concept to immersive VR using an HMD, Han et al.'s ``Haptic Around''~\cite{hanHapticMultipleTactile2018a} supports full weather simulation, including mist and wind, within a tracked 2×2m area; however it is difficult to replicate and limited to the upper body. Although IR lamps are effective in producing a sensation of heat, their low wattage (300~W) makes it difficult to convey the sensation of fire, which is critical for building evacuation simulations or firefighter training. In these cases, powerful IR heater (1-2~kW) were used~\cite{wareingUserPerceptionHeat2018, sosuke_ichihashi_high-speed_2021} but they were not directly controllable; instead, metal blinds had to be added in front and operated via the VE, resulting in significantly higher power consumption. More accessible solutions, such as conventional heaters or pedestal fans~\cite{narcisoStudyingInfluenceMultisensory2023, helfenstein-didierExploringCrossmodalInteraction2021}, that can be manually operated were also used to simulate ambient temperature changes. Although the systems presented above for providing ambient thermal feedback are voluminous and often limit the size of the VR play area because of the limited range of their thermal actuators, they offer genuine whole-body stimulation that is controllable by the VE. Their non-wearable nature makes them non-intrusive and convenient for VR user studies as they require no equipment to be mounted on the user.

Attempts to provide the sensation of ambient thermal feedback using strategically placed Peltier elements are promising since they minimize the spatial footprint of the thermal system. To mimic ambient thermal feedback, Peltier elements have either been integrated directly into the HMD~\cite{peiris_thermovr_2017, ragozin_thermoquest_2021, marquardt_temperature_2025} or within wearable necklaces~\cite{ranasinghe_ambiotherm_2017, ranasinghe_season_2018}. However, Peltier elements require high-performance cooling systems that cannot be comfortably integrated between the user's head and the HMD~\cite{peiris_thermovr_2017}, and often necessitate power tethers or battery packs that impair mobility and negatively impact the overall user experience. A recent integration of Peltier to augment controllers claims very low power consumption~\cite{mesnageStimulHeatClipOnLowEnergy2025a}, therefore requiring neither a large battery nor an imposing cooling system, but no user studies were available at the time of writing. To overcome the limitations of Peltier elements, localized but non-portable vortex tube solutions that are able to generate rapid cold sensations have been presented by Xu et al.~\cite{xu_non-contact_2019}. In later work, the same authors were able to provide lasting cold feedback by combining the vortex tube with light~\cite{xuIntegrationIndependentHeat2023}. An emerging approach to providing thermal sensations consists in directly stimulating the nerves responsible for the sensation, without creating an actual temperature change. Brooks et al.~\cite{brooks_trigeminal-based_2020} and later Jasmine et al.~\cite{jasmineluChemicalHapticsRendering2021} leveraged chemicals such as menthol and capsaicin, and Saito et al.~\cite{saito_thermal_2021} showed that electrical stimulation on the forehead was able to produce cold sensations. Nonetheless, while some studies suggest that stimulating specific body areas, such as the face or the neck, can create an illusion of global thermal change~\cite{liuEffectsLocalHeating2022}, to our knowledge no conclusive comparative studies have been conducted in VR to verify whether such localized feedback can truly replicate the overall environmental temperature of the VE.

For our study, we sought an ambient thermal feedback system that would integrate seamlessly into the VR experience, minimize intrusiveness, allow adaptive responses to user actions, and remain simple to implement within the development pipeline. To that end, we implemented the system presented in Villenave et al.~\cite{villenaveDynamicModularThermal2025}, prioritizing ergonomics over size of play area and hardware compactness.

%Attempt to provide global thermal feedback by integrating Peltier elements inside the HMD~\cite{peiris_thermovr_2017, ragozin_thermoquest_2021, marquardt_temperature_2025}, are interesting, as they minimize the space needed for the VR experience. Unfortunately in practice, Peltier elements require a high-performance cooling system that cannot be comfortably interfaced between helmet and Peltier, as well as tether for power or batteries that must be worn, all of which negatively impacts mobility and UX as a whole.

%Some wearable thermal feedback solutions claim to induce full-body thermal illusions through localized stimulation, by stimulating sensitive areas of the body that seem to have a greater influence on whole-body thermal comfort~\cite{liuEffectsLocalHeating2022} (e.g., neck or facial areas), to our knowledge, no conclusive comparative studies evaluate their ability to replicate a genuine sense of ambient temperature change when compared to spatial, global thermal systems. So we decided to implement a simple, low-cost, tried-and-tested solution that provided global stimulation~\cite{villenaveDynamicModularThermal2025}, as we favored non-intrusivity and agency over compactness.

\subsection{Effects of Thermal Feedback on User Experience}

Prior research has demonstrated that thermal feedback significantly influences various aspects of the VR experience. Han et al.~\cite{hanHapticMultipleTactile2018a, hanHapmosphereSimulatingWeathers2019a} found that adding multisensory stimuli, including thermal feedback, enhanced enjoyment, realism, and immersion, though maintaining coherence across stimuli was critical to avoid disrupting immersion. Günther et al.\cite{gunther_therminator_2020} demonstrated that thermal stimuli tend to dominate over visual stimuli when perceiving temperature. Similarly, they showed that congruent thermal cues enhanced user engagement, while incongruent ones diminished it. Ragozin et al.\cite{ragozin_thermoquest_2021} found the same effect on presence. Other studies explored the role of thermal feedback in specific contexts. Shaw et al.\cite{shaw_heat_2019}, using a previously developed system~\cite{wareingUserPerceptionHeat2018}, showed that thermal stimuli in a VR fire evacuation simulation increased realism, which led to heightened stress and more authentic behavior. Monteiro et al.\cite{monteiroDeliveringCriticalStimuli2021} argued that for firefighters training in VR, thermal stimuli are critical, though such stimuli slowed task performance and did not increase presence. Mazursky et al.\cite{mazursky_thermalgrasp_2024} found that their thermal feedback device boosted sensory engagement and realism in interactive VR scenarios.

Thermal comfort, which arises from the perception of the surrounding thermal environment, varies widely between individuals due to a range of physiological, psychological~\cite{schweiker_drivers_2018}, and cultural factors~\cite{naheed_review_2021}. Despite their relevance, these individual differences have received limited attention in VR research involving thermal feedback. In this regard, Philippe et al.~\cite{philippeCoolMeEffects2024} showed that users' thermal preferences shaped how thermal stimulation influenced emotional responses and thermal comfort during a seated VR experience; Han et al.~\cite{hanHapmosphereSimulatingWeathers2019a} also observed that participants' past experiences and thermal expectations can influence their immersion.

In summary, prior investigations into thermal feedback for VR have largely been restricted to on-off comparisons, and often implemented in simple VEs. Our study seeks to extend this research by identifying the optimal parameters for thermal feedback, thereby enhancing its applicability and impact in more immersive VR settings.

\subsection{Multimodal Analysis of User Experience}
Evaluation of UX in VR is most often conducted solely through post-experience questionnaires~\cite{alexandrovskyExaminingDesignChoices2020}. While essential to understand user experience, these questionnaires are asynchronous to the actual experience~\cite{kimSystematicReviewVirtual2020} and may be subject to biases~\cite{choiCatalogBiasesQuestionnaires2005a}, which makes them insufficient to fully characterize it, especially in complex scenarios involving structured tasks, narratives, and high levels of immersion. To complete questionnaires, protocols may include semi-structured interviews in which participants are asked to recall and reflect on their experience~\cite{kellyUsingInterpretativePhenomenological2023}. This qualitative feedback, although time-consuming to transcribe and analyze, is extremely valuable as it offers a rich and nuanced understanding of the lived experience of participants.
Objective and synchronous measures of user experience have been presented in recent studies, by leveraging data output from XR systems, as they offer low-cost opportunities for collecting rich behavioral data, including both discrete events (e.g., interactions, teleportations, event markers) and continuous streams (e.g., head and hand movements, eye tracking). These data sources can be used to infer physical, cognitive and affective states such as cybersickness~\cite{deguzmanReductionMotionComplexity2025}, cognitive load~\cite{reinhardtEntropyControllerMovements2019, poupardMovementLearningLeveraging2025}, curiosity~\cite{cenCuriosityShapesSpatial2024, poupardMovementLearningLeveraging2025}, presence~\cite{gehrkeExposingMovementCorrelates2024} and social presence~\cite{ochsMultimodalBehavioralCues2022}. Despite the apparent benefits of using such behavioral data to understand UX in XR, they are often discarded due to the complexity of analysis. However, recent advances in dedicated software tools~\cite{javerliatPLUMERecordReplay2024} have significantly lowered the barrier to recording and analyzing both discrete event and continuous signals from XR activity. In complement of behavioral data, XR studies may employ EDA, ECG, and EMG signals, which have been shown as valuable for assessing cognitive and emotional states~\cite{halbigSystematicReviewPhysiological2021b}. These strategies have their own strengths and limitations, and we argue that they should be employed in a complementary manner to enable multimodal analyses that can more accurately capture the complexity of user experience in XR. In the case of multisensory VR studies, analyses that combine results from questionnaires, interviews, as well as behavioral metrics and/or physiological data, such as Shaw et al. study~\cite{shaw_heat_2019}, are still scarce. By leveraging a powerful multimodal analysis for this work, we found that meaningful insights emerged when objective data are cross-analyzed and interpreted in the context of participants’ self-reported experiences, whether through questionnaires and/or semi-structured interviews.

\subsection{Motivations and Research Questions}
Previous VR studies that have used an ambient thermal feedback system agree that the addition of this thermal modality is an asset, as it enhances user experience, provided the stimuli are congruent with the VE. However, most of these studies were conducted within low visual immersion settings~\cite{gunther_therminator_2020}, in VEs with minimal or no interactivity~\cite{ragozin_thermoquest_2021}, with limited user mobility~\cite{lyuImmersiveMultisensoryVirtual2023}, or a combination of these constraints. Such limitations negatively impact the level of immersion, and reported benefits of ambient thermal feedback may not generalize when applied to more immersive VR experiences. Consequently, we argue there is a need to expand these findings to ascertain their applicability in highly immersive VR environments.

Previous work examined a single level of thermal feedback, comparing its presence against its absence. However, incorporating thermal feedback in VR involves considering its congruence with the VE, adaptability to interactions, spatialization and latency. While congruent feedback has been shown to positively influence user experience, the effects of other factors remain underexplored.

To address these gaps, we designed a study conducted in two visually rich environments that elicit distinct thermal expectations from users: a desert island and snowy mountains. In each scenario, participants engaged in tasks within a guided narrative and experienced three conditions: a control condition (audio-visual only), a static thermal feedback condition (fixed intensity), and a dynamic condition, where thermal intensity dynamically varied based on the user's actions. To evaluate user experience and specifically investigate the role of agency in thermal stimulation, we adopted a multimodal analysis approach combining self-reported questionnaires, semi-structured interviews, and behavioral metrics.

We address the following research questions:
\begin{itemize}
    \item \textbf{RQ1}: Does the addition of ambient thermal feedback in highly-immersive VR enhance UX further?
    \item \textbf{RQ2}: Does dynamic thermal feedback offer a superior UX compared to static feedback? 
    \item \textbf{RQ3}: Does thermal feedback quality have an influence on user behavior?
    \item \textbf{RQ4}: Does order of passage, inherent to within-subject design study, have an influence on UX and / or user behavior?
\end{itemize}
\section{User Study}
\label{User_Study}
To investigate the above-mentioned research questions, we conducted a multimodal analysis of a within-subject user study that aimed at comparing levels of quality of ambient thermal feedback in highly immersive and interactive VR scenarios. The following section details the experimental design, developed VR scenarios, conditions, and evaluation methods used to assess user experience.

\subsection{Participants}
We recruited 25 participants for the Desert Island scenario and 24 for Snowy Island, which theoretically allows us to detect effects of size 0.53 and 0.54 respectively as verified through a post-experiment power analysis conducted with G*Power~\cite{faulPowerFlexibleStatistical2007}. Participants were naive to the exact purpose of the experiment and had normal or corrected-to-normal vision and hearing. This study was approved by the research ethics committee of Université
de Lyon’s ComUE, under approval number 2023-09-21-005. All participants signed a consent form before participating in the study. After completion, they were instructed to remove outer layers of clothing to expose their arms, ideally wearing a t-shirt to homogenize clothing among participants as much as possible. Tab.~\ref{tab:demographics} presents demographic data among both groups.

\begin{table}[]
\centering
\begin{tabular}{@{}lcccc@{}}
\toprule
\textbf{Scenario} & \multicolumn{2}{c}{\textbf{Desert Island}} & \multicolumn{2}{c}{\textbf{Snowy Mountains}} \\ 
                             & Mean  & SD   & Mean  & SD   \\ \midrule
\textbf{Age}                 & 26.83 & 11.4 & 26.58 & 9.67 \\
\textbf{VR Proficiency}          & 1.71  & 1.23 & 1.92  & 1.10 \\
\textbf{Video Game Practice} & 2.38  & 1.50 & 2.58  & 1.56 \\
\textbf{Cold Sensitivity}    & 3.04  & 1.52 & 2.83  & 1.58 \\
\textbf{Warm Sensitivity}    & 2.96  & 1.68 & 3.167 & 1.66 \\ \midrule
\textbf{Gender}   & \multicolumn{2}{c}{11F; 12M; 1 Other}     & \multicolumn{2}{c}{8F; 15M; 1 Other}         \\ \bottomrule
\end{tabular}
\caption{Gender, Age, VR Proficiency (0 = Never Used, 4 = Expert), Video Game Practice (0 = Never, 4 = Regularly) and Thermal Sensitivity (0 = Completely Disagree, 6 = Completely Agree) of participants for each scenario.}
\label{tab:demographics}
\end{table}

\subsection{Apparatus}
\subsubsection{Thermal Feedback System}
% To provide an ambient thermal feedback, we opted for a non-wearable approach, fully described in previous work~\cite{villenaveDynamicModularThermal2025} and illustrated in ~\ref{fig:vr-setup}. Heat is delivered via 300W infrared lamps, which prioritize thermal stimulation while minimizing light output. Cold feedback is provided by axial fans, which create a cooling effect by replacing warm air around the skin with ambient air, enhancing evaporative cooling. Control is handled via the DMX-512 protocol, and we built a custom DMX controller connected to the VR computer via USB allowing the Unity application to control the thermal actuators. Infrared lamps are powered through a 4-channel DMX dimmer, while fans, equipped with built-in controllers, are daisy-chained via 3-pin XLR cables for synchronized operation. Our setup was shown to be able to provide controllable levels of warmth and cold sensations to the users. To ease replicability, hardware requirements and software are open-source on GitHub~\footnote{https://github.com/liris-xr/6dof-vr-thermal-feedback}.

\begin{figure}
    \centering
    \includegraphics[width=0.95\linewidth]{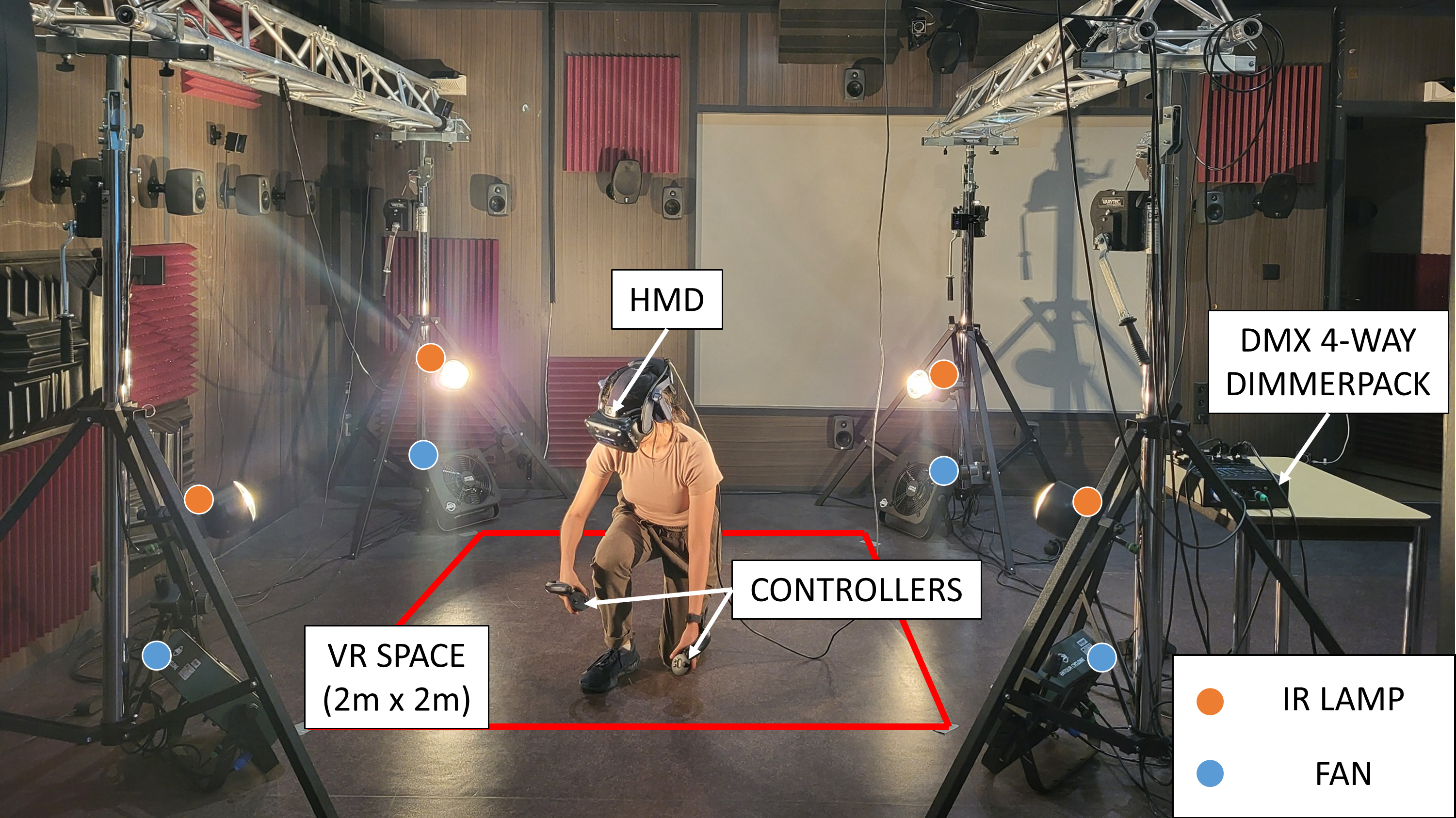}
    \caption{VR Setup with Thermal Feedback System}
    \label{fig:vr-setup}
\end{figure}

To provide ambient thermal feedback, we adopted a non-wearable solution previously developed and validated for use in 6DoF VR environments~\cite{villenaveDynamicModularThermal2025}, and illustrated in Figure~\ref{fig:vr-setup}. The system delivers warmth through 300\textit{W} infrared lamps that minimize light output while maximizing thermal effect. Cooling is achieved using axial fans that enhance evaporative heat loss by displacing warm air from the skin with ambient air. The entire setup is controlled via the DMX-512 protocol: a custom-built DMX controller interfaces with the VR computer via USB, enabling real-time actuation from the Unity environment. Infrared lamps are regulated through a 4-channel DMX dimmer, while fans equipped with integrated controllers are daisy-chained using 3-pin XLR cables for synchronized operation. This system enables reproducible and adjustable thermal sensations for both warmth and cold and perceptual characterization results (see Appendix. A in \textit{Supplementary Materials}). All hardware specifications and control software are open-source to facilitate reuse and adaptation by the research community\footnote{\url{https://github.com/liris-xr/6dof-vr-thermal-feedback}}.

\subsubsection{VR Setup}
Hardware was selected to provide a reliable and immersive VR experience. For the visual immersion we used a HTC Vive Pro 2 at Ultra settings (4896 x 2448 @ 90Hz) and interaction was enabled by the Valve Index Controllers, chosen for their ergonomics and intuitive design. We configured SteamVR in \textit{roomscale} mode (2\textit{m} x 2\textit{m}) allowing participants a freedom of movement within the real space but with limited real locomotion. By keeping the play area compact, we increase the reliability of the thermal feedback to stay effective and consistent across participants. Moreover, this setup is above the minimal recommendations for room-scale VR and can be easily replicated, ensuring broad accessibility of our findings. The computer used to render the virtual environment in the HMD featured a high-end CPU (ref. Intel-Core i9-9900KF) and GPU (ref. NVidia RTX 3080ti). Audio was rendered through active noise cancelling headphones (ref. Sony WH-1000XM5).

\subsubsection{Valve Index Controller Tutorial}
\label{subsection:Tutorial}
For this experiment, we used Valve Index controllers, which we see as a good compromise between standard Steam VR controllers (HTC Vive Wand) and hand tracking. Valve Index controllers are strapped to the user's hands and use a tactile zone instead of a trigger for the grab interaction. Thus, participants maintain a relaxed hand position at all times, reducing fatigue, and can use their entire hand to grasp virtual objects, instead of just the middle finger, enhancing both realism and intuitiveness of interaction. Compared to computer-vision based hand tracking, users do not need to have their hands in front of the HMD's sensors during the entire experience. Moreover, controllers provide generic passive and active haptic interfaces, as the controller in itself acts as a physical proxy for the virtual object, and vibrations when hovering provide a clear indicator of affordance. To ensure that participants are as familiar as possible with these controllers, and to limit the bias of VR proficiency on performance when performing our highly interactive scenarios, we designed a short dedicated tutorial. Beginning with instructions to walk within the virtual scene and reach four markers positioned near the boundaries of the play area. This task was designed to help participant detecting the virtual limits of the calibrated play space and adjusting to the immersive environment. Next, a visual representation of their hand appears. Participants are asked to move their real hand in space so as to link their hands' proprioception with the virtual hands' visuals. Following this, a 3D model of the controllers is introduced. Each button is highlighted in red one by one, prompting the user to press the corresponding button to deactivate the highlight. The virtual hand animates accordingly, reinforcing the link between physical input and visual feedback. The tutorial then introduces locomotion mechanics through tooltips above the virtual controller. Given the limited physical play space, but the need for participants to explore the virtual environment, they learn to perform snap rotation (i.e., quick 45° rotations using the joystick) and teleportation. Participants must traverse a virtual distance of 15 meters using this method. Finally, the tutorial concludes with interaction training. The subsequent scenarios only need for the participant to know how to grab objects to perform the required actions, they practice this by grabbing 3 virtual cubes and are encouraged to play with the cubes by stacking or throwing them. A video illustrating this tutorial is available in the \textit{Supplementary Materials}.

\subsection{Virtual Environments}
\label{subsection:VirtualEnvironments}
Virtual environments were developed with Unity 2022.3.21f and VR aspects of the experiment managed by OpenXR and XR Interaction Toolkit plugins. These VEs were designed to maximize the user’s sense of presence using high-quality visual and auditory stimuli. To ensure visual coherence, each environment was carefully designed with consistent mesh quality, detailed terrain, and realistic objects. Audio was implemented with non-spatialized ambient sound to reflect the overall atmosphere and localized spatialized sounds to enhance immersion and make user interactions feel more impactful. To align with previous studies investigating the influence of thermal stimuli on user experience~\cite{ranasinghe_season_2018, hanHapmosphereSimulatingWeathers2019a}, we created stereotypically thermal environments, i.e., environments that evoke universal thermal sensations. Specifically, each environment provides two distinct intensity levels of thermal feedback in the dynamic condition: one for when participants are exposed and another for when they are sheltered. For locomotion, participants could teleport over the entire surface of the terrain as it was marked as a teleportation area. Teleportation limits the risk of cybersickness by slightly reducing presence and the ability to orient oneself in the VE~\cite{alzayerVirtualLocomotionSurvey2020}, but was well-suited for guided scenario such as the ones proposed in this experiment. Finally, both scenarios were structured to ensure that the timing and sequence of thermal stimulations remained comparable between the two environments. The scenarios were intentionally designed as linear as possible, making them easy to understand while minimizing unnecessary exploration. Each step of the scenario was clearly displayed on a floating canvas at the bottom of the screen to guide the user.

\begin{figure}[ht]
  \captionsetup[subfloat]{labelformat=empty}
  \centering
  \subfloat[]{
    \hspace{-5pt}
    \includegraphics[width=0.2\linewidth]{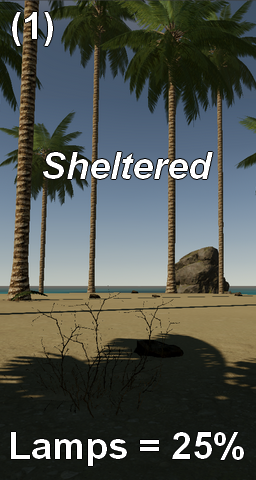}
    \hspace{-11pt}
  }
  \subfloat[]{
    \includegraphics[width=0.2\linewidth]{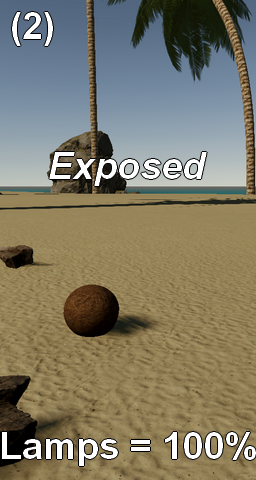}
    \hspace{-11pt}
  }
  \subfloat[]{
    \includegraphics[width=0.2\linewidth]{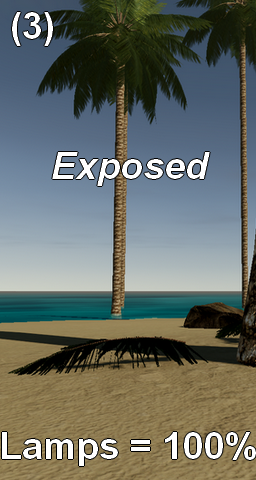}
    \hspace{-11pt}
  }
  \subfloat[]{
    \includegraphics[width=0.2\linewidth]{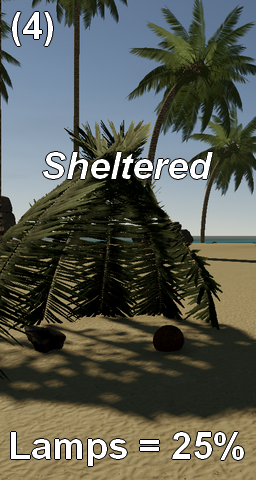}
    \hspace{-11pt}
  }
  \subfloat[]{
    \includegraphics[width=0.2\linewidth]{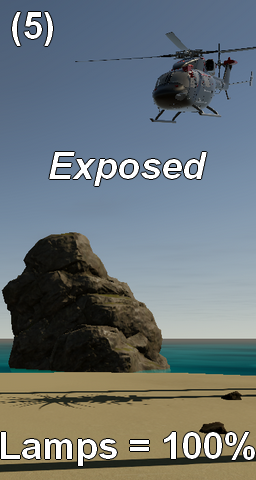}
  }
\caption{Walk-through the Desert Island Scenario.}
\label{fig:walkthrough_desertisland}
\end{figure}

\textbf{Desert Island Scenario}
Participants begin the experience by waking up stranded on a deserted island following a plane crash (1 - sheltered). As the sun intensifies, they are prompted to search for hydration. The scenario points out the presence of coconuts, and instructs participants to use one of the surrounding stone with force to break them and drink the water inside (2 - exposed). Once hydrated, participants are instructed to build a small shelter to protect themselves from the increasing heat. To this end, they have to gather large palm leaves scattered around the trees and plant them in the sand in the indicated place points (3 - exposed). After successfully constructing the hut, they are instructed to hide from the sun inside, where they need to open a second coconut and drink from it while waiting for rescue (4 - sheltered). Eventually, a rescue helicopter approaches the island, and participants are required to make their way toward it to complete the scenario (5 - exposed). Fig.~\ref{fig:walkthrough_desertisland} depicts this scenario.

\begin{figure}[ht]
  \captionsetup[subfloat]{labelformat=empty}
  \centering
  \subfloat[]{
    \hspace{-5pt}
    \includegraphics[width=0.2\linewidth]{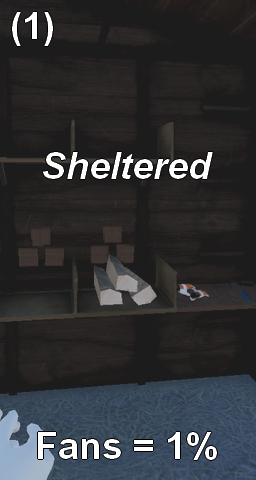}
    \hspace{-11pt}
  }
  \subfloat[]{
    \includegraphics[width=0.2\linewidth]{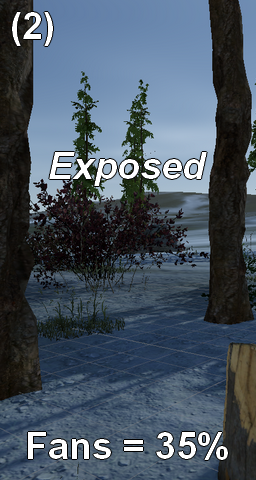}
    \hspace{-11pt}
  }
  \subfloat[]{
    \includegraphics[width=0.2\linewidth]{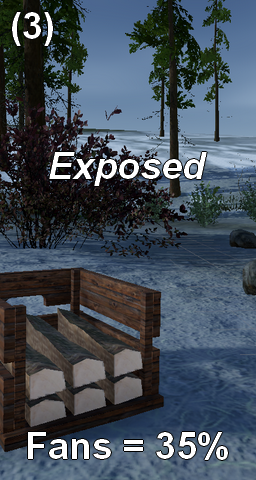}
    \hspace{-11pt}
  }
  \subfloat[]{
    \includegraphics[width=0.2\linewidth]{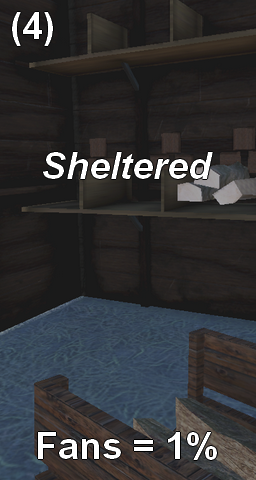}
    \hspace{-11pt}
  }
  \subfloat[]{
    \includegraphics[width=0.2\linewidth]{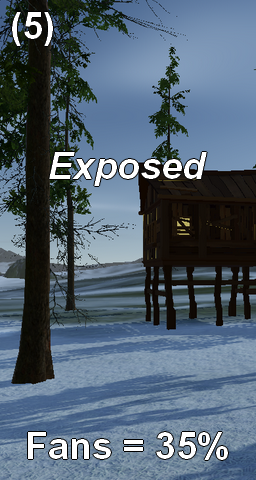}
  }
\caption{Walk-through the Snowy Mountains scenario.}
\label{fig:walkthrough_snowymountains}
\end{figure}

\textbf{Snowy Mountains Scenario}
As they enter the VE, participants find themselves immersed in a cabin surrounded by a snow-covered forest, indicative of the fast-approaching winter. The scenario prompts them to collect burning wood in preparation for the coming cold, and require them to put on gloves and retrieve their axe (1 - sheltered). Once equipped, participants get out of cabin, head for the trees and strike them with multiple axe blows to bring them down (2 - exposed). The fallen trees leave behind logs, which participants must bring inside a crate to ease transport (3 - exposed). They then carry the crate full of wood inside the hut, and stack the logs in piles on the shelves (4 - sheltered). After completing the piles, participants have to go outside the cabin and return to their home in order to conclude the experience (5 - exposed). Fig.~\ref{fig:walkthrough_snowymountains} depicts this scenario.

\textbf{Thermal Stimuli Selection}
In preparation for this study, we calibrated our thermal feedback system to propose stimulus intensities consistent with our VEs. For this, 10 participants were immersed in the VEs described above. They were asked to fine-tune the intensity of our thermal feedback system using sliders going from 0 to 50 for the fans and 0 to 100 for the lamps. For each scenario, they were asked to determine their ideal thermal feedback for two positions: ``exposed'' and ``sheltered'' (see Section~\ref{subsection:VirtualEnvironments}). Using this calibration as a guide we established the different intensity levels of the devices as shown in Tab.~\ref{tab:intensity_levels}.

\begin{table}[]
\centering
\begin{tabular}{@{}lcccc@{}}
\toprule
 &
  \multicolumn{2}{c}{\textbf{\begin{tabular}[c]{@{}c@{}}Snowy Mountains (fans)\end{tabular}}} &
  \multicolumn{2}{c}{\textbf{\begin{tabular}[c]{@{}c@{}}Desert Island (lamps)\end{tabular}}} \\
 & Sheltered   & Exposed  & Sheltered   & Exposed   \\ \midrule
\textbf{Static}   & \multicolumn{2}{c}{35} & \multicolumn{2}{c}{100} \\ \midrule
\textbf{Dynamic}  & 1           & 35       & 25          & 100       \\ \bottomrule
\end{tabular}
\caption{Intensity levels (in \%) of the system depending on the VE, the experimental condition and the user position within the VE.}
\label{tab:intensity_levels}
\end{table}

\subsection{Experience Design}
Our study employed a within-subjects experimental design to control the influence of individual differences. Each participant was exposed to the three different conditions for one scenario. The order in which participants experienced these conditions was counterbalanced using a balanced latin square to mitigate potential order effects. Experimental conditions were as follows:

\begin{itemize}
\item \textbf{Control}: audio-visual + vibrations in controllers
\item \textbf{Static}: audio-visual + vibrations in controllers + congruent thermal feedback where the intensity remains the same throughout the scenario.
\item \textbf{Dynamic}: audio-visual + vibrations in controllers + congruent thermal feedback where intensity varies depending on user's actions in the VE.
\end{itemize}

Tab.~\ref{tab:intensity_levels} specifies thermal feedback intensity levels used for each VEs and for both the Static and Dynamic conditions.

\subsection{Procedure}
\begin{figure*}[ht]
    \centering
    \includegraphics[width=\linewidth]{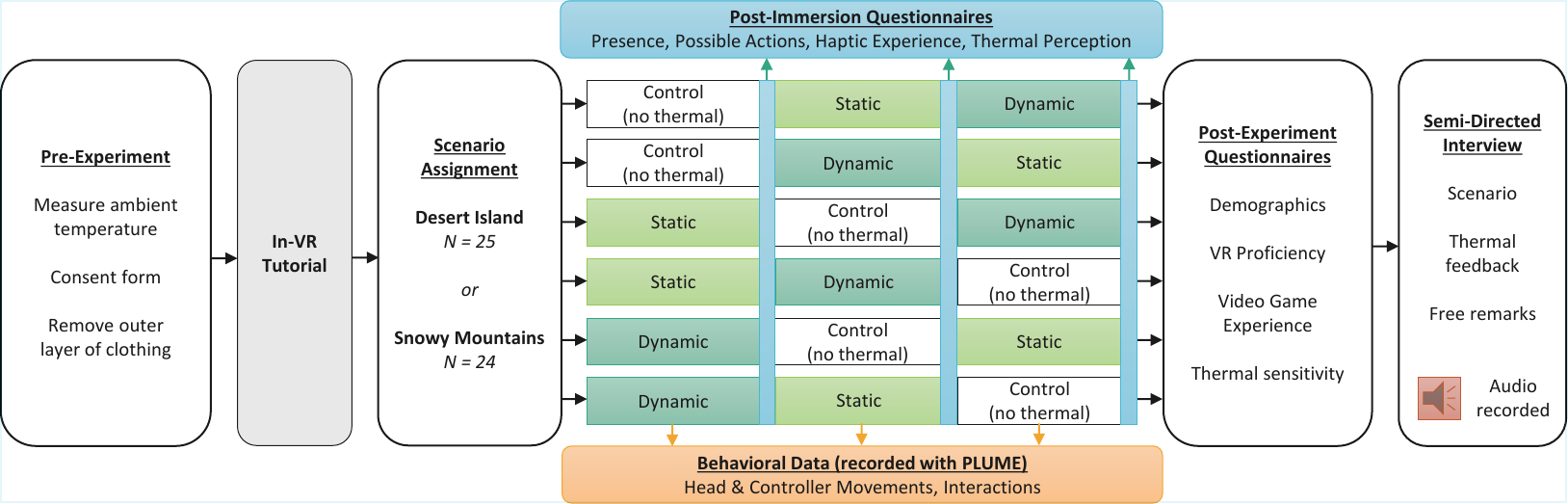}
        \caption{Experimental protocol.}
        \label{fig:experimental_protocol}
\end{figure*}

The experimental procedure is summarized in Figure~\ref{fig:experimental_protocol}. Upon arrival, participants first completed an informed consent form. They then engaged in a guided VR tutorial designed to familiarize them with the Valve Index Controllers and the interaction mechanisms implemented in the scenarios—specifically, object grabbing and teleportation (see Subsection~\ref{subsection:Tutorial}). Participants were randomly assigned to one of the two scenarios and were subsequently immersed in each of the three experimental conditions. During each VR session, interactions and movements were recorded using PLUME~\cite{javerliatPLUMERecordReplay2024}. After each condition, participants exited the VE to complete a set of post-immersion questionnaires assessing presence, sense of possible actions, haptic experience, and thermal perception. Upon completion of the third and final condition, participants answered post-experiment questionnaires where we collected participant profile data, including demographics (age and gender), prior experience with VR and video games (rated on a 5-point scale: 1 = Very Low to 5 = Very High), and thermal sensitivity. The latter was measured using 2 items from Van Someren et al.’s Thermal Sensitivity and Regulation Questionnaire~\cite{vansomerenExperiencedTemperatureSensitivity2015}:
\textit{“People differ in how quickly or intensely they experience warmth / cold. Indicate how quickly you experience warmth / cold compared to others.”}
(rated from -3 = Much Later to +3 = Much Quicker). To conclude the experiment, participants took part in a recorded semi-structured interview. This interview aimed to gather richer qualitative feedback regarding their overall experience, including perceived enjoyment, usability, cybersickness, and general quality of experience (QoE). The total duration of the experiment was approximately 55 minutes per participant.

\subsection{Post-Immersion Questionnaires}
To measure user experience depending on the thermal stimulation level of quality, participants completed a series of questionnaires after each condition. Thermal haptic experience was measured using the haptic experience questionnaire by Anwar et al.~\cite{anwarFactorsHapticExperience2023} alongside the ASHRAE55 7-point Likert scale~\cite{ashrae2004} that measure perceived thermal sensation (ranging from -3 = Very Cold to +3 = Very Hot). Participants also rated the intensity of thermal variation using a custom 4-point scale (0 = None, 1 = Small, 2 = Moderate, 3 = Large). Given that adding thermal feedback to the VR experience is meant to enhance immersion which subsequently improves presence in VR~\cite{skarbezSurveyPresenceRelated2017}, we measured spatial and general presence to relate to previous results using the general presence item of the IPQ~\cite{schubertExperiencePresenceFactor2001} and the Self-Location subscale of the SPES~\cite{hartmannSpatialPresenceExperience2015}. Moreover, because the primary distinction between the Static and Dynamic conditions lies in the user's control over the thermal stimulus, we hypothesized that sense of agency would differ across conditions. To evaluate this, we included the Possible Actions subscale of the SPES~\cite{hartmannSpatialPresenceExperience2015}.

\subsection{Behavioural Indicators}
Using PLUME~\cite{javerliatPLUMERecordReplay2024}, we recorded the experimental sessions to perform comprehensive VR scene logging, including but not limited to the transforms (position, rotation, and scale) of all scene objects for each frame, user input actions, and interaction events as defined by Unity’s XR Interaction Toolkit (XRITK). From these recordings, we extracted user interactions with virtual objects as well as the physical and virtual positions and rotations of the headset and controllers. Positions are expressed in meters and rotations in quaternions. These data were used to compute behavioral described in the following paragraphs.

\subsubsection{Number of Teleportations}
To quantify locomotion within the virtual environment, we recorded the number of teleportations performed by each participant. In our implementation, teleportation is triggered by selecting designated teleportation areas (such as the terrains in our scenes) using the ray-cast interaction system. Therefore, the total number of teleportation events corresponds to the number of successful terrain selections during the experience.

\subsubsection{Physical and virtual Distances}
To quantify users' physical movement in the real world as well as their virtual locomotion within the VE, we extracted the position of the HMD using the Transform of the virtual camera GameObject. Physical movement within the calibrated play area (e.g., walking, crouching) was assessed using the camera's local Transform, defined relative to the VR setup origin during calibration. In contrast, virtual locomotion within the VE which includes physical movement combined with teleportation and snap rotation, was measured using the world Transform, defined relative to the origin of the VE. For each session, we computed both the physical and virtual total distances traveled by summing the Euclidean distances between consecutive frames. We also computed the physical and virtual total rotation by summing angular distances between consecutive frames.

\subsubsection{Roaming Entropy}
\begin{figure}[]
    \captionsetup[subfigure]{justification=centering}
    \centering
    \subfloat[Low Roaming Entropy \\ (Participant 4, RE = 0.71)]{
        \includegraphics[width=0.45\linewidth, trim={3cm 4cm 6cm 2cm}, clip]{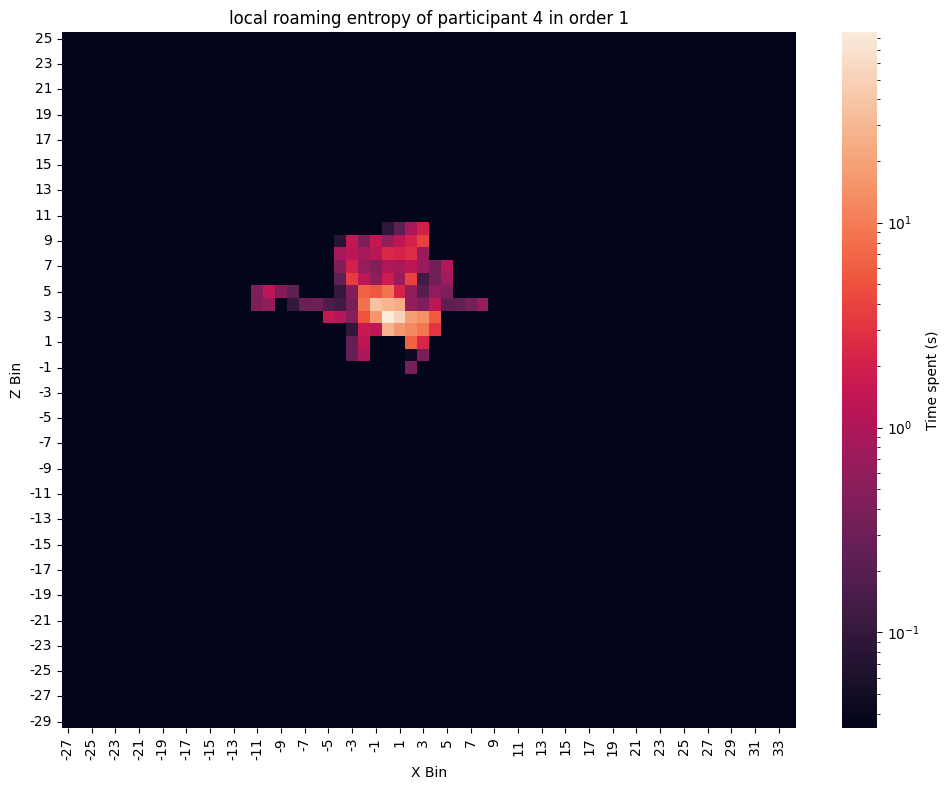}
        \label{fig:low_roaming_entropy}
    }
    \subfloat[High Roaming Entropy \\ (Participant 25, RE = 0.88)]{
        \includegraphics[width=0.45\linewidth, trim={3cm 4cm 6cm 2cm}, clip]{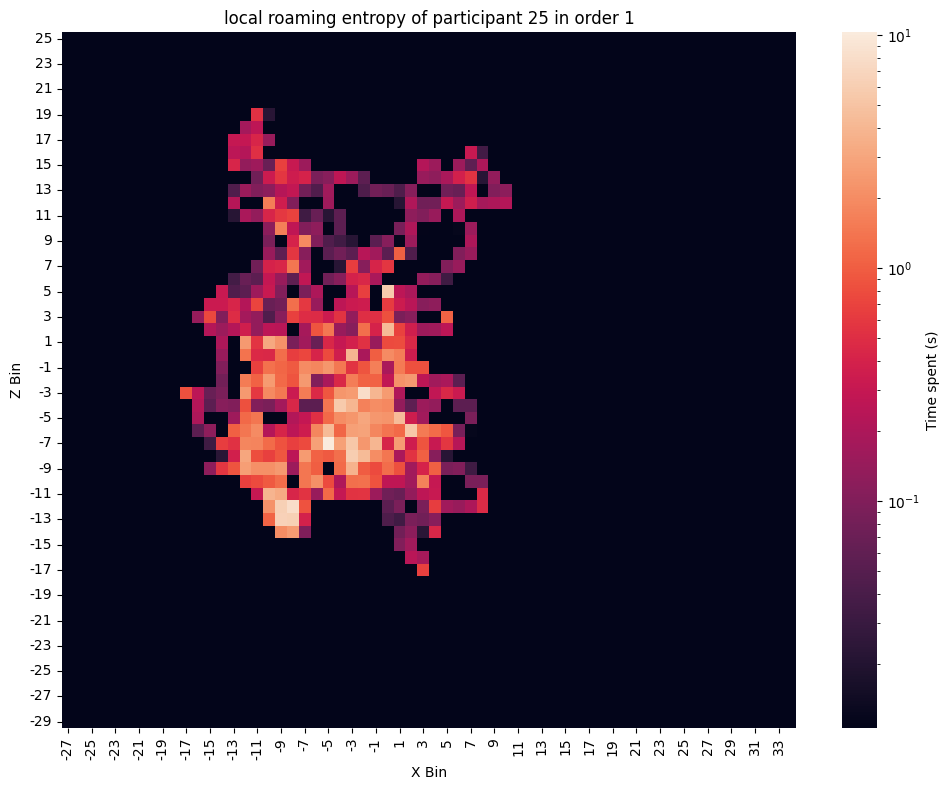}
        \label{fig:high_roaming_entropy}
    }
    \caption{Projection heatmaps (on X-Z plane) of probability of presence in the physical space for 2 different users.}
    \label{fig:physical_roaming_entropy}
\end{figure}

Roaming Entropy (RE) as defined by Freund et al.~\cite{freund_emergence_2013} is a quantitative measure of the spatial diversity of exploration. It is calculated by discretizing the explored space into a 3D grid, with each cell representing a fixed volume of the virtual environment. The time spent in each cell is computed, then transformed into a probability of presence by dividing this time by the total duration of immersion. These probabilities are then used to calculate the Shannon entropy as:
\[
RE = -\sum_{i=1}^{N} {p}_{i} log({p}_{i})
\]

where $p_i$ is the probability of occurrence (presence) in cell $i$, and $N$ the total number of cells. As illustrated in Figure~\ref{fig:physical_roaming_entropy}, a low RE reflects a restricted and repetitive exploration of certain areas, while a high RE indicates a more extensive and homogeneous exploration of the environment. The cell size was defined as follows: for the physical space, one cell measures 1~cm × 1~cm × 1~cm, and for the virtual space, one cell measures 1~m × 1~m × 1~m. The grid size was determined as the bounding box encompassing all recorded positions, to ensure complete spatial coverage while avoiding unnecessary computational overhead, given the vast area available to participants. Respectively for the physical and virtual spaces of Desert Island, the resulting grid dimensions are 3.1~m x 2.1~m x 2.8~m and 134~m x 8~m x 121~m, and for Snowy Mountains the resulting grid dimensions are 3.1~m x 1.9~m x 2.4~m and 142~m x 5~m x 52~m.

\subsubsection{Head Movement Entropy}
Analogous to roaming entropy, movement entropy captures the diversity of head orientations throughout the experience. It is computed based on head rotations, focusing on the pitch (vertical axis) and yaw (horizontal axis), which were discretized into 36 and 18 bins, respectively, each spanning 10°, thereby covering the full accessible range of 360° and 180°. Unlike spatial position, head orientation is bounded within absolute limits, making binning straightforward and independent of the recorded data distribution. Each recorded head rotation expressed as a quaternion is converted into Euler angles and the frame time added to its corresponding bin. Entropy was then computed using the same formulation described above for roaming entropy.
\section{Results}
\label{Results}
For each scenario, we study the influence of two factors on post-immersion questionnaires and behavioral indicators :
\begin{itemize}
    \item Thermal condition (Control, Static, Dynamic).
    \item Order of passage (1st, 2nd, 3rd).
\end{itemize}

Statistical analyses were performed using R~\cite{r2024} and \(\alpha = 0.05\). The normality assumption was not met for any dependent variable or behavioral indicator, so we processed data using non-parametric Friedman tests. If the Friedman test proved significant, post-hoc pairwise and paired Wilcoxon tests with Holm corrections were used. We also computed Spearman’s rank-order correlations to explore monotonic relationships between demographics and questionnaire results. This section highlights significant results; complementary statistical results are available in the supplementary materials.

\begin{figure}[]
    \vspace{-0.5cm}
    \subfloat[Thermal Sensation]{
        \includegraphics[width=0.49\linewidth, trim=20pt 0 0 0]{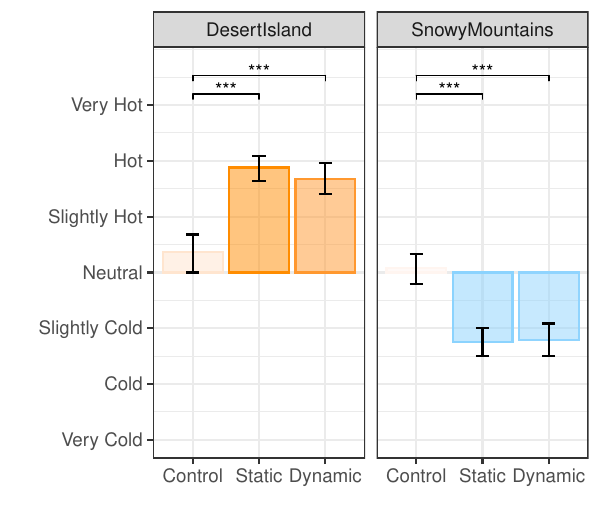}
        \label{thermal_sensation}
    }
    \subfloat[Thermal Variation]{
        \includegraphics[width=0.49\linewidth, trim=20pt 0 0 0]{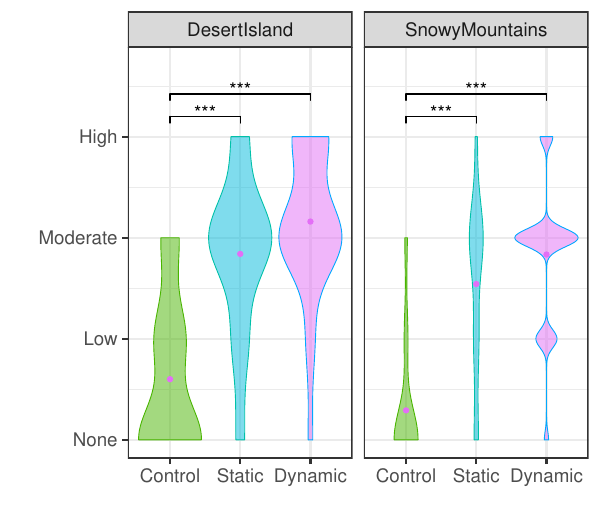}
        \label{thermal_variation}
    }
    \caption{Thermal Perception}
    \label{fig:thermals}
\end{figure}

\subsection{Thermal Perception (Fig.~\ref{fig:thermals})}
\textbf{Thermal Sensation.} For Desert Island, participants felt hot in both thermal conditions, since paired tests revealed significant differences between Control and Static (\scalebox{0.8}{\begin{math}p = 1.3 \cdot 10^{-4}\end{math}}), as well as between Control and Dynamic (\scalebox{0.8}{\begin{math}p = 3.2 \cdot 10^{-4}\end{math}}). We obtained similar results for Snowy Mountains: participants felt slightly cold as paired tests revealed significant differences between Control and Static (\scalebox{0.8}{\begin{math}p = 1.9\cdot 10^{-4}\end{math}}), as well as between Control and Dynamic (\scalebox{0.8}{\begin{math}p = 3.3 \cdot 10^{-4}\end{math}}).
Order of passage had no effect on thermal sensation for both scenarios.

\textbf{Thermal Variation.} For Desert Island, participants felt moderate variation in both thermal conditions, since paired tests revealed significant differences between Control and Static (\scalebox{0.8}{\begin{math}p = 6.6 \cdot 10^{-4}\end{math}}), as well as between Control and Dynamic (\scalebox{0.8}{\begin{math}p = 6.6 \cdot 10^{-4}\end{math}}). We obtained similar results for Snowy Mountains: paired tests revealed significant differences between Control and Static (\scalebox{0.8}{\begin{math}p = 2.8 \cdot 10^{-4}\end{math}}), as well as between Control and Dynamic (\scalebox{0.8}{\begin{math}p = 2.5 \cdot 10^{-4}\end{math}}). Order of passage had no effect on thermal variation for both scenarios.

\begin{figure}[ht]
\vspace{-0.5cm}
  \centering
  \subfloat[Realism]{
    \includegraphics[width=0.49\linewidth, trim=20pt 0 0 0]{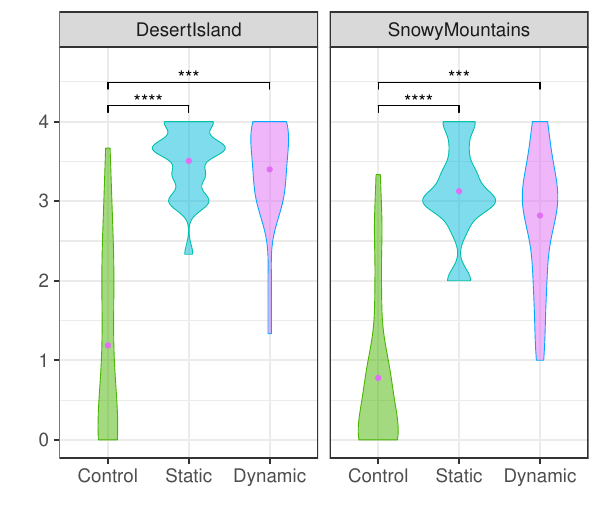}
    \label{hx:realism}
  }
  \subfloat[Expressivity]{
    \includegraphics[width=0.49\linewidth, trim=20pt 0 0 0]{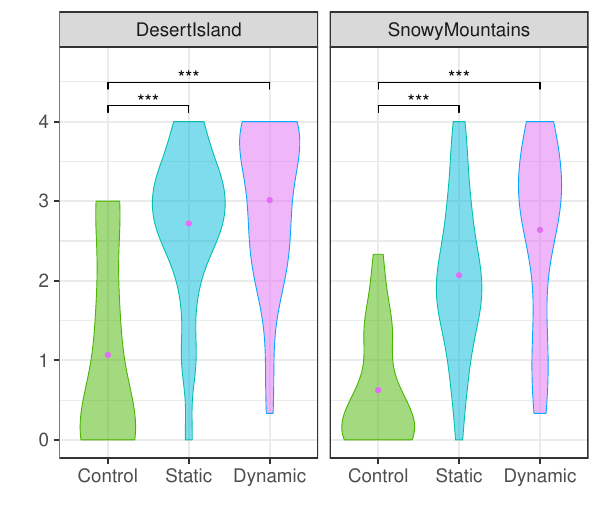}
    \label{hx:realism}
  }
  
  \subfloat[Harmony]{
    \includegraphics[width=0.49\linewidth, trim=20pt 0 0 0]{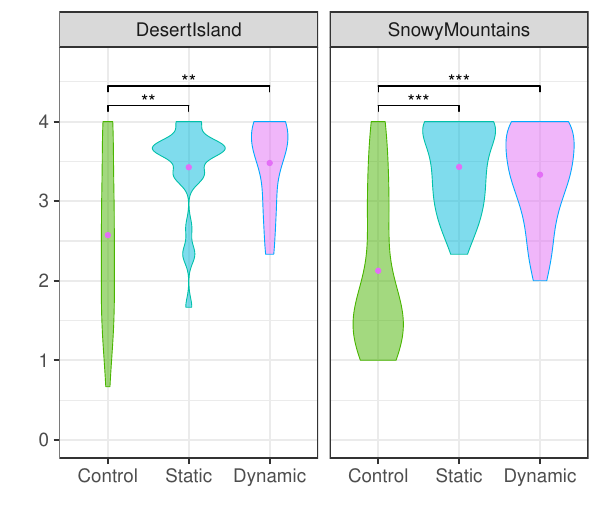}
    \label{hx:realism}
  }
  \subfloat[Involvement]{
    \includegraphics[width=0.49\linewidth, trim=20pt 0 0 0]{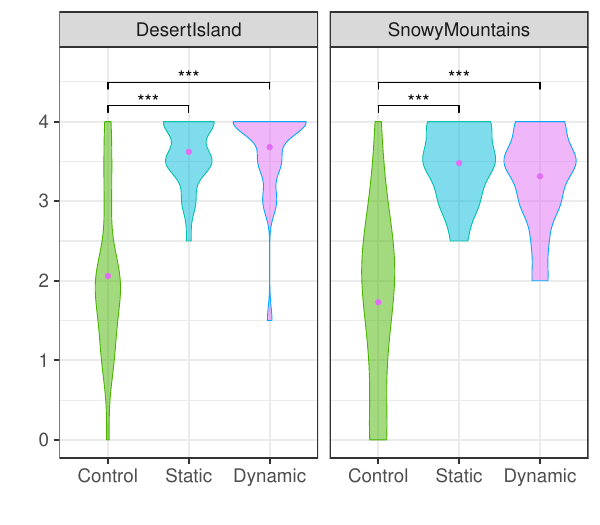}
    \label{hx:realism}
  }
\caption{Haptic Experience}
\label{fig:hx}
\end{figure}

\subsection{Haptic Experience (Fig.~\ref{fig:hx})}

\textbf{Realism.} For Desert Island, participants felt the realism of thermal feedback was higher in thermal conditions as paired tests revealed significant differences between Control and Static (\scalebox{0.8}{\begin{math}p = 5.7 \cdot 10^{-5}\end{math}}) and between Control and Dynamic (\scalebox{0.8}{\begin{math}p = 1.0 \cdot 10^{-4}\end{math}}). Similar results for Snowy Mountains were obtained as paired tests revealed significant differences between Control and Static (\scalebox{0.8}{\begin{math}p = 5.6 \cdot 10^{-5}\end{math}}), as well as between Control and Dynamic (\scalebox{0.8}{\begin{math}p = 1.8 \cdot 10^{-3}\end{math}}). Order of passage had no effect on realism for both scenarios.

\textbf{Expressivity.} For Desert Island, participants felt the expressivity of thermal feedback was higher in thermal conditions as paired tests revealed significant differences between Control and Static (\scalebox{0.8}{\begin{math}p = 5.3 \cdot 10^{-4}\end{math}}), as well as between Control and Dynamic (\scalebox{0.8}{\begin{math}p = 5.3 \cdot 10^{-4}\end{math}}). Similar results for Snowy Mountains were obtained as paired tests revealed significant differences between Control and Static (\scalebox{0.8}{\begin{math}p = 1.9 \cdot 10^{-4}\end{math}}), as well as between Control and Dynamic (\scalebox{0.8}{\begin{math}p = 1.9 \cdot 10^{-4}\end{math}}). Order of passage had no effect on expressivity for both scenarios.

\textbf{Harmony.} For Desert Island, participants felt the harmony of thermal feedback with the rest of the VE was higher in thermal conditions as paired tests revealed significant differences between Control and Static (\scalebox{0.8}{\begin{math}p = 5.0 \cdot 10^{-3}\end{math}}), as well as between Control and Dynamic (\scalebox{0.8}{\begin{math}p = 5.0 \cdot 10^{-3}\end{math}}). Similar results for Snowy Mountains were obtained as paired tests revealed significant differences between Control and Static (\scalebox{0.8}{\begin{math}p = 4.4 \cdot 10^{-4}\end{math}}), as well as between Control and Dynamic (\scalebox{0.8}{\begin{math}p = 6.7 \cdot 10^{-4}\end{math}}). Order of passage had no effect on harmony for both scenarios.

\textbf{Involvement.} For Desert Island, participants felt the involvement created by thermal feedback was higher in thermal conditions as paired tests revealed significant differences between Control and Static (\scalebox{0.8}{\begin{math}p = 1.1 \cdot 10^{-4}\end{math}}), as well as between Control and Dynamic (\scalebox{0.8}{\begin{math}p = 1.1 \cdot 10^{-4}\end{math}}). Similar results for Snowy Mountains were obtained as paired tests revealed significant differences between Control and Static (\scalebox{0.8}{\begin{math}p = 1.7 \cdot 10^{-4}\end{math}}), as well as between Control and Dynamic (\scalebox{0.8}{\begin{math}p = 3.6 \cdot 10^{-4}\end{math}}). Order of passage had no effect on involvement for both scenarios.

\begin{figure}[ht]
\vspace{-0.5cm}
  \centering
  \subfloat[Spatial Presence]{
    \includegraphics[width=0.49\linewidth, trim=20pt 0 0 0]{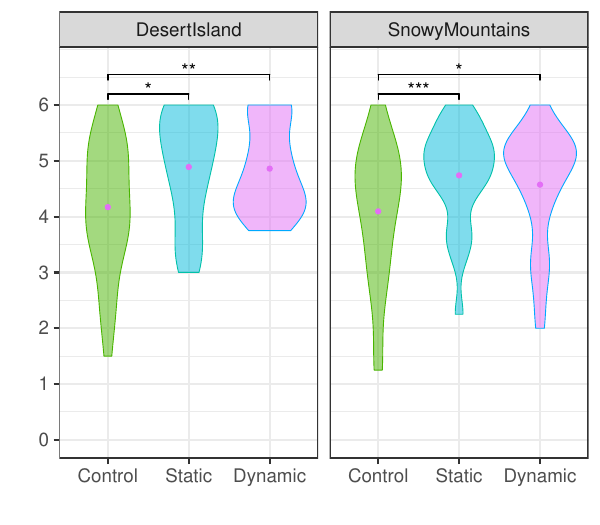}
    \label{spes:spatial_location}
  }
  \subfloat[Possible Actions]{
    \includegraphics[width=0.49\linewidth, trim=20pt 0 0 0]{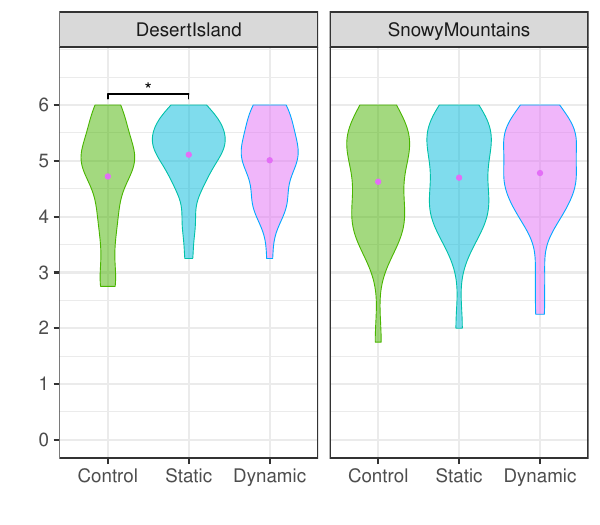}
    \label{spes:possible_actions}
  }
\caption{Spatial Presence Experience Scale}
\label{fig:spes}
\end{figure}

\subsection{VR experience (Fig.~\ref{fig:spes})}

\textbf{General Presence.} For Desert Island, participants felt more present in thermal conditions as paired tests revealed significant differences between Control and Static (\scalebox{0.8}{\begin{math}p = 1.0 \cdot 10^{-3}\end{math}}), as well as between Control and Dynamic (\scalebox{0.8}{\begin{math}p = 2.0 \cdot 10^{-3}\end{math}}). Order of passage had no effect on General Presence for Desert Island. Similar results for Snowy Mountains were obtained as paired tests revealed significant differences between Control and Static (\scalebox{0.8}{\begin{math}p = 4.0 \cdot 10^{-3}\end{math}}), as well as between Control and Dynamic (\scalebox{0.8}{\begin{math}p = 4.0 \cdot 10^{-3}\end{math}}). While the Friedman test for order of passage is significant (\scalebox{0.8}{\begin{math}p = .04 \end{math}}), no post-hoc tests revealed a significant difference between conditions.

\textbf{Self-Location.} For Desert Island, participant's sense of self-location was higher in thermal conditions as paired tests revealed significant differences between Control and Static (\scalebox{0.8}{\begin{math}p = 1.3 \cdot 10^{-2}\end{math}}), as well as between Control and Dynamic (\scalebox{0.8}{\begin{math}p = 7.0 \cdot 10^{-3}\end{math}}). Similar results for Snowy Mountains were obtained as paired tests revealed significant differences between Control and Static (\scalebox{0.8}{\begin{math}p = 1.0 \cdot 10^{-3}\end{math}}), as well as between Control and Dynamic (\scalebox{0.8}{\begin{math}p = 2.7 \cdot 10^{-2}\end{math}}). Order of passage had no effect on sense of self-location for both scenarios.

\textbf{Possible Actions.} For Desert Island, participant's sense of possible actions was higher in Static condition as paired tests revealed significant differences between Control and Static (\scalebox{0.8}{\begin{math}p = 1.8 \cdot 10^{-2}\end{math}}). No other tests proved significant for this questionnaire. Moreover, the order of passage had no effect on sense of possible actions for both scenarios.

\subsection{Interpersonal influences (Tab.~\ref{tab:correlations})}
For Desert Island scenario, videogame practice was negatively correlated with General Presence (\scalebox{0.8}{\begin{math}r = -0.31, p = 7.0 \cdot 10^{-3}\end{math}}) and Possible Actions (\scalebox{0.8}{\begin{math}r = -0.26, p = 2.2 \cdot 10^{-2}\end{math}}).
For Snowy Mountains scenario, age was negatively correlated with Thermal Variation (\scalebox{0.8}{\begin{math}r = -0.25, p = 3.6 \cdot 10^{-2}\end{math}}). Videogame practice was positively correlated with Thermal Variation (\scalebox{0.8}{\begin{math}r = 0.27, p = 2.1 \cdot 10^{-2}\end{math}}) and negatively correlated with General Presence (\scalebox{0.8}{\begin{math}r = -0.25, p = 3.7 \cdot 10^{-2}\end{math}}). Cold sensitivity was positively correlated with Realism (\scalebox{0.8}{\begin{math}r = 0.33, p = 2.2 \cdot 10^{-2}\end{math}}) and Harmony (\scalebox{0.8}{\begin{math}r = 0.32, p = 2.5 \cdot 10^{-2}\end{math}}).

\begin{table}[]
\centering
\resizebox{\linewidth}{!}{%
\begin{tabular}{@{}llllcc@{}}
\toprule
Scenario        & n  & Demographic & Questionnaire             & Correlation & p-value \\ \midrule
Desert Island   & 25 & Videogame Practice & General Presence   & -0.31       & 0.007 \\
Desert Island   & 25 & Videogame Practice & Possible Actions   & -0.26       & 0.022   \\ \midrule
Snowy Mountains & 24 & Age & Thermal Variation                 & -0.25       & 0.036   \\
Snowy Mountains & 24 & Cold Sensitivity & Realism              & 0.33        & 0.022   \\
Snowy Mountains & 24 & Cold Sensitivity & Harmony              & 0.32        & 0.025   \\
Snowy Mountains & 24 & Videogame Practice & Thermal Variation  & 0.27        & 0.021   \\
Snowy Mountains & 24 & Videogame Practice & General Presence   & -0.25       & 0.037   \\ \bottomrule

\end{tabular}%
}
\caption{Reports of significative correlations (\(p < 0.05\)) between post-immersion questionnaires and individual characteristics. Cold and Warmth sensitivity have been tested against questionnaire answers in thermal conditions only (i.e., Static and Dynamic).}
\label{tab:correlations}
\end{table}

\subsection{Behavioral Indicators (Tab.~\ref{tab:behavioral_metrics})}
\textbf{Number of Teleportations}
For Desert Island, participants teleported more during the 1st passage than during the 2nd and 3rd, as paired tests revealed significant differences between 1st and 2nd order (\scalebox{0.8}{\begin{math}p = 3.3 \cdot 10^{-2}\end{math}}) as well as between the 1st and 3rd (\scalebox{0.8}{\begin{math}p = 2.0 \cdot 10^{-4}\end{math}}).

\begin{figure}[]
    \captionsetup[subfigure]{justification=centering}
    \centering
    \subfloat[High Traveled Distance \\ (1st passage, TD = 745.4)]{
        \includegraphics[width=0.45\linewidth]{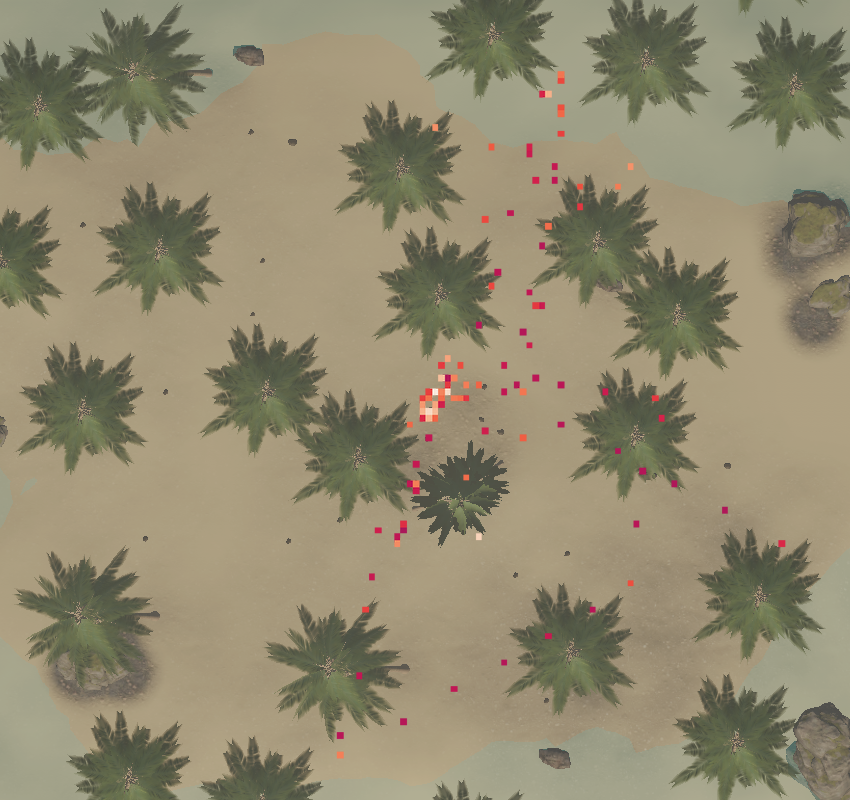}
        \label{fig:desert_island_high_roaming_entropy}
    }
    \subfloat[Low Traveled Distance \\ (3rd passage, TD = 211.0)]{
        \includegraphics[width=0.45\linewidth]{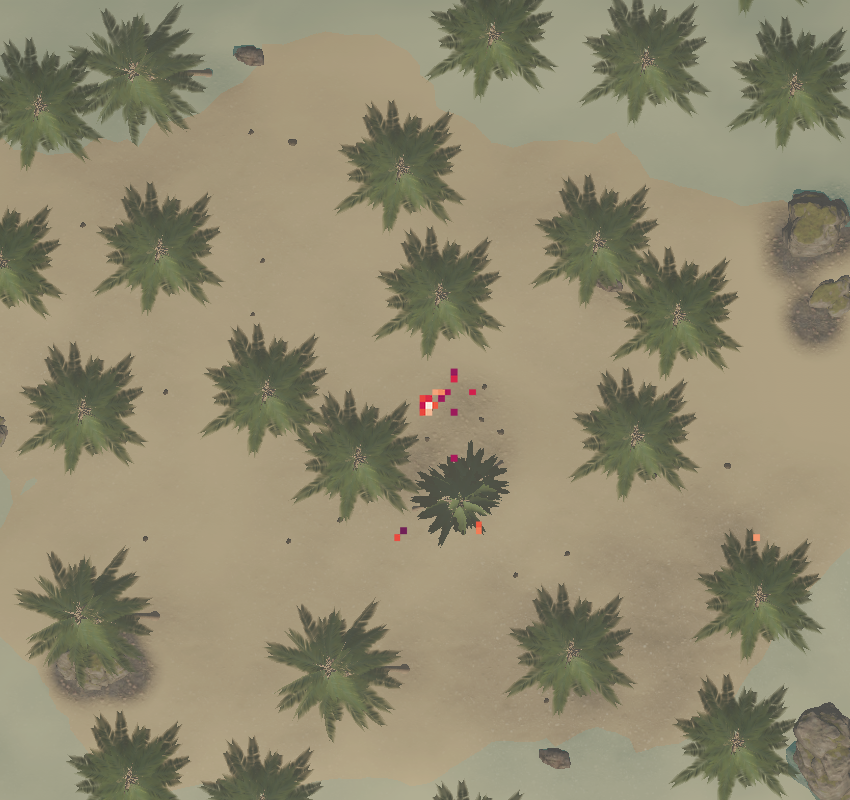}
        \label{fig:desert_island_low_roaming_entropy}
    }
    \caption{Projection heatmaps (on X-Z plane) of probability of presence in the virtual space for 2 different passages of participant 4 in Desert Island.}
    \label{fig:virtual_traveled_distance}
\end{figure}

\textbf{Distance}
For Desert Island, participant moved more in the real space and traveled more in the virtual space during the 1st passage than during the 2nd and more in the 2nd than during the 3rd. Fig.~\ref{fig:virtual_traveled_distance} provides an illustration of these results for Virtual Distance in Desert Island. Paired tests revealed significant differences between 1st and 2nd order for Physical Distance (\scalebox{0.8}{\begin{math}p = 9.1 \cdot 10^{-5}\end{math}}) and Virtual Distance (\scalebox{0.8}{\begin{math}p = 3.0 \cdot 10^{-3}\end{math}}), as well as between 2nd and 3rd order for Physical Distance (\scalebox{0.8}{\begin{math}p = 1.8 \cdot 10^{-6}\end{math}}) and Virtual Distance (\scalebox{0.8}{\begin{math}p = 5.6 \cdot 10^{-4}\end{math}}).
For Snowy Mountains, participant moved more in the real space during the 1st passage than during the 2nd and more in the 2nd than during the 3rd. Paired tests revealed significant differences for Physical Distance between 1st and 2nd (\scalebox{0.8}{\begin{math}p = 8.6 \cdot 10^{-4}\end{math}}) as well as between 2nd and 3rd (\scalebox{0.8}{\begin{math}p = 1.7 \cdot 10^{-2}\end{math}}). Thermal condition had no effect on distance metrics for both scenarios.

\textbf{Rotation}
For Desert Island, participant turned their head more in the real space and oriented themselves more in the virtual space during the 1st passage than during the 2nd and more in the 2nd than during the 3rd. Paired tests revealed significant differences between 1st and 2nd order for Physical Rotation (\scalebox{0.8}{\begin{math}p = 9.1 \cdot 10^{-5}\end{math}}), and Virtual Rotation (\scalebox{0.8}{\begin{math}p = 1.3 \cdot 10^{-4}\end{math}}), as well as between 2nd and 3rd order for Physical Rotation (\scalebox{0.8}{\begin{math}p = 1.8 \cdot 10^{-7}\end{math}}), and Virtual Rotation (\scalebox{0.8}{\begin{math}p = 3.6 \cdot 10^{-7}\end{math}}).
For Snowy Mountains, participant turned their head more in the real space and oriented themselves more in the virtual space during the 1st passage than during the 2nd and 3rd. Paired tests revealed significant differences between 1st and 2nd for Physical Rotation (\scalebox{0.8}{\begin{math}p = 4.5 \cdot 10^{-6}\end{math}}), and Virtual Rotation (\scalebox{0.8}{\begin{math}p = 4.5 \cdot 10^{-6}\end{math}}), as well as between 1st and 3rd order for Physical Rotation (\scalebox{0.8}{\begin{math}p = 1.1 \cdot 10^{-6}\end{math}}), and Virtual Rotation (\scalebox{0.8}{\begin{math}p = 3.5 \cdot 10^{-6}\end{math}}).
Thermal condition had no effect on rotation metrics for both scenarios.

\textbf{Roaming Entropy}
For Desert Island, participants expressed more exploratory behavior during the 1st passage than during the 2nd and the 3rd, as significant differences were found in Physical Roaming Entropy between 1st and 2nd order (\scalebox{0.8}{\begin{math}p = 7.0 \cdot 10^{-3}\end{math}}), as well as between 1st and 3rd order (\scalebox{0.8}{\begin{math}p = 5.0 \cdot 10^{-3}\end{math}}). Condition had no effect on Roaming Entropy for Desert Island. For Snowy Mountains, neither Condition or Order of Passage had any effect on Roaming Entropy.

\textbf{Head Movement Entropy}
For Desert Island, participants expressed more orientation / research behavior during the 1st and 2nd passage than during the 3rd, as significant difference were found between 1st and 3rd order for Physical Movement Entropy (\scalebox{0.8}{\begin{math}p = 1.8 \cdot 10^{-2}\end{math}}), as well as Virtual Head Movement Entropy (\scalebox{0.8}{\begin{math}p = 1.1 \cdot 10^{-2}\end{math}}). 
For Snowy Mountains, neither Condition or Order of Passage had any effect on Head Movement Entropy.

\begin{table*}[ht]
\centering
\label{tab:behavioral_metrics}
    \subfloat[Desert Island]{%
        \resizebox{0.5\linewidth}{!}{%
        \begin{threeparttable}
        \begin{tabular}{llcccccc}
        \toprule
        \textbf{Space} & \textbf{Metric} & \multicolumn{2}{c}{\textbf{First}} & \multicolumn{2}{c}{\textbf{Second}} & \multicolumn{2}{c}{\textbf{Third}} \\
                                         & & Mean & SD & Mean & SD & Mean & SD \\
        \midrule
         Physical & Distance (m)       & 47.4\textsubscript{a}  & 17.5 & 32.0\textsubscript{b}  & 10.4 & 27.7\textsubscript{c}  & 7.38 \\
         Virtual & Distance (m)       & 321.0\textsubscript{a} & 145  & 245.0\textsubscript{b} & 63.5 & 220.0\textsubscript{c} & 44.1 \\
         Physical & Rotation (°)       & 170.0\textsubscript{a} & 60.9 & 113.0\textsubscript{b} & 42.3 & 92.7\textsubscript{c}  & 24.4 \\
         Virtual & Rotation (°)       &  205.0\textsubscript{a} & 72.2 & 140.0\textsubscript{b} & 40.9 & 116.0\textsubscript{c} & 31.0 \\
         \midrule
         Physical & RE& 0.798\textsubscript{a} & 0.058 & 0.765\textsubscript{b} & 0.061 & 0.751\textsubscript{b} & 0.070 \\
         Virtual & RE &  0.807 & 0.043 & 0.802 & 0.052 & 0.796 & 0.082 \\
         Physical & HME & 0.883\textsubscript{a} & 0.027 & 0.872\textsubscript{a} & 0.028 & 0.866\textsubscript{b} & 0.031 \\
         Virtual & HME & 0.893\textsubscript{a} & 0.024 & 0.882\textsubscript{a} & 0.027 & 0.874\textsubscript{b} & 0.035 \\
         \midrule
         & Teleportation & 61.3\textsubscript{a} & 26.1 & 42.8\textsubscript{b} & 21.5 & 36.6\textsubscript{b} & 13.0 \\
        \bottomrule
        \end{tabular}
        \end{threeparttable}
        }
    }
    \subfloat[Snowy Mountains]{%
        \resizebox{0.5\linewidth}{!}{%
        \begin{threeparttable}
        \begin{tabular}{llcccccc}
        \toprule
        \textbf{Space} & \textbf{Metric} & \multicolumn{2}{c}{\textbf{First}} & \multicolumn{2}{c}{\textbf{Second}} & \multicolumn{2}{c}{\textbf{Third}} \\
                                         & & Mean & SD & Mean & SD & Mean & SD \\
        \midrule
         Physical & Distance (m)       & 50.8\textsubscript{a}  & 12.6 & 39.9\textsubscript{b}  & 16.0 & 35.4\textsubscript{c}  & 10.4 \\
         Virtual & Distance (m)       & 315.0 & 89.9 & 280.0 & 30.3 & 285.0 & 36.0 \\
         Physical & Rotation (°)       & 173.0\textsubscript{a} & 50.0 & 112.0\textsubscript{b} & 43.1 & 105.0\textsubscript{b} & 34.7 \\
         Virtual & Rotation (°)       & 221.0\textsubscript{a} & 62.2 & 148.0\textsubscript{b} & 49.0 & 147.0\textsubscript{b} & 43.9 \\
         \midrule
         Physical & RE & 0.830 & 0.052 & 0.829 & 0.042 & 0.820 & 0.046 \\
         Virtual & RE & 0.823 & 0.069 & 0.848 & 0.029 & 0.858 & 0.026 \\
         Physical & HME & 0.869 & 0.031 & 0.876 & 0.018 & 0.878 & 0.021 \\
         Virtual & HME & 0.874 & 0.033 & 0.885 & 0.018 & 0.887 & 0.016 \\
         \midrule
         & Teleportation & 55.4 & 17.8 & 45.4 & 12.0 & 48.9 & 12.9 \\
        \bottomrule
        \end{tabular}
        \end{threeparttable}
        }
    }
\caption{Behavioral metrics (Mean ± SD) depending on order of passage. Different superscripts ($^{a,b,c}$) in the same row indicate significant differences in the post hoc-tests \(p < 0.05\).}
\label{tab:behavioral_metrics}
\end{table*}

\subsection{Qualitative Feedback}
To complete the questionnaires and behavioral indicators presented above, we produced a qualitative analysis of the experience using theme analysis~\cite{braunUsingThematicAnalysis2006} on recorded semi-directed interviews. During them, participants described the tasks within the scenario, then elaborated on the thermal feedback they felt in each condition and evaluated the impact it had on their experience as a VR user. Participants were also free to comment on the VE, the proposed scenario and the associated interactions. These interviews have been transcribed and diarized automatically using WhisperX~\cite{bain2022whisperx} and PyAnnote~\cite{Bredin23pyannote, Plaquet23pyannote}. A verification and cleaning pass was made by the authors after transcription.

\textbf{Thermal Perception and Comfort.} 
Every participant perceived and appreciated the thermal feedback provided by the lamps in Desert Island, in at least one of the two conditions where it was present (P1: \textit{``It was even rather pleasant because it was not too hot!''}). A majority (17/25) perceived the difference between static and dynamic conditions, managing to associate warm thermal feedback with moments when they were exposed to the sun. Some noted slight thermal variations during the static condition, particularly when bending down in the hut (P21: \textit{``I felt more or less warmth depending on the moment, especially when I sometimes moved or bent down to pick things up.''}). However, 7 participants reported a thermal discomfort caused by either the prolonged exposure to overly intense stimuli in the static condition, or the excessive heating of the HMD especially during the third session. Participants' expectations of the actual thermal experience on the desert island differed, and although the majority agreed that the thermal feedback was coherent with the experience, some participants would have made alterations to the thermal feedback to improve realism (P24: \textit{``You can feel it is hot, but it is not as hot as you would expect on a normal beach.''}; P9: \textit{``Coconuts that are supposed to quench my thirst did not cool me down.''}; P4: \textit{``It needs a bit of wind to make it even more immersive''}; P20: \textit{``I expected that when the helicopter arrived, I would feel a movement of air.''}).

In Snowy Mountains, a majority (19/24) of participants perceived and appreciated the thermal feedback provided by the fans, in at least one of the two conditions where it was present. The remaining participants either did not perceive the fans' thermal effect, or considered that it was not significant enough to alter their user experience. Half of participants did not perceive any difference between static and dynamic conditions, and three of them felt they were too focused on the scenario and tasks at hand during their first run to pay attention to their thermal sensations. Unexpectedly, P58 did not perceive any thermal stimulation during the experiment and explains: \textit{``I think I was hyper-concentrated on my task. All I could think about was trying to pick up these logs, putting them in the right place, not dropping my crate, going to the right place. I think, in fact, my other senses were blocked, especially the thermal sense.''}. Regarding the consistency of the thermal stimulus with the VE, some felt that it was not cold enough during the dynamic condition (P53: \textit{``The best would have been the [dynamic condition] with the temperature of the [static condition].''}).

\textbf{Immersion and Presence.} Participants appreciated both VEs and emphasized their audiovisual quality (P15 about Desert Island: \textit{``It is pretty, and right away, with the waves sound, I thought I was really transported there.''}; P50 about Snowy Mountains: \textit{``I found that the visuals were very realistic.''}). When participants compare the control condition with the thermal conditions, they often mention an improvement in ``realism'' and ``immersion'', which gave them a real sense of being there (P18: \textit{``I really felt something and I really felt an added value for immersion''}; P43: \textit{``I think it adds a little something to believe we are really there''}). Minor breaks in the feeling of presence occurred for a restricted number of participants: P2 and P13 felt the HMD cable distracted them, P59 was always a little wary of available space, which hindered their agentivity. P2, P37 and P48 encountered small issues regarding collisions in the VE, P41 got pleasantly surprised and distracted by the dynamism of the thermal stimulus, P45 found repeating the same scenario took them out of the experience. P6, P21 and P47, which did the Control condition after a Thermal condition, felt disconnected from the experience in this condition because it lacked thermal feedback.

\textbf{Interactions and Locomotion.} Across both scenarios, 19 participants mentioned they enjoyed the available interactions, especially breaking the coconut (see Fig.~\ref{fig:walkthrough_desertisland}) in Desert Island (P15: \textit{``I liked grabbing the coconut and breaking it against the stone, it seemed very realistic.''}) and cutting the tree in Snowy Mountains (P47: \textit{``The interaction to cut the tree was great!''}). The majority of participants had no issue with teleportation, except P7, P23 and P43, which had difficulties mastering teleportation, despite the tutorial. This difficulty largely disappeared during the 2nd and 3rd runs.

\textbf{Behavior and Flow.} During the Desert Island scenario, 6 participants perceived that their intentions and behavior had been influenced by the presence of thermal feedback, prompting them to react more ``correctly''. Compared to textual instructions, thermal feedback is regarded as more intuitive, as P8 mentioned: \textit{``The fact that the thermal stimulus is dynamic, it pushes you into the shade and implicitly tells you to go into the cabin, so it's more unconscious.''}. Since the experimental design involved repeating the same scenario 3 times, some participants mentioned that during their 2nd and 3rd runs, their actions flowed more smoothly. (P2: ``I was more familiar with the scenario, so I chained the actions, bim-bam-boom !''). Despite the repetition, the overwhelming majority of participants did not feel bored, as it was a rare experience that they enjoyed (P42: \textit{``I really liked it, because VR is not something I do every day.''}).
\section{Discussion}

\textbf{RQ1: Addition of ambient thermal feedback significantly enhanced presence in our highly immersive VR scenarios.}
According to Melo et al.'s~\cite{meloHowMuchPresence2023} qualitative scale for general presence measured by the IPQ, the following presence thresholds apply: \textit{Very Good} Presence (\(> 4.07\)) and \textit{Excellent Presence} (\(> 4.41\)). In our study, presence ratings from the general presence question of the IPQ for both scenarios indicate that the audio-visual conditions alone achieve \textit{Very Good} presence scores. Similarly, the sense of possible actions yielded high mean values across all conditions (\(>4.6/6\)), demonstrating that our scenarios and apparatus provided a high sense of possible actions. Nevertheless, the addition of congruent thermal feedback, regardless of its quality, significantly enhances presence, elevating it to \textit{Excellent} levels for both scenarios while also reducing variance. This suggests that in metaphorically thermal virtual environments, incorporating congruent thermal feedback holds genuine value, even when visual and auditory immersion is already high. Interestingly, the inclusion of this modality raises participants' expectations regarding immersion. Some participants, such as P6, P21, and P47, reported feeling disappointed when thermal feedback was removed, indicating that once experienced, thermal feedback becomes an integral part of the expected immersive experience.

\textbf{RQ2: The quality difference between Static and Dynamic thermal feedback was not perceived by all participants, highlighting the heterogeneity in thermal perception.}
Answers to post-immersion questionnaires revealed no significant difference between Static and Dynamic conditions which is unexpected, as we hypothesized that dynamism would improve the haptic experience as well as the sense of possible actions. Nonetheless for Snowy Mountains, results show tendencies for differences in Realism and Expressivity between Static and Dynamic. Realism is marginally higher in Static condition than in Dynamic (\(p = 0.094\)), aligned with some comments stating that it's not cold enough in Dynamic, so not realistic enough compared to Static. However, expressivity is marginally higher in the Dynamic condition (\(p = 0.085\)).

Both Dynamic conditions were created on the basis of a prior calibration and the results of the psychophysical study of the thermal feedback system used (see Tab. I \& II in \textit{Supplementary Materials}), which demonstrated that the two levels presented provided significantly different thermal sensations. However, qualitative feedback showed that not all participants perceived the difference between the static and dynamic conditions. While a majority was able to distinguish both conditions in Desert Island, only half were able to make the distinction in Snowy Mountains. The causes for heterogeneity of perception of this difference are undoubtedly multiple as revealed by our results: 1) Thermal perception across people is heterogeneous and modulated by the physiological and psychological state~\cite{schweiker_drivers_2018} as well as the cultural background~\cite{naheed_review_2021} and indeed participants to this study show diversity in the thermal sensitivity questionnaire reported in Fig.~\ref{tab:demographics}. Ultimately, the chosen stimuli (see Tab.~\ref{tab:intensity_levels}) were too weak for less sensitive participants while being too strong for highly sensitive participants. Moreover, correlations between participants' sensitivity to cold and their answers to the haptic experience questionnaires showed that it is positively correlated to realism and harmony in the Snowy Mountains scenario (see Tab.~\ref{tab:correlations}). Thus, it seems important to take into account the thermal sensitivities of each individual before designing thermal stimuli in order to maximize the haptic experience. 2) Although we did not measure cognitive load using quantitative methods, a handful of users reported high cognitive load when doing a scenario for the first time and explained it altered their thermal perception while in VR. 3) As our apparatus allowed movement within the VR zone, when participants got closer or further away from the thermal actuators, they reported thermal variations while rating the Static condition. 4) Using ambient thermal feedback in VR for long sessions leads to an accumulation of heat in the clothing that prevented cooling down and lead to thermal discomfort, which negatively impacted UX as a whole.

\textbf{RQ3: Thermal feedback had no influence on measured exploratory behavior.}
Our measurements indicated no significant effect of thermal conditions on exploratory behavior. However, this does not contradict the findings of Shaw et al.~\cite{shaw_heat_2019}, who observed behavioral differences between audio-visual and multisensory (including thermal feedback) conditions, during a fire evacuation VR simulation. It is important to note that our measures aggregate each run into a single value. This aligns in part with Shaw et al.'s results, as their aggregated measure of total evacuation time, also was not influenced by the immersion condition.

That said, Shaw et al. study broke down behavior into smaller units, revealing differences influenced by immersion at a more granular level. Our study did not explore behavior with such detail. Subtle behavioral changes, such as the time spent in the shade on Desert Island or inside the cabin in Snowy Mountains, could potentially illustrate the influence of thermal feedback on behavior. However, in our case, these nuances would likely be obscured by the time constraints of tasks and heavily guided scenario. 

\textbf{RQ4: Order of passage has a significant influence on exploratory behavior.}
Our study used a within-subject design, in which conditions were counterbalanced to try and eliminate order bias when comparing conditions. While we found that order had no measurable influence on the answers to the post-immersion questionnaires, our results show that behavior is modified according to the order of passage. In fact, for both scenarios, participants have significantly higher values for real movement, travel in the VE, head turn and orientation in the VE during their first passage, and these metrics decrease during the second passage and again in the third. This can be explained as during their first run, users are discovering the environment, and if they are curious to explore it, they will. This curiosity greatly diminishes during the second and third runs. It's interesting to note that both entropy metrics show significant differences only for Desert Island and not for Snowy Mountains. To understand this phenomenon, we visually compared the longer trajectories against the shorter ones (see Appendix D in \textit{Supplementary Materials}). While both scenarios were designed to be as similar as possible, we found that by design, the Desert Island scenario leaves more freedom of movement to perform the first task (see Fig~\ref{fig:walkthrough_desertisland}), as users appear in the middle of the island. They are correctly oriented towards the objective, but with no visible constraints. In comparison with Snowy Mountains, users are constrained by the door of the shed and do not feel the need to explore to find the location of the 2nd task.

\textbf{Recommendations}
% This suggests that subjective measures alone may overlook order effects that emerge in user behavior. A previous study by Lavoué et al.~\cite{lavoue_influence_2023} similarly noted that condition order influenced the subjective flow experience in VR. Taken together, these observations underscore that within-subject designs, while statistically powerful and useful for accounting for inter-individual variability, require methodological adaptation when used to evaluate UX within multisensory VR. Actually, participants should perform a context-relevant learning phase, as we observed that generic VR tutorials like the one we proposed seem insufficient. Conversely, if the objective of the study is to assess effects related to novelty or surprise, a between-subject design might be preferable.

% Considering the use of thermal feedback within VR scenario, while our level of thermal stimulation were calibrated for our VEs based on the mean of answers of a preliminary panel, our findings highlight significant inter-individual variability in thermal perception. This suggests that personalized calibration may substantially improve feedback effectiveness and user comfort. However, we need to implement and test calibration protocol to ensure their effectiveness and usability, as the need for calibration might hinder the use of thermal feedback due to impracticality.

These findings suggest that subjective measures alone may overlook subtle order effects that manifest in user behavior. A previous study by Lavoué et al.~\cite{lavoue_influence_2023} similarly noted that the order of conditions influenced the subjective flow experience in VR. Together, these observations underscore the complexities inherent in within-subject designs. Although such designs are statistically powerful and effective for managing inter-individual variability in perception, which is a major concern in multisensory VR studies, they require careful methodological adaptations to limit learning effects. For instance, participants might benefit from a context-relevant learning phase, as generic VR tutorials, like the one we implemented, appear insufficient. In addition, studies should consistently verify through statistical tests that the order of conditions does not influence the outcomes. If an order effect is detected, it should be thoroughly explained as it might overpower the primary factors under investigation. Conversely, if the study aims to assess effects related to novelty or surprise, a between-subject design might be more appropriate, despite requiring a larger sample size to achieve statistical relevance. 

When considering the use of thermal feedback within VR scenarios, it is important to note that although the presented levels of thermal stimulation were calibrated based on preliminary panel feedback, significant inter-individual variability in thermal perception was observed. This underscores the potential benefits of personalized calibration to enhance feedback effectiveness and user comfort. However, practical implementation and testing of calibration protocols are essential to ensure their efficacy and usability, as the necessity for calibration could impede the practical application of thermal feedback.
% La perception thermique est multifactorielle, avec des variables dépendantes de l'environnement mais aussi des variables dépendantes des individus. Pour optimiser l'expérience utilisateur lors d'experience d'XR qui impliquent des stimulations thermiques, il conviendrait d'estimer les sensibilités des participants en amont pour calibrer la stimulation. Cependant, ce n'est pas chose facile, ni pratique.

% Si la stimulation d'ambiance est congruente, sa qualité semble peu importer dans un VE de haute qualité et engageant. Cependant, il faut faire attention à l'accumulation de chaleur (ou de froid) qui peut venir perturber l'experience. Pour cette raison, nous faisons quand même la recommendation d'avoir une stimulation dynamique. 

% Dans les protocoles within-subject implicant un VE haute qualité et un scenario complexe, le tutoriel doit non-seulement apprendre en faisant (et pas juste une vidéo) les contrôles, mais aussi integrer toutes les tâches qui pourront être effectuée, decontextualisées, voire même avec des visuels differents. Dans le même sens pour les experiences de XR multisensorielles, l'effet de surprise bien que vraimenblablement bénéfique aux réponses aux questionnaires, doit être éliminé en présentant les stimulation en amont, hors contexte encore une fois.

% \textcolor{blue}{je verrai bien, à voir où la placer dans ce paragraphe, une recommendation en mode "ben du coup on recommende de ne pas faire de within-subject pour comprendre des effets liés au dynamisme (ici le thermique controlé par actions utilisateurs) en VR}

\section{Limitations and Perspectives}

This study focused on a single modality of thermal stimuli, either heat or cold (induced by wind), but did not investigate their combination. Yet, as suggested by da Silveira et al.~\cite{dasilveiraThermalWindDevices2023b}, there is a need for studies to understand the interactions between hot and cold stimuli when used in the same scenario.

Although the thermal system used in this study can modulate the intensity of stimuli based on the user's position relative to the thermal actuators, our study did not use this feature in order to focus on dynamism as a quality level. As a result, participant movement led to unintended variations in stimulus perception. Similar to the impact of spatialized audio on immersion, future work should assess the added value of spatially coherent thermal cues for enhancing presence and user experience. Going further, adding thermal actuators to the thermal feedback system, could enable the precise spatialization of ambient thermal sources that are localized within the VE, such as a fire or a waterfall.

Finally, our results indicate a potential relationship between cognitive load and thermal perception in VR. A recent study from Philippe et al., has explored the effects of thermal stimuli on cognitive performance~\cite{philippeCoolMeEffects2024}; however qualitative feedback from our study suggests that cognitive demands may also modulate thermal sensitivity. As the present study did not measure cognitive load directly, this remains an open research question. Future investigations within ecologically valid VR scenario should incorporate cognitive load assessments to answer this question, such as post-immersion questionnaire, dual-task paradigms, or physiological indicators like pupil dilation~\cite{lee_measuring_2024}.

In summary, future work should focus on evaluating the impact of spatializion of thermal feedback on UX, develop a rapid user-specific thermal feedback calibration, and integrate cognitive state measures to investigate the relationship between cognitive load and thermal perception in VR. These directions are essential for advancing the integration of thermal feedback within highly-immersive and adaptive VR experiences.
\section{Conclusion}
In this study, we investigated the effects of dynamic thermal feedback on UX within highly immersive VR environments. Our results indicate that incorporating congruent ambient thermal feedback significantly enhances the sense of presence, even in settings characterized by high audio-visual fidelity. However, the differentiation between static and dynamic thermal feedback was not consistently perceived across all participants, highlighting the significance of individual variations in thermal sensitivity. Although the quality of thermal feedback did not markedly affect user behavior in our experiments, the sequence of exposure had a pronounced impact on exploratory behaviors, and we propose solutions to limit the learning effect in within-subject designs. 

Our findings underscore key challenges and opportunities in the development of multisensory VR experiences. Notably, individual differences in thermal sensitivity, the need for future studies to incorporate spatialized ambient thermal cues, and the interaction between cognitive load and thermal perception emerged as pivotal considerations. Addressing these issues will necessitate advancements in methodologies for ambient thermal feedback calibration and real-time monitoring of cognitive states in VR.

\section*{Acknowledgments}
The authors wish to thank every participant for their involvement in our study. This work was supported in part by a grant from the French National Agency for Research for the RENFORCE project (ANR-22-CE38-0008).

\bibliographystyle{IEEEtran}
\bibliography{bibliography}

\begin{IEEEbiography}[{\includegraphics[width=1in,height=1.25in,clip,keepaspectratio]{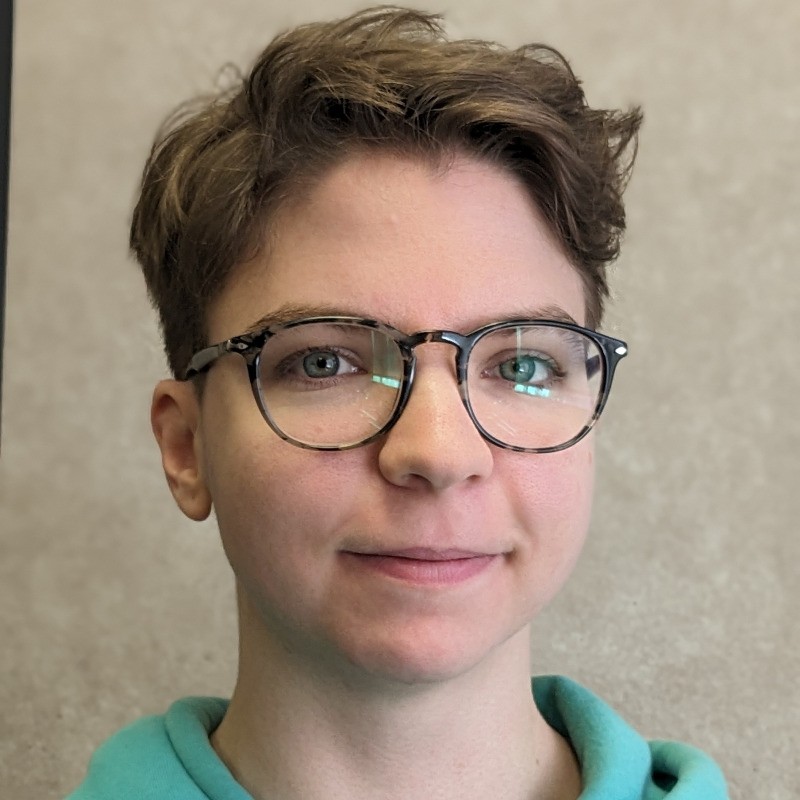}}]{Sophie Villenave}
Sophie Villenave is a PhD student at École Centrale de Lyon - ENISE, France. They received their diploma in computer science from INSA Lyon in 2021, before starting their PhD at the LIRIS laboratory. Their current work focuses on the implementation and evaluation of user interfaces to provide thermal feedback in highly immersive virtual reality applications. 
\end{IEEEbiography}

\vspace{11pt}

\begin{IEEEbiography}[{\includegraphics[width=1in,height=1.25in,clip,keepaspectratio]{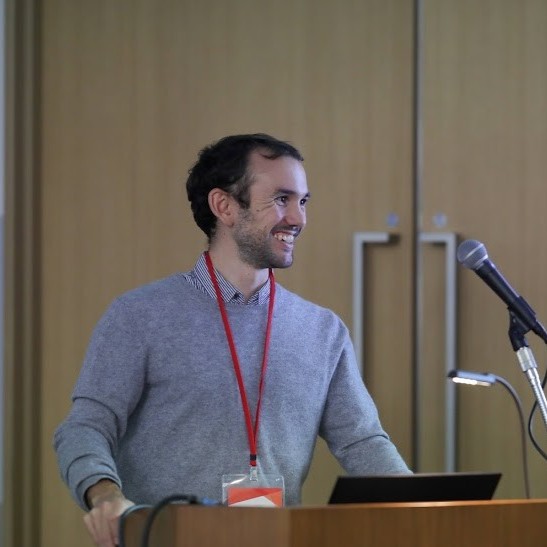}}]{Pierre Raimbaud}
Pierre Raimbaud is an associate professor at École Centrale de Lyon - ENISE, France. His research revolves around the study of virtual reality users. He focused on theoretical approaches to user interaction design during his PhD, then studied human-agent interactions in virtual environments. His current research seeks to understand human behavior in virtual reality, from the user's and society's perspective. 
\end{IEEEbiography}

\vspace{11pt}

\begin{IEEEbiography}[{\includegraphics[width=1in,height=1.25in,clip,keepaspectratio]{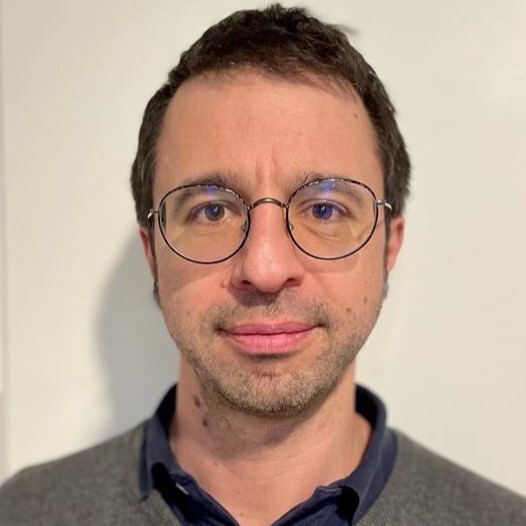}}]{Guillaume Lavoué}
Guillaume Lavoué is professor at École Centrale de Lyon - ENISE, France. Member of the LIRIS laboratory since 2006, his research focuses on computer graphics, virtual reality and perception in the broadest sense. Head of the LIRIS XR team, he has co-authored some 100 international publications in peer-reviewed journals and conferences.
\end{IEEEbiography}

\vfill

\end{document}

\title{Supplementary Materials for "Dynamic Thermal Feedback in Highly Immersive VR Scenarios : a Multimodal Analysis of User Experience"}

\author{Sophie Villenave{},
        Pierre Raimbaud{},
       and Guillaume Lavoué{}

\IEEEcompsocitemizethanks{\IEEEcompsocthanksitem S. Villenave, P. Raimbaud, G. Lavoué are affiliated to Ecole Centrale de Lyon, CNRS, INSA Lyon, Universite Claude Bernard Lyon 1, LIRIS, UMR5205, ENISE, 42023 Saint Etienne, France. E-mail: name.surname@liris.cnrs.fr.\protect}}

\maketitle

\appendix
\subsection{Apparatus}
\begin{figure}[H]
  \centering
  \includegraphics[alt={Schematic of the system, the PC is linked via USB to the arduino which acts as the DMX controller. The arduino is linked to the 4-Way DMX dimmerpack. The dimmerpack controls 4 lamps and is linked a first DMX Fan. Other DMX fans are chained to the first DMX fan. Actuators are placed around the 2m by 2m VR zone.}, width=\textwidth]{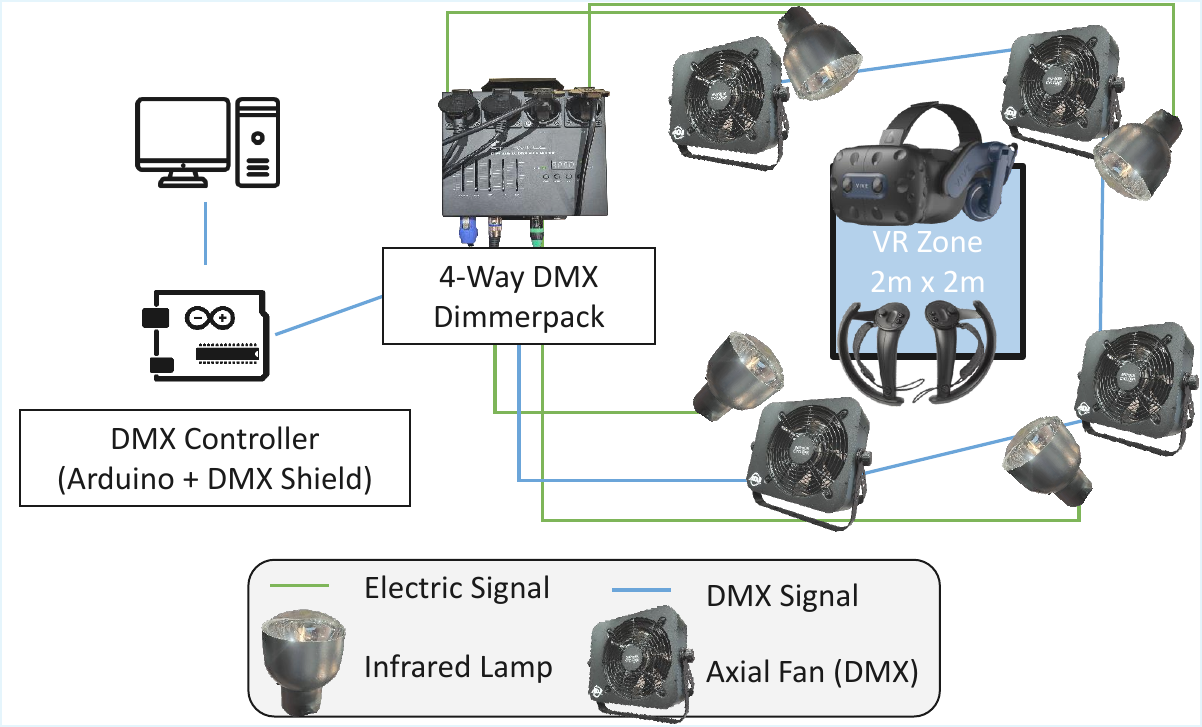}
  \caption{Schematic overview of implemented thermal feedback system. From Villenave et al.~\cite{villenaveDynamicModularThermal2025}.}
  \label{fig:system_schematics}
\end{figure}

\begin{figure}[]
  \centering
  \begin{minipage}[t]{\linewidth}
    \centering
    \includegraphics[width =0.35\linewidth]{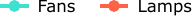}
  \end{minipage}
  \begin{minipage}[t]{0.49\linewidth}
    \centering
    \includegraphics[alt={y = Perceived Thermal Sensation going from 0 to 6 vs x = Actuator Intensity, normalized between 0 to 1. Lamps Measured Points: (0, 0); (0.125, 0.12); (0.25, 0.6); (0.38, 1.2); (0.5, 2.2); (0.63, 3.2); (0.75, 3.6); (0.88, 3.7); (1, 4.1). Fans Measured Points: (0.02, 1.5); (0.1, 2.2); (0.2, 1.8); (0.3, 1.7); (0.4, 2.7); (0.6, 2.9); (0.8, 3.5); (1, 4.2)}, width=\linewidth]{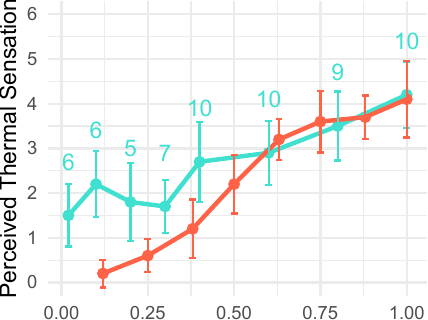}
    \label{fig:perceived_intensity}
  \end{minipage}
  \begin{minipage}[t]{0.49\linewidth}
    \centering
    \includegraphics[alt={y = Reaction Time going from 0s to 15s vs x = Actuator Intensity, normalized between 0 to 1. Lamps Measured Points: (0.125, 5.32); (0.25, 3.52); (0.38, 2.79); (0.5, 2.45); (0.63, 1.65); (0.75, 2.07); (0.88, 1.73); (1, 2.01). Fans Measured Points: (0.02, 9.84); (0.1, 10.93); (0.2, 9.24); (0.3, 12.09); (0.4, 10.96); (0.6, 9.27); (0.8, 9.33); (1, 8.91)}, width=\linewidth]{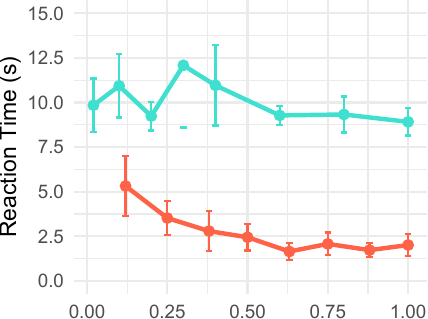}
    \label{fig:reaction_time}
  \end{minipage}
  \begin{minipage}[t]{0.5\linewidth}
    \centering
    \includegraphics[alt={y = Sound Level going from 35dBA to 55dBA vs x = Actuator Intensity, normalized between 0 to 1. Reference Sound Level = 39.0dBA. Lamps Measured Points: (0.01, 39.9); (0.2, 40.9); (0.4, 41.3); (0.6, 41.3); (0.8, 40.4); (1, 39.0). Fans Measured Points: (0.02, 43.0); (0.2, 43.9); (0.4, 45.5); (0.6, 48.2); (0.8, 52.3); (1, 55.0)},width=\linewidth]{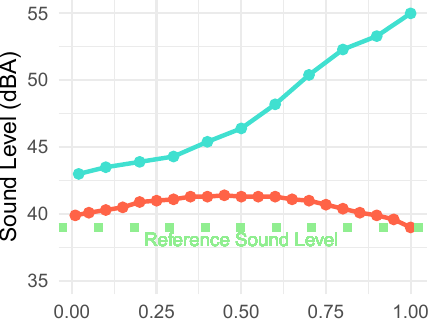}
    \label{fig:sound_level}
  \end{minipage}
  \vspace{-10pt}
  \caption{Psychophysical and sound level measures vs. thermal actuator intensity. Fan intensity (0–50\%) and lamp intensity (0–100\%) are normalized to [0, 1]. N° of participants (out of 10) perceiving wind is shown on the perceived thermal sensation graph. From Villenave et al.~\cite{villenaveDynamicModularThermal2025}}
  \label{fig:psychophysical_results}
\end{figure}

\clearpage

\begin{table}[H]
\caption{Student tests comparing intensity levels (in \%) of lamps actuation.}
\centering
\begin{tabular}{llrrrrrrl}
\toprule
group1 & group2 & n1 & n2 & statistic & df & p & p.adj & p.adj.signif\\
\midrule
12 & 25 & 10 & 10 & -2.4494897 & 9 & 3.70e-02 & 4.90e-02 & *\\
12 & 38 & 10 & 10 & -4.7434165 & 9 & 1.00e-03 & 2.00e-03 & **\\
12 & 50 & 10 & 10 & -6.7082039 & 9 & 8.77e-05 & 2.23e-04 & ***\\
12 & 63 & 10 & 10 & -11.6189500 & 9 & 1.00e-06 & 1.30e-05 & ****\\
12 & 75 & 10 & 10 & -9.1598704 & 9 & 7.40e-06 & 3.45e-05 & ****\\
12 & 88 & 10 & 10 & -15.6524758 & 9 & 1.00e-07 & 2.20e-06 & ****\\
12 & 100 & 10 & 10 & -10.3012756 & 9 & 2.80e-06 & 1.95e-05 & ****\\
\addlinespace
25 & 38 & 10 & 10 & -2.2500000 & 9 & 5.10e-02 & 6.20e-02 & ns\\
25 & 50 & 10 & 10 & -4.7067872 & 9 & 1.00e-03 & 2.00e-03 & **\\
25 & 63 & 10 & 10 & -9.7500000 & 9 & 4.40e-06 & 2.48e-05 & ****\\
25 & 75 & 10 & 10 & -8.2158384 & 9 & 1.79e-05 & 5.43e-05 & ****\\
25 & 88 & 10 & 10 & -11.1958843 & 9 & 1.40e-06 & 1.30e-05 & ****\\
25 & 100 & 10 & 10 & -8.1742389 & 9 & 1.86e-05 & 5.43e-05 & ****\\
\addlinespace
38 & 50 & 10 & 10 & -2.3717082 & 9 & 4.20e-02 & 5.30e-02 & ns\\
38 & 63 & 10 & 10 & -6.0000000 & 9 & 2.02e-04 & 4.35e-04 & ***\\
38 & 75 & 10 & 10 & -6.0000000 & 9 & 2.02e-04 & 4.35e-04 & ***\\
38 & 88 & 10 & 10 & -8.1348922 & 9 & 1.94e-05 & 5.43e-05 & ****\\
38 & 100 & 10 & 10 & -8.3330887 & 9 & 1.60e-05 & 5.43e-05 & ****\\
\addlinespace
50 & 63 & 10 & 10 & -2.7386128 & 9 & 2.30e-02 & 3.40e-02 & *\\
50 & 75 & 10 & 10 & -2.8062430 & 9 & 2.00e-02 & 3.20e-02 & *\\
50 & 88 & 10 & 10 & -5.5815631 & 9 & 3.42e-04 & 6.84e-04 & ***\\
50 & 100 & 10 & 10 & -3.9427724 & 9 & 3.00e-03 & 6.00e-03 & **\\
\addlinespace
63 & 75 & 10 & 10 & -1.3093073 & 9 & 2.23e-01 & 2.40e-01 & ns\\
63 & 88 & 10 & 10 & -1.6269784 & 9 & 1.38e-01 & 1.55e-01 & ns\\
63 & 100 & 10 & 10 & -2.5861310 & 9 & 2.90e-02 & 4.10e-02 & *\\
\addlinespace
75 & 88 & 10 & 10 & -0.2307692 & 9 & 8.23e-01 & 8.23e-01 & ns\\
75 & 100 & 10 & 10 & -1.6269784 & 9 & 1.38e-01 & 1.55e-01 & ns\\
\addlinespace
88 & 100 & 10 & 10 & -1.0000000 & 9 & 3.43e-01 & 3.56e-01 & ns\\
\bottomrule
\end{tabular}
\end{table}

\begin{table}[H]
\caption{Student tests comparing intensity levels (in \%) of fans actuation.}
\centering
\begin{tabular}[t]{llrrrrrrl}
\toprule
group1 & group2 & n1 & n2 & statistic & df & p & p.adj & p.adj.signif\\
\midrule
1 & 5 & 10 & 10 & -1.5609177 & 9 & 1.53e-01 & 0.204000 & ns\\
1 & 10 & 10 & 10 & -0.5797410 & 9 & 5.76e-01 & 0.613000 & ns\\
1 & 15 & 10 & 10 & -0.5570860 & 9 & 5.91e-01 & 0.613000 & ns\\
1 & 20 & 10 & 10 & -3.3425161 & 9 & 9.00e-03 & 0.020000 & *\\
1 & 30 & 10 & 10 & -3.0962811 & 9 & 1.30e-02 & 0.024000 & *\\
1 & 40 & 10 & 10 & -4.0451992 & 9 & 3.00e-03 & 0.011000 & *\\
1 & 50 & 10 & 10 & -6.0206826 & 9 & 1.97e-04 & 0.002000 & **\\
\addlinespace
5 & 10 & 10 & 10 & 0.7682213 & 9 & 4.62e-01 & 0.539000 & ns\\
5 & 15 & 10 & 10 & 1.3416408 & 9 & 2.13e-01 & 0.271000 & ns\\
5 & 20 & 10 & 10 & -1.0476454 & 9 & 3.22e-01 & 0.392000 & ns\\
5 & 30 & 10 & 10 & -1.7685190 & 9 & 1.11e-01 & 0.155000 & ns\\
5 & 40 & 10 & 10 & -2.6234239 & 9 & 2.80e-02 & 0.048000 & *\\
5 & 50 & 10 & 10 & -3.8729833 & 9 & 4.00e-03 & 0.012000 & *\\
\addlinespace
10 & 15 & 10 & 10 & 0.2307692 & 9 & 8.23e-01 & 0.823000 & ns\\
10 & 20 & 10 & 10 & -2.2119262 & 9 & 5.40e-02 & 0.089000 & ns\\
10 & 30 & 10 & 10 & -3.1608267 & 9 & 1.20e-02 & 0.023000 & *\\
10 & 40 & 10 & 10 & -4.0193631 & 9 & 3.00e-03 & 0.011000 & *\\
10 & 50 & 10 & 10 & -4.8107024 & 9 & 9.59e-04 & 0.005000 & **\\
\addlinespace
15 & 20 & 10 & 10 & -3.3541020 & 9 & 8.00e-03 & 0.020000 & *\\
15 & 30 & 10 & 10 & -4.8107024 & 9 & 9.59e-04 & 0.005000 & **\\
15 & 40 & 10 & 10 & -7.2160535 & 9 & 5.00e-05 & 0.000700 & ***\\
15 & 50 & 10 & 10 & -9.3026051 & 9 & 6.50e-06 & 0.000182 & ***\\
\addlinespace
20 & 30 & 10 & 10 & -0.6882472 & 9 & 5.09e-01 & 0.570000 & ns\\
20 & 40 & 10 & 10 & -1.9215378 & 9 & 8.70e-02 & 0.135000 & ns\\
20 & 50 & 10 & 10 & -3.3084658 & 9 & 9.00e-03 & 0.020000 & *\\
\addlinespace
30 & 40 & 10 & 10 & -1.7650452 & 9 & 1.11e-01 & 0.155000 & ns\\
30 & 50 & 10 & 10 & -3.2843925 & 9 & 9.00e-03 & 0.020000 & *\\
\addlinespace
40 & 50 & 10 & 10 & -4.5825757 & 9 & 1.00e-03 & 0.006000 & **\\
\bottomrule
\end{tabular}
\end{table}

\subsection{Results - Additional figures}
This section presents additional figures for the \textit{Results} section of the paper. These figures were not included in the main manuscript as they illustrate comparisons that are not significant. 
\begin{figure}[H]
    \vspace{-0.5cm}
    \subfloat[Thermal Sensation]{
        \includegraphics[width=0.49\linewidth]{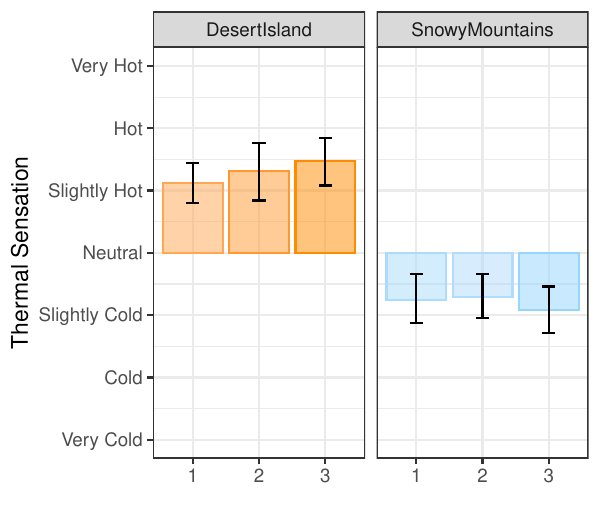}
        \label{thermal_sensation}
    }
    \subfloat[Thermal Variation]{
        \includegraphics[width=0.49\linewidth]{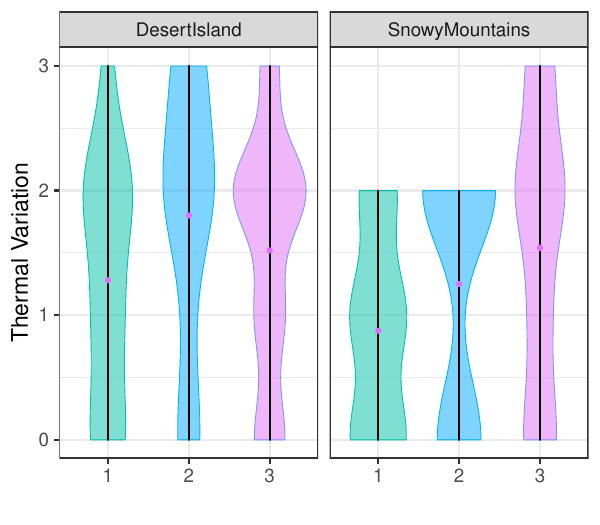}
        \label{thermal_variation}
    }
    \caption{Influence of Order of Passage on Thermal Perception}
    \label{fig:thermals}
\end{figure}

\begin{figure}[H]
\vspace{-0.5cm}
  \centering
  \subfloat[Realism]{
    \includegraphics[width=0.49\linewidth]{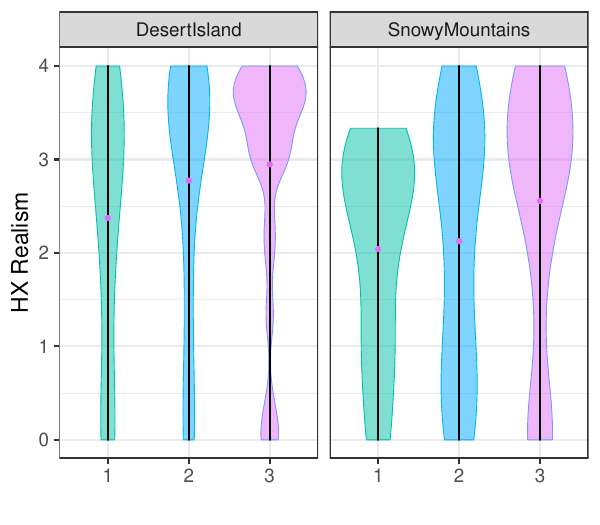}
    \label{hx:realism}
  }
  \subfloat[Expressivity]{
    \includegraphics[width=0.49\linewidth]{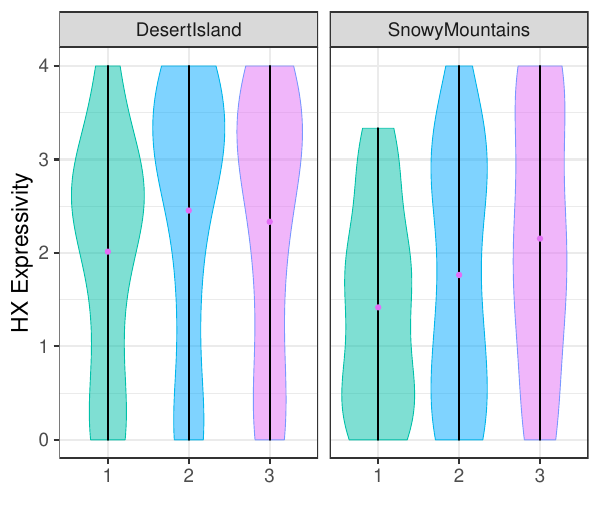}
    \label{hx:realism}
  }
  
  \subfloat[Harmony]{
    \includegraphics[width=0.49\linewidth]{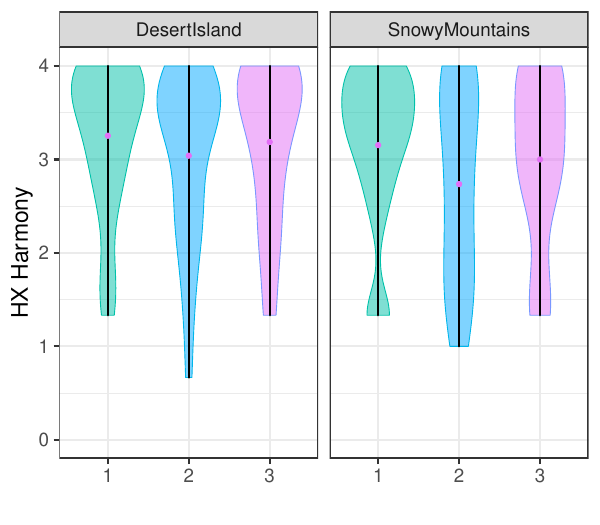}
    \label{hx:realism}
  }
  \subfloat[Involvement]{
    \includegraphics[width=0.49\linewidth]{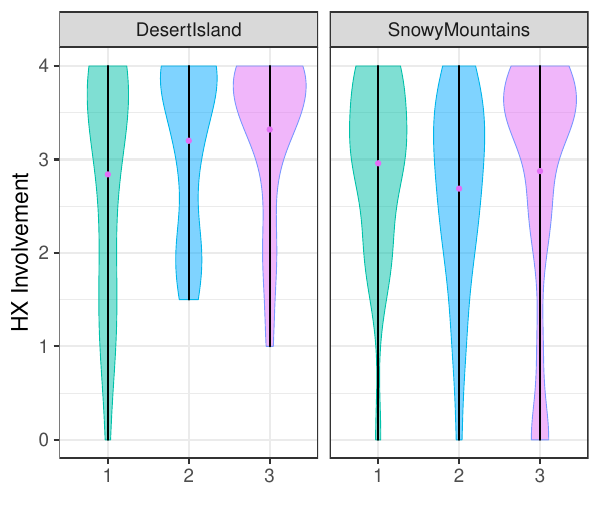}
    \label{hx:realism}
  }
\caption{Influence of Order of Passage on Haptic Experience}
\label{fig:hx}
\end{figure}

\begin{figure}[H]
\vspace{-0.5cm}
  \centering
  \subfloat[Spatial Presence]{
    \includegraphics[width=0.49\linewidth]{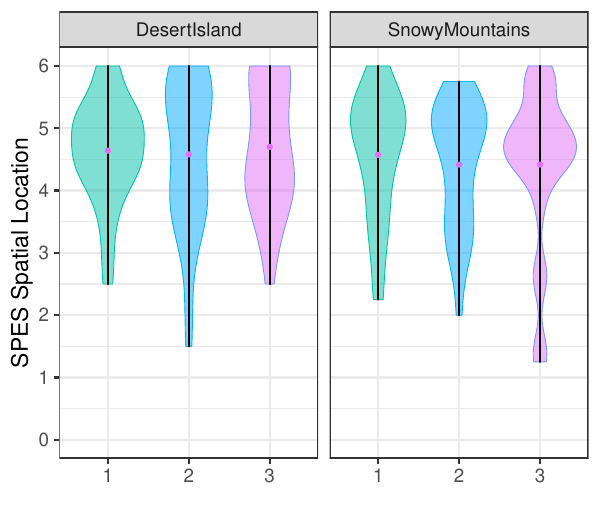}
    \label{spes:spatial_location}
  }
  \subfloat[Possible Actions]{
    \includegraphics[width=0.49\linewidth]{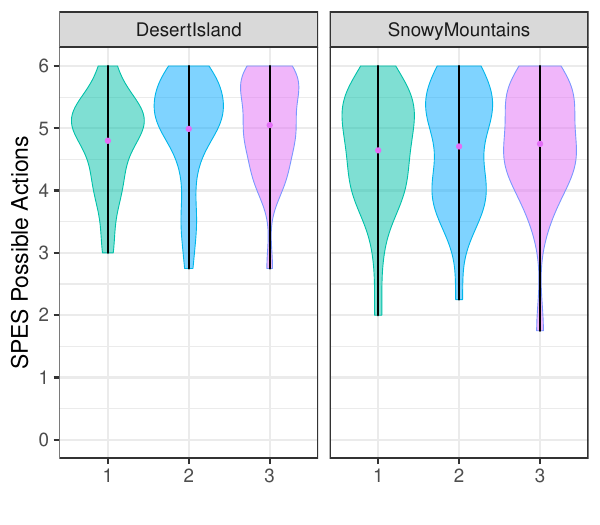}
    \label{spes:possible_actions}
  }
\caption{Influence of Order of Passage on Spatial Presence Experience}
\label{fig:spes}
\end{figure}

\clearpage

\subsection{Results - Additional tables}
This section presents additional tables for the \textit{Results} section of the paper. These tables report in details results for every Friedman tests and Wilcoxon tests performed as well as Spearman's correlations computed between demographics description of participants and answers to questionnaires.  

\subsubsection{Statistical summary of post-immersion questionnaire by experimental condition}

\phantom{*}

\begin{table}[H]

\caption{Statistical summary by experimental condition for HX Expressivity}
\centering
\begin{tabular}[t]{llrrrrrrrrrrrr}
\toprule
Experiment & Condition & n & min & max & median & q1 & q3 & iqr & mean & sd & se & ci95\_lower & ci95\_upper\\
\midrule
DesertIsland & Control & 25 & 0.000 & 3.00 & 0.667 & 0.00 & 1.67 & 1.67 & 1.067 & 1.194 & 0.239 & 0.640 & 1.533\\
DesertIsland & Static & 25 & 0.000 & 4.00 & 3.000 & 2.33 & 3.33 & 1.00 & 2.720 & 0.985 & 0.197 & 2.333 & 3.093\\
DesertIsland & Dynamic & 25 & 0.333 & 4.00 & 3.333 & 2.33 & 4.00 & 1.67 & 3.013 & 1.007 & 0.201 & 2.600 & 3.373\\
SnowyMountains & Control & 24 & 0.000 & 2.33 & 0.333 & 0.00 & 1.33 & 1.33 & 0.625 & 0.718 & 0.147 & 0.361 & 0.903\\
SnowyMountains & Static & 24 & 0.000 & 4.00 & 2.000 & 1.58 & 2.75 & 1.17 & 2.069 & 1.002 & 0.205 & 1.694 & 2.486\\

SnowyMountains & Dynamic & 24 & 0.333 & 4.00 & 3.000 & 1.92 & 3.33 & 1.42 & 2.639 & 1.167 & 0.238 & 2.167 & 3.056\\
\bottomrule
\end{tabular}
\end{table}
\begin{table}[H]

\caption{Statistical summary by experimental condition for HX Realism}
\centering
\begin{tabular}[t]{llrrrrrrrrrrrr}
\toprule
Experiment & Condition & n & min & max & median & q1 & q3 & iqr & mean & sd & se & ci95\_lower & ci95\_upper\\
\midrule
DesertIsland & Control & 25 & 0.00 & 3.67 & 1.00 & 0.00 & 2.00 & 2.000 & 1.187 & 1.217 & 0.243 & 0.720 & 1.64\\
DesertIsland & Static & 25 & 2.33 & 4.00 & 3.67 & 3.33 & 3.67 & 0.333 & 3.507 & 0.432 & 0.086 & 3.320 & 3.67\\
DesertIsland & Dynamic & 25 & 1.33 & 4.00 & 3.67 & 3.00 & 4.00 & 1.000 & 3.400 & 0.660 & 0.132 & 3.107 & 3.64\\
SnowyMountains & Control & 24 & 0.00 & 3.33 & 0.50 & 0.00 & 1.00 & 1.000 & 0.778 & 0.991 & 0.202 & 0.417 & 1.19\\
SnowyMountains & Static & 24 & 2.00 & 4.00 & 3.00 & 3.00 & 3.42 & 0.417 & 3.125 & 0.596 & 0.122 & 2.875 & 3.35\\

SnowyMountains & Dynamic & 24 & 1.00 & 4.00 & 3.00 & 2.25 & 3.33 & 1.083 & 2.819 & 0.845 & 0.173 & 2.458 & 3.14\\
\bottomrule
\end{tabular}
\end{table}
\begin{table}[H]

\caption{Statistical summary by experimental condition for HX Involvement}
\centering
\begin{tabular}[t]{llrrrrrrrrrrrr}
\toprule
Experiment & Condition & n & min & max & median & q1 & q3 & iqr & mean & sd & se & ci95\_lower & ci95\_upper\\
\midrule
DesertIsland & Control & 25 & 0.0 & 4 & 2.0 & 1.5 & 2.50 & 1.000 & 2.06 & 0.993 & 0.199 & 1.66 & 2.44\\
DesertIsland & Static & 25 & 2.5 & 4 & 3.5 & 3.5 & 4.00 & 0.500 & 3.62 & 0.415 & 0.083 & 3.46 & 3.76\\
DesertIsland & Dynamic & 25 & 1.5 & 4 & 4.0 & 3.5 & 4.00 & 0.500 & 3.68 & 0.575 & 0.115 & 3.44 & 3.88\\
SnowyMountains & Control & 24 & 0.0 & 4 & 2.0 & 1.0 & 2.50 & 1.500 & 1.73 & 1.161 & 0.237 & 1.27 & 2.17\\
SnowyMountains & Static & 24 & 2.5 & 4 & 3.5 & 3.0 & 4.00 & 1.000 & 3.48 & 0.477 & 0.097 & 3.27 & 3.67\\

SnowyMountains & Dynamic & 24 & 2.0 & 4 & 3.5 & 3.0 & 3.62 & 0.625 & 3.31 & 0.605 & 0.123 & 3.06 & 3.54\\
\bottomrule
\end{tabular}
\end{table}
\begin{table}[H]

\caption{Statistical summary by experimental condition for HX Harmony}
\centering
\begin{tabular}[t]{llrrrrrrrrrrrr}
\toprule
Experiment & Condition & n & min & max & median & q1 & q3 & iqr & mean & sd & se & ci95\_lower & ci95\_upper\\
\midrule
DesertIsland & Control & 25 & 0.667 & 4 & 2.67 & 1.67 & 3.33 & 1.667 & 2.57 & 1.025 & 0.205 & 2.17 & 2.97\\
DesertIsland & Static & 25 & 1.667 & 4 & 3.67 & 3.33 & 3.67 & 0.333 & 3.43 & 0.582 & 0.116 & 3.20 & 3.63\\
DesertIsland & Dynamic & 25 & 2.333 & 4 & 3.67 & 3.00 & 4.00 & 1.000 & 3.48 & 0.545 & 0.109 & 3.27 & 3.69\\
SnowyMountains & Control & 24 & 1.000 & 4 & 1.67 & 1.33 & 2.75 & 1.417 & 2.12 & 0.977 & 0.199 & 1.75 & 2.51\\
SnowyMountains & Static & 24 & 2.333 & 4 & 3.50 & 3.00 & 4.00 & 1.000 & 3.43 & 0.534 & 0.109 & 3.22 & 3.65\\

SnowyMountains & Dynamic & 24 & 2.000 & 4 & 3.33 & 3.00 & 3.75 & 0.750 & 3.33 & 0.598 & 0.122 & 3.10 & 3.57\\
\bottomrule
\end{tabular}
\end{table}
\begin{table}[H]

\caption{Statistical summary by experimental condition for General Presence}
\centering
\begin{tabular}[t]{llrrrrrrrrrrrr}
\toprule
Experiment & Condition & n & min & max & median & q1 & q3 & iqr & mean & sd & se & ci95\_lower & ci95\_upper\\
\midrule
DesertIsland & Control & 25 & 2 & 6 & 4 & 4 & 5.00 & 1.00 & 4.32 & 1.108 & 0.222 & 3.88 & 4.76\\
DesertIsland & Static & 25 & 4 & 6 & 5 & 5 & 6.00 & 1.00 & 5.20 & 0.707 & 0.141 & 4.92 & 5.48\\
DesertIsland & Dynamic & 25 & 4 & 6 & 5 & 5 & 6.00 & 1.00 & 5.36 & 0.700 & 0.140 & 5.08 & 5.60\\
SnowyMountains & Control & 24 & 0 & 6 & 5 & 4 & 5.25 & 1.25 & 4.42 & 1.501 & 0.306 & 3.79 & 4.96\\
SnowyMountains & Static & 24 & 4 & 6 & 5 & 5 & 6.00 & 1.00 & 5.25 & 0.676 & 0.138 & 4.96 & 5.50\\

SnowyMountains & Dynamic & 24 & 2 & 6 & 5 & 5 & 6.00 & 1.00 & 5.17 & 1.007 & 0.206 & 4.71 & 5.54\\
\bottomrule
\end{tabular}
\end{table}
\begin{table}[H]

\caption{Statistical summary by experimental condition for SPES Spatial Location}
\centering
\begin{tabular}[t]{llrrrrrrrrrrrr}
\toprule
Experiment & Condition & n & min & max & median & q1 & q3 & iqr & mean & sd & se & ci95\_lower & ci95\_upper\\
\midrule
DesertIsland & Control & 25 & 1.50 & 6 & 4.25 & 3.50 & 5.25 & 1.750 & 4.17 & 1.179 & 0.236 & 3.71 & 4.61\\
DesertIsland & Static & 25 & 3.00 & 6 & 5.00 & 4.25 & 5.75 & 1.500 & 4.89 & 0.995 & 0.199 & 4.51 & 5.28\\
DesertIsland & Dynamic & 25 & 3.75 & 6 & 4.75 & 4.25 & 5.50 & 1.250 & 4.86 & 0.788 & 0.158 & 4.59 & 5.18\\
SnowyMountains & Control & 24 & 1.25 & 6 & 4.38 & 3.44 & 5.00 & 1.562 & 4.09 & 1.229 & 0.251 & 3.58 & 4.58\\
SnowyMountains & Static & 24 & 2.25 & 6 & 4.88 & 4.44 & 5.25 & 0.812 & 4.74 & 0.907 & 0.185 & 4.38 & 5.07\\

SnowyMountains & Dynamic & 24 & 2.00 & 6 & 5.00 & 4.19 & 5.25 & 1.062 & 4.57 & 1.052 & 0.215 & 4.13 & 4.96\\
\bottomrule
\end{tabular}
\end{table}
\begin{table}[H]

\caption{Statistical summary by experimental condition for SPES Possible Actions}
\centering
\begin{tabular}[t]{llrrrrrrrrrrrr}
\toprule
Experiment & Condition & n & min & max & median & q1 & q3 & iqr & mean & sd & se & ci95\_lower & ci95\_upper\\
\midrule
DesertIsland & Control & 25 & 2.75 & 6 & 5.00 & 4.25 & 5.25 & 1.00 & 4.72 & 0.966 & 0.193 & 4.35 & 5.08\\
DesertIsland & Static & 25 & 3.25 & 6 & 5.25 & 4.75 & 5.50 & 0.75 & 5.11 & 0.693 & 0.139 & 4.85 & 5.37\\
DesertIsland & Dynamic & 25 & 3.25 & 6 & 5.00 & 4.25 & 5.50 & 1.25 & 5.01 & 0.716 & 0.143 & 4.72 & 5.28\\
SnowyMountains & Control & 24 & 1.75 & 6 & 4.75 & 3.94 & 5.31 & 1.38 & 4.62 & 1.011 & 0.206 & 4.24 & 5.00\\
SnowyMountains & Static & 24 & 2.00 & 6 & 4.88 & 4.19 & 5.50 & 1.31 & 4.70 & 0.944 & 0.193 & 4.31 & 5.04\\

SnowyMountains & Dynamic & 24 & 2.25 & 6 & 4.75 & 4.44 & 5.50 & 1.06 & 4.78 & 0.907 & 0.185 & 4.41 & 5.12\\
\bottomrule
\end{tabular}
\end{table}
\begin{table}[H]

\caption{Statistical summary by experimental condition for Thermal Sensation}
\centering
\begin{tabular}[t]{llrrrrrrrrrrrr}
\toprule
Experiment & Condition & n & min & max & median & q1 & q3 & iqr & mean & sd & se & ci95\_lower & ci95\_upper\\
\midrule
DesertIsland & Control & 25 & -2 & 2 & 0 & 0 & 1.00 & 1.00 & 0.360 & 0.860 & 0.172 & 0.000 & 0.720\\
DesertIsland & Static & 25 & 1 & 3 & 2 & 2 & 2.00 & 0.00 & 1.880 & 0.600 & 0.120 & 1.640 & 2.120\\
DesertIsland & Dynamic & 25 & 0 & 3 & 2 & 1 & 2.00 & 1.00 & 1.680 & 0.748 & 0.150 & 1.400 & 1.960\\
SnowyMountains & Control & 24 & -2 & 1 & 0 & 0 & 0.25 & 0.25 & 0.083 & 0.717 & 0.146 & -0.208 & 0.375\\
SnowyMountains & Static & 24 & -3 & 0 & -1 & -2 & -1.00 & 1.00 & -1.250 & 0.676 & 0.138 & -1.501 & -1.000\\

SnowyMountains & Dynamic & 24 & -2 & 0 & -1 & -2 & -1.00 & 1.00 & -1.208 & 0.721 & 0.147 & -1.500 & -0.917\\
\bottomrule
\end{tabular}
\end{table}
\begin{table}[H]

\caption{Statistical summary by experimental condition for Thermal Variation}
\centering
\begin{tabular}[t]{llrrrrrrrrrrrr}
\toprule
Experiment & Condition & n & min & max & median & q1 & q3 & iqr & mean & sd & se & ci95\_lower & ci95\_upper\\
\midrule
DesertIsland & Control & 25 & 0 & 2 & 0 & 0.00 & 1 & 1.00 & 0.600 & 0.764 & 0.153 & 0.320 & 0.920\\
DesertIsland & Static & 25 & 0 & 3 & 2 & 2.00 & 2 & 0.00 & 1.840 & 0.800 & 0.160 & 1.520 & 2.120\\
DesertIsland & Dynamic & 25 & 0 & 3 & 2 & 2.00 & 3 & 1.00 & 2.160 & 0.746 & 0.149 & 1.840 & 2.401\\
SnowyMountains & Control & 24 & 0 & 2 & 0 & 0.00 & 0 & 0.00 & 0.292 & 0.624 & 0.127 & 0.083 & 0.542\\
SnowyMountains & Static & 24 & 0 & 3 & 2 & 1.00 & 2 & 1.00 & 1.542 & 0.884 & 0.180 & 1.167 & 1.875\\

SnowyMountains & Dynamic & 24 & 0 & 3 & 2 & 1.75 & 2 & 0.25 & 1.833 & 0.702 & 0.143 & 1.542 & 2.083\\
\bottomrule
\end{tabular}
\end{table}

\subsubsection{Statistical summary of post-immersion questionnaire by order of passage}

\phantom{*}

\begin{table}[H]

\caption{Statistical summary by order of passage for HX Expressivity}
\centering
\begin{tabular}[t]{llrrrrrrrrrrrr}
\toprule
Experiment & Order & n & min & max & median & q1 & q3 & iqr & mean & sd & se & ci95\_lower & ci95\_upper\\
\midrule
DesertIsland & 1 & 25 & 0 & 4.00 & 2.33 & 1.000 & 2.67 & 1.67 & 2.01 & 1.27 & 0.254 & 1.51 & 2.49\\
DesertIsland & 2 & 25 & 0 & 4.00 & 3.00 & 1.667 & 3.33 & 1.67 & 2.45 & 1.42 & 0.284 & 1.89 & 2.97\\
DesertIsland & 3 & 25 & 0 & 4.00 & 3.00 & 1.000 & 3.33 & 2.33 & 2.33 & 1.40 & 0.280 & 1.76 & 2.88\\
SnowyMountains & 1 & 24 & 0 & 3.33 & 1.50 & 0.333 & 2.00 & 1.67 & 1.42 & 1.07 & 0.219 & 1.01 & 1.86\\
SnowyMountains & 2 & 24 & 0 & 4.00 & 1.67 & 0.583 & 3.00 & 2.42 & 1.76 & 1.33 & 0.272 & 1.28 & 2.33\\

SnowyMountains & 3 & 24 & 0 & 4.00 & 2.33 & 1.250 & 3.42 & 2.17 & 2.15 & 1.39 & 0.283 & 1.60 & 2.69\\
\bottomrule
\end{tabular}
\end{table}
\begin{table}[H]

\caption{Statistical summary by order of passage for HX Realism}
\centering
\begin{tabular}[t]{llrrrrrrrrrrrr}
\toprule
Experiment & Order & n & min & max & median & q1 & q3 & iqr & mean & sd & se & ci95\_lower & ci95\_upper\\
\midrule
DesertIsland & 1 & 25 & 0 & 4.00 & 3.00 & 1.333 & 3.33 & 2.000 & 2.37 & 1.41 & 0.282 & 1.79 & 2.89\\
DesertIsland & 2 & 25 & 0 & 4.00 & 3.33 & 2.000 & 3.67 & 1.667 & 2.77 & 1.39 & 0.279 & 2.19 & 3.27\\
DesertIsland & 3 & 25 & 0 & 4.00 & 3.33 & 3.000 & 3.67 & 0.667 & 2.95 & 1.25 & 0.251 & 2.45 & 3.40\\
SnowyMountains & 1 & 24 & 0 & 3.33 & 2.50 & 1.000 & 3.00 & 2.000 & 2.04 & 1.13 & 0.230 & 1.57 & 2.46\\
SnowyMountains & 2 & 24 & 0 & 4.00 & 2.50 & 0.667 & 3.33 & 2.667 & 2.12 & 1.40 & 0.286 & 1.62 & 2.67\\

SnowyMountains & 3 & 24 & 0 & 4.00 & 3.00 & 1.917 & 3.67 & 1.750 & 2.56 & 1.44 & 0.293 & 1.97 & 3.10\\
\bottomrule
\end{tabular}
\end{table}
\begin{table}[H]

\caption{Statistical summary by order of passage for HX Involvement}
\centering
\begin{tabular}[t]{llrrrrrrrrrrrr}
\toprule
Experiment & Order & n & min & max & median & q1 & q3 & iqr & mean & sd & se & ci95\_lower & ci95\_upper\\
\midrule
DesertIsland & 1 & 25 & 0.0 & 4 & 3.5 & 2.00 & 4.00 & 2.00 & 2.84 & 1.231 & 0.246 & 2.36 & 3.28\\
DesertIsland & 2 & 25 & 1.5 & 4 & 3.5 & 2.00 & 4.00 & 2.00 & 3.20 & 0.924 & 0.185 & 2.86 & 3.52\\
DesertIsland & 3 & 25 & 1.0 & 4 & 3.5 & 3.00 & 4.00 & 1.00 & 3.32 & 0.865 & 0.173 & 2.96 & 3.62\\
SnowyMountains & 1 & 24 & 0.0 & 4 & 3.0 & 2.38 & 3.62 & 1.25 & 2.96 & 0.988 & 0.202 & 2.56 & 3.29\\
SnowyMountains & 2 & 24 & 0.0 & 4 & 3.0 & 2.00 & 3.50 & 1.50 & 2.69 & 1.051 & 0.215 & 2.27 & 3.06\\

SnowyMountains & 3 & 24 & 0.0 & 4 & 3.5 & 2.50 & 3.62 & 1.12 & 2.88 & 1.329 & 0.271 & 2.33 & 3.33\\
\bottomrule
\end{tabular}
\end{table}
\begin{table}[H]

\caption{Statistical summary by order of passage for HX Harmony}
\centering
\begin{tabular}[t]{llrrrrrrrrrrrr}
\toprule
Experiment & Order & n & min & max & median & q1 & q3 & iqr & mean & sd & se & ci95\_lower & ci95\_upper\\
\midrule
DesertIsland & 1 & 25 & 1.333 & 4 & 3.67 & 3.00 & 4.00 & 1.00 & 3.25 & 0.801 & 0.160 & 2.96 & 3.56\\
DesertIsland & 2 & 25 & 0.667 & 4 & 3.33 & 2.33 & 3.67 & 1.33 & 3.04 & 0.909 & 0.182 & 2.68 & 3.39\\
DesertIsland & 3 & 25 & 1.333 & 4 & 3.67 & 2.67 & 4.00 & 1.33 & 3.19 & 0.856 & 0.171 & 2.87 & 3.49\\
SnowyMountains & 1 & 24 & 1.333 & 4 & 3.33 & 2.92 & 3.67 & 0.75 & 3.15 & 0.840 & 0.171 & 2.81 & 3.46\\
SnowyMountains & 2 & 24 & 1.000 & 4 & 2.83 & 1.67 & 3.67 & 2.00 & 2.74 & 1.040 & 0.212 & 2.29 & 3.14\\

SnowyMountains & 3 & 24 & 1.333 & 4 & 3.17 & 2.67 & 3.75 & 1.08 & 3.00 & 0.906 & 0.185 & 2.64 & 3.33\\
\bottomrule
\end{tabular}
\end{table}
\begin{table}[H]

\caption{Statistical summary by order of passage for General Presence}
\centering
\begin{tabular}[t]{llrrrrrrrrrrrr}
\toprule
Experiment & Order & n & min & max & median & q1 & q3 & iqr & mean & sd & se & ci95\_lower & ci95\_upper\\
\midrule
DesertIsland & 1 & 25 & 4 & 6 & 5 & 5 & 6 & 1 & 5.12 & 0.781 & 0.156 & 4.84 & 5.40\\
DesertIsland & 2 & 25 & 2 & 6 & 5 & 5 & 5 & 0 & 4.84 & 1.028 & 0.206 & 4.44 & 5.20\\
DesertIsland & 3 & 25 & 3 & 6 & 5 & 4 & 6 & 2 & 4.92 & 1.077 & 0.215 & 4.48 & 5.32\\
SnowyMountains & 1 & 24 & 4 & 6 & 5 & 5 & 6 & 1 & 5.29 & 0.690 & 0.141 & 5.04 & 5.58\\
SnowyMountains & 2 & 24 & 2 & 6 & 5 & 4 & 6 & 2 & 4.88 & 1.035 & 0.211 & 4.46 & 5.25\\

SnowyMountains & 3 & 24 & 0 & 6 & 5 & 4 & 6 & 2 & 4.67 & 1.551 & 0.317 & 4.04 & 5.25\\
\bottomrule
\end{tabular}
\end{table}
\begin{table}[H]

\caption{Statistical summary by order of passage for SPES Spatial Location}
\centering
\begin{tabular}[t]{llrrrrrrrrrrrr}
\toprule
Experiment & Order & n & min & max & median & q1 & q3 & iqr & mean & sd & se & ci95\_lower & ci95\_upper\\
\midrule
DesertIsland & 1 & 25 & 2.50 & 6.00 & 4.75 & 4.25 & 5.25 & 1.000 & 4.64 & 0.854 & 0.171 & 4.31 & 4.96\\
DesertIsland & 2 & 25 & 1.50 & 6.00 & 5.00 & 3.75 & 5.50 & 1.750 & 4.58 & 1.231 & 0.246 & 4.09 & 5.03\\
DesertIsland & 3 & 25 & 2.50 & 6.00 & 4.50 & 4.00 & 6.00 & 2.000 & 4.70 & 1.046 & 0.209 & 4.31 & 5.10\\
SnowyMountains & 1 & 24 & 2.25 & 6.00 & 4.88 & 3.94 & 5.25 & 1.312 & 4.57 & 1.031 & 0.210 & 4.20 & 4.96\\
SnowyMountains & 2 & 24 & 2.00 & 5.75 & 4.75 & 3.50 & 5.25 & 1.750 & 4.42 & 1.010 & 0.206 & 4.00 & 4.82\\

SnowyMountains & 3 & 24 & 1.25 & 6.00 & 4.75 & 4.25 & 5.06 & 0.812 & 4.42 & 1.255 & 0.256 & 3.91 & 4.92\\
\bottomrule
\end{tabular}
\end{table}
\begin{table}[H]

\caption{Statistical summary by order of passage for SPES Possible Actions}
\centering
\begin{tabular}[t]{llrrrrrrrrrrrr}
\toprule
Experiment & Order & n & min & max & median & q1 & q3 & iqr & mean & sd & se & ci95\_lower & ci95\_upper\\
\midrule
DesertIsland & 1 & 25 & 3.00 & 6 & 5.00 & 4.25 & 5.25 & 1.00 & 4.80 & 0.764 & 0.153 & 4.49 & 5.08\\
DesertIsland & 2 & 25 & 2.75 & 6 & 5.25 & 4.75 & 5.75 & 1.00 & 4.99 & 0.876 & 0.175 & 4.63 & 5.31\\
DesertIsland & 3 & 25 & 2.75 & 6 & 5.00 & 4.75 & 5.75 & 1.00 & 5.05 & 0.791 & 0.158 & 4.71 & 5.35\\
SnowyMountains & 1 & 24 & 2.00 & 6 & 4.75 & 4.00 & 5.50 & 1.50 & 4.65 & 0.947 & 0.193 & 4.27 & 5.00\\
SnowyMountains & 2 & 24 & 2.25 & 6 & 5.00 & 4.00 & 5.50 & 1.50 & 4.71 & 0.980 & 0.200 & 4.31 & 5.06\\

SnowyMountains & 3 & 24 & 1.75 & 6 & 4.75 & 4.25 & 5.50 & 1.25 & 4.75 & 0.941 & 0.192 & 4.37 & 5.12\\
\bottomrule
\end{tabular}
\end{table}
\begin{table}[H]

\caption{Statistical summary by order of passage for Thermal Sensation}
\centering
\begin{tabular}[t]{llrrrrrrrrrrrr}
\toprule
Experiment & Order & n & min & max & median & q1 & q3 & iqr & mean & sd & se & ci95\_lower & ci95\_upper\\
\midrule
DesertIsland & 1 & 25 & 0 & 2 & 1 & 0 & 2 & 2 & 1.120 & 0.833 & 0.167 & 0.84 & 1.440\\
DesertIsland & 2 & 25 & -2 & 3 & 1 & 1 & 2 & 1 & 1.320 & 1.145 & 0.229 & 0.92 & 1.760\\
DesertIsland & 3 & 25 & -1 & 3 & 2 & 1 & 2 & 1 & 1.480 & 1.005 & 0.201 & 1.08 & 1.840\\
SnowyMountains & 1 & 24 & -3 & 1 & -1 & -1 & 0 & 1 & -0.750 & 0.944 & 0.193 & -1.08 & -0.417\\
SnowyMountains & 2 & 24 & -2 & 1 & -1 & -1 & 0 & 1 & -0.708 & 0.908 & 0.185 & -1.08 & -0.375\\

SnowyMountains & 3 & 24 & -2 & 1 & -1 & -2 & 0 & 2 & -0.917 & 0.974 & 0.199 & -1.29 & -0.542\\
\bottomrule
\end{tabular}
\end{table}
\begin{table}[H]

\caption{Statistical summary by order of passage for Thermal Variation}
\centering
\begin{tabular}[t]{llrrrrrrrrrrrr}
\toprule
Experiment & Order & n & min & max & median & q1 & q3 & iqr & mean & sd & se & ci95\_lower & ci95\_upper\\
\midrule
DesertIsland & 1 & 25 & 0 & 3 & 1 & 0.00 & 2.00 & 2.00 & 1.280 & 0.980 & 0.196 & 0.880 & 1.68\\
DesertIsland & 2 & 25 & 0 & 3 & 2 & 1.00 & 3.00 & 2.00 & 1.800 & 1.080 & 0.216 & 1.400 & 2.20\\
DesertIsland & 3 & 25 & 0 & 3 & 2 & 1.00 & 2.00 & 1.00 & 1.520 & 0.963 & 0.193 & 1.120 & 1.88\\
SnowyMountains & 1 & 24 & 0 & 2 & 1 & 0.00 & 1.25 & 1.25 & 0.875 & 0.797 & 0.163 & 0.583 & 1.21\\
SnowyMountains & 2 & 24 & 0 & 2 & 2 & 0.00 & 2.00 & 2.00 & 1.250 & 0.989 & 0.202 & 0.833 & 1.67\\

SnowyMountains & 3 & 24 & 0 & 3 & 2 & 0.75 & 2.00 & 1.25 & 1.542 & 1.103 & 0.225 & 1.125 & 1.96\\
\bottomrule
\end{tabular}
\end{table}

\subsubsection{Statistical summary of behavioral metrics by experimental condition}

\phantom{*}

\begin{table}[H]

\caption{Statistical summary by experimental condition for TotalLocalDistance}
\centering
\begin{tabular}[t]{llrrrrrrrrrrrr}
\toprule
Experiment & Condition & n & min & max & median & q1 & q3 & iqr & mean & sd & se & ci95\_lower & ci95\_upper\\
\midrule
DesertIsland & Control & 25 & 16.6 & 86.5 & 32.8 & 23.9 & 37.8 & 13.9 & 35.2 & 16.1 & 3.22 & 29.6 & 42.1\\
DesertIsland & Static & 25 & 19.0 & 71.6 & 34.1 & 24.6 & 44.9 & 20.2 & 37.5 & 15.7 & 3.13 & 31.9 & 43.7\\
DesertIsland & Dynamic & 25 & 14.5 & 64.9 & 32.3 & 24.9 & 44.5 & 19.6 & 34.4 & 13.5 & 2.69 & 29.4 & 39.7\\
SnowyMountains & Control & 24 & 24.4 & 85.1 & 40.5 & 32.9 & 45.8 & 12.9 & 43.2 & 15.6 & 3.18 & 37.9 & 49.1\\
SnowyMountains & Static & 24 & 25.1 & 60.9 & 40.8 & 31.3 & 49.4 & 18.1 & 40.3 & 10.4 & 2.11 & 36.1 & 44.0\\

SnowyMountains & Dynamic & 24 & 20.5 & 105.3 & 40.5 & 34.0 & 47.4 & 13.4 & 42.6 & 17.3 & 3.52 & 36.6 & 50.7\\
\bottomrule
\end{tabular}
\end{table}
\begin{table}[H]

\caption{Statistical summary by experimental condition for TotalWorldDistance}
\centering
\begin{tabular}[t]{llrrrrrrrrrrrr}
\toprule
Experiment & Condition & n & min & max & median & q1 & q3 & iqr & mean & sd & se & ci95\_lower & ci95\_upper\\
\midrule
DesertIsland & Control & 25 & 112 & 532 & 239 & 197 & 278 & 81.7 & 259 & 97.6 & 19.5 & 226 & 299\\
DesertIsland & Static & 25 & 164 & 616 & 232 & 204 & 272 & 68.7 & 260 & 97.9 & 19.6 & 228 & 299\\
DesertIsland & Dynamic & 25 & 155 & 745 & 244 & 195 & 302 & 107.3 & 268 & 117.2 & 23.4 & 230 & 318\\
SnowyMountains & Control & 24 & 193 & 500 & 280 & 254 & 320 & 66.1 & 292 & 61.6 & 12.6 & 270 & 316\\
SnowyMountains & Static & 24 & 135 & 453 & 284 & 264 & 319 & 55.3 & 287 & 68.4 & 14.0 & 261 & 315\\

SnowyMountains & Dynamic & 24 & 252 & 465 & 291 & 264 & 315 & 50.8 & 300 & 49.9 & 10.2 & 282 & 321\\
\bottomrule
\end{tabular}
\end{table}
\begin{table}[H]

\caption{Statistical summary by experimental condition for TotalLocalRotationDistance}
\centering
\begin{tabular}[t]{llrrrrrrrrrrrr}
\toprule
Experiment & Condition & n & min & max & median & q1 & q3 & iqr & mean & sd & se & ci95\_lower & ci95\_upper\\
\midrule
DesertIsland & Control & 25 & 53.6 & 303 & 106 & 87.1 & 156 & 68.5 & 126 & 60.9 & 12.18 & 104 & 151\\
DesertIsland & Static & 25 & 62.9 & 243 & 124 & 84.7 & 160 & 75.0 & 127 & 49.2 & 9.85 & 109 & 147\\
DesertIsland & Dynamic & 25 & 45.3 & 288 & 103 & 84.1 & 168 & 83.4 & 123 & 57.7 & 11.54 & 101 & 148\\
SnowyMountains & Control & 24 & 73.4 & 300 & 110 & 92.2 & 141 & 48.3 & 131 & 57.9 & 11.82 & 108 & 156\\
SnowyMountains & Static & 24 & 56.9 & 236 & 124 & 94.9 & 153 & 57.8 & 128 & 46.6 & 9.51 & 109 & 147\\

SnowyMountains & Dynamic & 24 & 55.6 & 282 & 124 & 100.6 & 147 & 46.2 & 132 & 54.5 & 11.13 & 112 & 153\\
\bottomrule
\end{tabular}
\end{table}
\begin{table}[H]

\caption{Statistical summary by experimental condition for TotalWorldRotationDistance}
\centering
\begin{tabular}[t]{llrrrrrrrrrrrr}
\toprule
Experiment & Condition & n & min & max & median & q1 & q3 & iqr & mean & sd & se & ci95\_lower & ci95\_upper\\
\midrule
DesertIsland & Control & 25 & 82.6 & 372 & 128 & 113 & 185 & 72.0 & 156 & 68.3 & 13.66 & 133 & 184\\
DesertIsland & Static & 25 & 85.4 & 275 & 154 & 109 & 187 & 77.8 & 154 & 49.1 & 9.83 & 137 & 174\\
DesertIsland & Dynamic & 25 & 62.5 & 367 & 128 & 101 & 176 & 75.3 & 150 & 71.5 & 14.29 & 125 & 180\\
SnowyMountains & Control & 24 & 87.8 & 407 & 163 & 128 & 211 & 83.5 & 176 & 73.2 & 14.94 & 149 & 207\\
SnowyMountains & Static & 24 & 74.2 & 298 & 156 & 128 & 223 & 94.3 & 173 & 61.0 & 12.46 & 151 & 195\\

SnowyMountains & Dynamic & 24 & 100.7 & 293 & 154 & 129 & 176 & 46.2 & 167 & 53.1 & 10.83 & 147 & 188\\
\bottomrule
\end{tabular}
\end{table}
\begin{table}[H]

\caption{Statistical summary by experimental condition for LocalRoamingEntropy}
\centering
\begin{tabular}[t]{llrrrrrrrrrrrr}
\toprule
Experiment & Condition & n & min & max & median & q1 & q3 & iqr & mean & sd & se & ci95\_lower & ci95\_upper\\
\midrule
DesertIsland & Control & 25 & 0.661 & 0.883 & 0.784 & 0.751 & 0.821 & 0.070 & 0.791 & 0.050 & 0.010 & 0.771 & 0.810\\
DesertIsland & Static & 25 & 0.715 & 0.867 & 0.818 & 0.788 & 0.838 & 0.050 & 0.806 & 0.044 & 0.009 & 0.789 & 0.822\\
DesertIsland & Dynamic & 25 & 0.694 & 0.850 & 0.803 & 0.751 & 0.824 & 0.073 & 0.785 & 0.047 & 0.009 & 0.767 & 0.803\\
SnowyMountains & Control & 24 & 0.763 & 0.901 & 0.843 & 0.823 & 0.861 & 0.038 & 0.840 & 0.031 & 0.006 & 0.827 & 0.851\\
SnowyMountains & Static & 24 & 0.730 & 0.885 & 0.836 & 0.812 & 0.866 & 0.054 & 0.830 & 0.045 & 0.009 & 0.813 & 0.846\\

SnowyMountains & Dynamic & 24 & 0.750 & 0.899 & 0.844 & 0.813 & 0.863 & 0.050 & 0.838 & 0.039 & 0.008 & 0.823 & 0.852\\
\bottomrule
\end{tabular}
\end{table}
\begin{table}[H]

\caption{Statistical summary by experimental condition for WorldRoamingEntropy}
\centering
\begin{tabular}[t]{llrrrrrrrrrrrr}
\toprule
Experiment & Condition & n & min & max & median & q1 & q3 & iqr & mean & sd & se & ci95\_lower & ci95\_upper\\
\midrule
DesertIsland & Control & 25 & 0.680 & 0.905 & 0.845 & 0.823 & 0.864 & 0.041 & 0.836 & 0.050 & 0.010 & 0.814 & 0.852\\
DesertIsland & Static & 25 & 0.765 & 0.888 & 0.836 & 0.816 & 0.856 & 0.041 & 0.833 & 0.032 & 0.006 & 0.821 & 0.845\\
DesertIsland & Dynamic & 25 & 0.724 & 0.908 & 0.843 & 0.821 & 0.859 & 0.038 & 0.838 & 0.041 & 0.008 & 0.821 & 0.853\\
SnowyMountains & Control & 24 & 0.770 & 0.912 & 0.851 & 0.844 & 0.875 & 0.031 & 0.854 & 0.029 & 0.006 & 0.841 & 0.865\\
SnowyMountains & Static & 24 & 0.604 & 0.894 & 0.855 & 0.841 & 0.870 & 0.030 & 0.838 & 0.068 & 0.014 & 0.809 & 0.861\\

SnowyMountains & Dynamic & 24 & 0.777 & 0.902 & 0.861 & 0.845 & 0.881 & 0.035 & 0.860 & 0.029 & 0.006 & 0.849 & 0.871\\
\bottomrule
\end{tabular}
\end{table}
\begin{table}[H]

\caption{Statistical summary by experimental condition for LocalMovementEntropy}
\centering
\begin{tabular}[t]{llrrrrrrrrrrrr}
\toprule
Experiment & Condition & n & min & max & median & q1 & q3 & iqr & mean & sd & se & ci95\_lower & ci95\_upper\\
\midrule
DesertIsland & Control & 25 & 0.737 & 0.918 & 0.879 & 0.856 & 0.895 & 0.039 & 0.871 & 0.039 & 0.008 & 0.854 & 0.884\\
DesertIsland & Static & 25 & 0.817 & 0.912 & 0.882 & 0.869 & 0.898 & 0.029 & 0.879 & 0.025 & 0.005 & 0.868 & 0.888\\
DesertIsland & Dynamic & 25 & 0.832 & 0.916 & 0.870 & 0.859 & 0.889 & 0.030 & 0.872 & 0.022 & 0.004 & 0.864 & 0.881\\
SnowyMountains & Control & 24 & 0.844 & 0.909 & 0.878 & 0.870 & 0.890 & 0.020 & 0.878 & 0.017 & 0.003 & 0.871 & 0.885\\
SnowyMountains & Static & 24 & 0.755 & 0.908 & 0.874 & 0.857 & 0.886 & 0.029 & 0.868 & 0.031 & 0.006 & 0.855 & 0.879\\

SnowyMountains & Dynamic & 24 & 0.824 & 0.905 & 0.884 & 0.867 & 0.893 & 0.025 & 0.877 & 0.022 & 0.004 & 0.868 & 0.885\\
\bottomrule
\end{tabular}
\end{table}
\begin{table}[H]

\caption{Statistical summary by experimental condition for WorldMovementEntropy}
\centering
\begin{tabular}[t]{llrrrrrrrrrrrr}
\toprule
Experiment & Condition & n & min & max & median & q1 & q3 & iqr & mean & sd & se & ci95\_lower & ci95\_upper\\
\midrule
DesertIsland & Control & 25 & 0.729 & 0.926 & 0.896 & 0.872 & 0.904 & 0.032 & 0.880 & 0.043 & 0.009 & 0.863 & 0.895\\
DesertIsland & Static & 25 & 0.822 & 0.921 & 0.886 & 0.873 & 0.900 & 0.027 & 0.885 & 0.023 & 0.005 & 0.876 & 0.893\\
DesertIsland & Dynamic & 25 & 0.840 & 0.922 & 0.881 & 0.874 & 0.897 & 0.023 & 0.885 & 0.020 & 0.004 & 0.878 & 0.892\\
SnowyMountains & Control & 24 & 0.856 & 0.918 & 0.891 & 0.875 & 0.900 & 0.025 & 0.888 & 0.016 & 0.003 & 0.881 & 0.894\\
SnowyMountains & Static & 24 & 0.783 & 0.914 & 0.886 & 0.863 & 0.892 & 0.030 & 0.876 & 0.032 & 0.007 & 0.863 & 0.887\\

SnowyMountains & Dynamic & 24 & 0.829 & 0.910 & 0.889 & 0.869 & 0.896 & 0.027 & 0.882 & 0.020 & 0.004 & 0.873 & 0.889\\
\bottomrule
\end{tabular}
\end{table}
\begin{table}[H]

\caption{Statistical summary by experimental condition for NumberOfTeleportations}
\centering
\begin{tabular}[t]{llrrrrrrrrrrrr}
\toprule
Experiment & Condition & n & min & max & median & q1 & q3 & iqr & mean & sd & se & ci95\_lower & ci95\_upper\\
\midrule
DesertIsland & Control & 25 & 16 & 117 & 37.0 & 32.0 & 65.0 & 33.0 & 49.0 & 24.7 & 4.94 & 40.0 & 58.9\\
DesertIsland & Static & 25 & 24 & 114 & 35.0 & 31.0 & 50.0 & 19.0 & 45.4 & 23.6 & 4.72 & 37.7 & 55.5\\
DesertIsland & Dynamic & 25 & 19 & 105 & 44.0 & 26.0 & 63.0 & 37.0 & 46.3 & 22.0 & 4.40 & 38.2 & 54.6\\
SnowyMountains & Control & 24 & 29 & 85 & 47.5 & 37.0 & 64.2 & 27.2 & 51.2 & 15.6 & 3.19 & 44.8 & 56.8\\
SnowyMountains & Static & 24 & 28 & 86 & 50.5 & 36.8 & 58.5 & 21.8 & 50.2 & 16.4 & 3.35 & 43.9 & 56.8\\

SnowyMountains & Dynamic & 24 & 27 & 75 & 46.0 & 40.2 & 58.2 & 18.0 & 48.2 & 12.7 & 2.60 & 43.0 & 53.1\\
\bottomrule
\end{tabular}
\end{table}

\subsubsection{Statistical summary of behavioral metrics by order of passage}

\phantom{*}

\begin{table}[H]

\caption{Statistical summary by order of passage for TotalLocalDistance}
\centering
\begin{tabular}[t]{llrrrrrrrrrrrr}
\toprule
Experiment & Order & n & min & max & median & q1 & q3 & iqr & mean & sd & se & ci95\_lower & ci95\_upper\\
\midrule
DesertIsland & 1 & 25 & 16.6 & 86.5 & 45.9 & 34.1 & 62.5 & 28.5 & 47.4 & 17.47 & 3.49 & 41.1 & 54.5\\
DesertIsland & 2 & 25 & 16.4 & 53.3 & 32.7 & 23.7 & 40.9 & 17.2 & 32.0 & 10.36 & 2.07 & 28.1 & 35.9\\
DesertIsland & 3 & 25 & 14.5 & 41.8 & 25.8 & 22.7 & 34.0 & 11.3 & 27.7 & 7.38 & 1.48 & 24.9 & 30.8\\
SnowyMountains & 1 & 24 & 24.0 & 85.1 & 49.2 & 44.3 & 58.2 & 13.9 & 50.8 & 12.58 & 2.57 & 46.2 & 55.7\\
SnowyMountains & 2 & 24 & 24.6 & 105.3 & 40.1 & 30.2 & 42.6 & 12.4 & 39.9 & 15.98 & 3.26 & 34.5 & 46.9\\

SnowyMountains & 3 & 24 & 20.5 & 69.1 & 34.0 & 28.3 & 39.4 & 11.1 & 35.4 & 10.43 & 2.13 & 31.4 & 39.7\\
\bottomrule
\end{tabular}
\end{table}
\begin{table}[H]

\caption{Statistical summary by order of passage for TotalWorldDistance}
\centering
\begin{tabular}[t]{llrrrrrrrrrrrr}
\toprule
Experiment & Order & n & min & max & median & q1 & q3 & iqr & mean & sd & se & ci95\_lower & ci95\_upper\\
\midrule
DesertIsland & 1 & 25 & 112 & 745 & 270 & 266 & 339 & 72.9 & 321 & 145.3 & 29.07 & 268 & 381\\
DesertIsland & 2 & 25 & 155 & 442 & 231 & 204 & 266 & 62.1 & 245 & 63.5 & 12.70 & 222 & 271\\
DesertIsland & 3 & 25 & 164 & 319 & 204 & 188 & 231 & 43.2 & 220 & 44.1 & 8.82 & 204 & 238\\
SnowyMountains & 1 & 24 & 135 & 500 & 310 & 279 & 343 & 63.8 & 315 & 89.9 & 18.36 & 282 & 349\\
SnowyMountains & 2 & 24 & 193 & 341 & 277 & 267 & 292 & 25.0 & 280 & 30.3 & 6.18 & 268 & 291\\

SnowyMountains & 3 & 24 & 246 & 360 & 275 & 255 & 313 & 57.9 & 285 & 36.0 & 7.34 & 273 & 301\\
\bottomrule
\end{tabular}
\end{table}
\begin{table}[H]

\caption{Statistical summary by order of passage for TotalLocalRotationDistance}
\centering
\begin{tabular}[t]{llrrrrrrrrrrrr}
\toprule
Experiment & Order & n & min & max & median & q1 & q3 & iqr & mean & sd & se & ci95\_lower & ci95\_upper\\
\midrule
DesertIsland & 1 & 25 & 87.1 & 303 & 169.8 & 128.9 & 208 & 79.2 & 170.2 & 60.9 & 12.19 & 147.6 & 194\\
DesertIsland & 2 & 25 & 52.6 & 201 & 100.8 & 73.4 & 148 & 75.1 & 112.8 & 42.3 & 8.46 & 97.5 & 129\\
DesertIsland & 3 & 25 & 45.3 & 125 & 95.0 & 76.0 & 116 & 40.1 & 92.7 & 24.4 & 4.88 & 83.1 & 102\\
SnowyMountains & 1 & 24 & 107.4 & 300 & 157.4 & 134.9 & 202 & 67.4 & 173.1 & 50.0 & 10.21 & 153.1 & 193\\
SnowyMountains & 2 & 24 & 55.6 & 282 & 104.3 & 90.4 & 123 & 33.0 & 112.4 & 43.1 & 8.81 & 97.8 & 133\\

SnowyMountains & 3 & 24 & 56.9 & 211 & 97.4 & 82.0 & 116 & 33.8 & 104.6 & 34.7 & 7.09 & 93.1 & 118\\
\bottomrule
\end{tabular}
\end{table}
\begin{table}[H]

\caption{Statistical summary by order of passage for TotalWorldRotationDistance}
\centering
\begin{tabular}[t]{llrrrrrrrrrrrr}
\toprule
Experiment & Order & n & min & max & median & q1 & q3 & iqr & mean & sd & se & ci95\_lower & ci95\_upper\\
\midrule
DesertIsland & 1 & 25 & 102.1 & 372 & 193 & 167.0 & 242 & 74.7 & 205 & 72.2 & 14.44 & 178 & 233\\
DesertIsland & 2 & 25 & 75.7 & 230 & 129 & 109.2 & 165 & 56.0 & 140 & 40.9 & 8.18 & 126 & 157\\
DesertIsland & 3 & 25 & 62.5 & 185 & 111 & 98.9 & 132 & 33.4 & 116 & 31.0 & 6.20 & 104 & 129\\
SnowyMountains & 1 & 24 & 139.9 & 407 & 210 & 169.5 & 258 & 88.6 & 221 & 62.2 & 12.69 & 196 & 245\\
SnowyMountains & 2 & 24 & 88.0 & 288 & 135 & 116.5 & 158 & 41.5 & 148 & 49.0 & 9.99 & 131 & 168\\

SnowyMountains & 3 & 24 & 74.2 & 243 & 146 & 119.3 & 166 & 46.2 & 147 & 43.9 & 8.96 & 130 & 164\\
\bottomrule
\end{tabular}
\end{table}
\begin{table}[H]

\caption{Statistical summary by order of passage for LocalRoamingEntropy}
\centering
\begin{tabular}[t]{llrrrrrrrrrrrr}
\toprule
Experiment & Order & n & min & max & median & q1 & q3 & iqr & mean & sd & se & ci95\_lower & ci95\_upper\\
\midrule
DesertIsland & 1 & 25 & 0.745 & 0.883 & 0.821 & 0.773 & 0.845 & 0.073 & 0.814 & 0.042 & 0.008 & 0.799 & 0.830\\
DesertIsland & 2 & 25 & 0.694 & 0.871 & 0.790 & 0.743 & 0.830 & 0.087 & 0.788 & 0.051 & 0.010 & 0.768 & 0.807\\
DesertIsland & 3 & 25 & 0.661 & 0.836 & 0.791 & 0.751 & 0.815 & 0.064 & 0.780 & 0.044 & 0.009 & 0.763 & 0.796\\
SnowyMountains & 1 & 24 & 0.730 & 0.899 & 0.849 & 0.815 & 0.868 & 0.054 & 0.840 & 0.040 & 0.008 & 0.825 & 0.855\\
SnowyMountains & 2 & 24 & 0.752 & 0.901 & 0.838 & 0.809 & 0.852 & 0.043 & 0.834 & 0.037 & 0.007 & 0.819 & 0.848\\

SnowyMountains & 3 & 24 & 0.736 & 0.888 & 0.830 & 0.816 & 0.861 & 0.045 & 0.833 & 0.040 & 0.008 & 0.816 & 0.847\\
\bottomrule
\end{tabular}
\end{table}
\begin{table}[H]

\caption{Statistical summary by order of passage for WorldRoamingEntropy}
\centering
\begin{tabular}[t]{llrrrrrrrrrrrr}
\toprule
Experiment & Order & n & min & max & median & q1 & q3 & iqr & mean & sd & se & ci95\_lower & ci95\_upper\\
\midrule
DesertIsland & 1 & 25 & 0.735 & 0.888 & 0.848 & 0.831 & 0.862 & 0.032 & 0.843 & 0.031 & 0.006 & 0.829 & 0.853\\
DesertIsland & 2 & 25 & 0.724 & 0.908 & 0.834 & 0.802 & 0.858 & 0.056 & 0.829 & 0.045 & 0.009 & 0.811 & 0.847\\
DesertIsland & 3 & 25 & 0.680 & 0.905 & 0.843 & 0.816 & 0.858 & 0.042 & 0.834 & 0.046 & 0.009 & 0.816 & 0.850\\
SnowyMountains & 1 & 24 & 0.604 & 0.883 & 0.849 & 0.838 & 0.859 & 0.021 & 0.828 & 0.066 & 0.013 & 0.801 & 0.850\\
SnowyMountains & 2 & 24 & 0.806 & 0.898 & 0.860 & 0.844 & 0.879 & 0.035 & 0.858 & 0.025 & 0.005 & 0.848 & 0.867\\

SnowyMountains & 3 & 24 & 0.799 & 0.912 & 0.866 & 0.851 & 0.890 & 0.039 & 0.866 & 0.028 & 0.006 & 0.855 & 0.876\\
\bottomrule
\end{tabular}
\end{table}
\begin{table}[H]

\caption{Statistical summary by order of passage for LocalMovementEntropy}
\centering
\begin{tabular}[t]{llrrrrrrrrrrrr}
\toprule
Experiment & Order & n & min & max & median & q1 & q3 & iqr & mean & sd & se & ci95\_lower & ci95\_upper\\
\midrule
DesertIsland & 1 & 25 & 0.807 & 0.918 & 0.889 & 0.871 & 0.903 & 0.031 & 0.883 & 0.027 & 0.005 & 0.872 & 0.893\\
DesertIsland & 2 & 25 & 0.817 & 0.916 & 0.871 & 0.862 & 0.894 & 0.032 & 0.872 & 0.028 & 0.006 & 0.860 & 0.883\\
DesertIsland & 3 & 25 & 0.737 & 0.898 & 0.868 & 0.858 & 0.884 & 0.026 & 0.866 & 0.031 & 0.006 & 0.853 & 0.876\\
SnowyMountains & 1 & 24 & 0.755 & 0.903 & 0.878 & 0.867 & 0.884 & 0.017 & 0.869 & 0.031 & 0.006 & 0.855 & 0.880\\
SnowyMountains & 2 & 24 & 0.844 & 0.908 & 0.873 & 0.867 & 0.891 & 0.024 & 0.876 & 0.018 & 0.004 & 0.869 & 0.884\\

SnowyMountains & 3 & 24 & 0.824 & 0.909 & 0.886 & 0.862 & 0.893 & 0.031 & 0.878 & 0.021 & 0.004 & 0.870 & 0.886\\
\bottomrule
\end{tabular}
\end{table}
\begin{table}[H]

\caption{Statistical summary by order of passage for WorldMovementEntropy}
\centering
\begin{tabular}[t]{llrrrrrrrrrrrr}
\toprule
Experiment & Order & n & min & max & median & q1 & q3 & iqr & mean & sd & se & ci95\_lower & ci95\_upper\\
\midrule
DesertIsland & 1 & 25 & 0.815 & 0.926 & 0.899 & 0.879 & 0.908 & 0.029 & 0.893 & 0.024 & 0.005 & 0.884 & 0.903\\
DesertIsland & 2 & 25 & 0.822 & 0.922 & 0.887 & 0.867 & 0.900 & 0.033 & 0.882 & 0.027 & 0.005 & 0.872 & 0.893\\
DesertIsland & 3 & 25 & 0.729 & 0.910 & 0.879 & 0.872 & 0.895 & 0.024 & 0.874 & 0.035 & 0.007 & 0.858 & 0.886\\
SnowyMountains & 1 & 24 & 0.783 & 0.909 & 0.886 & 0.867 & 0.896 & 0.029 & 0.874 & 0.033 & 0.007 & 0.860 & 0.887\\
SnowyMountains & 2 & 24 & 0.841 & 0.918 & 0.888 & 0.878 & 0.898 & 0.020 & 0.885 & 0.018 & 0.004 & 0.877 & 0.891\\

SnowyMountains & 3 & 24 & 0.858 & 0.914 & 0.888 & 0.874 & 0.898 & 0.024 & 0.887 & 0.016 & 0.003 & 0.880 & 0.893\\
\bottomrule
\end{tabular}
\end{table}
\begin{table}[H]

\caption{Statistical summary by order of passage for NumberOfTeleportations}
\centering
\begin{tabular}[t]{llrrrrrrrrrrrr}
\toprule
Experiment & Order & n & min & max & median & q1 & q3 & iqr & mean & sd & se & ci95\_lower & ci95\_upper\\
\midrule
DesertIsland & 1 & 25 & 28 & 117 & 64.0 & 38.0 & 73.0 & 35.0 & 61.3 & 26.1 & 5.22 & 52.4 & 71.9\\
DesertIsland & 2 & 25 & 19 & 104 & 34.0 & 28.0 & 59.0 & 31.0 & 42.8 & 21.5 & 4.29 & 35.4 & 52.0\\
DesertIsland & 3 & 25 & 16 & 71 & 32.0 & 30.0 & 44.0 & 14.0 & 36.6 & 13.0 & 2.61 & 31.9 & 42.2\\
SnowyMountains & 1 & 24 & 27 & 86 & 56.0 & 44.0 & 66.0 & 22.0 & 55.4 & 17.8 & 3.64 & 48.6 & 62.5\\
SnowyMountains & 2 & 24 & 28 & 65 & 45.5 & 36.8 & 57.2 & 20.5 & 45.4 & 12.0 & 2.45 & 41.0 & 49.9\\

SnowyMountains & 3 & 24 & 32 & 73 & 45.5 & 39.0 & 61.0 & 22.0 & 48.9 & 12.9 & 2.63 & 44.1 & 53.6\\
\bottomrule
\end{tabular}
\end{table}

\subsubsection{Comparisons of questionnaire answers between experimental conditions}

\phantom{*}

\begin{table}[H]

\caption{Friedman tests comparing experimental conditions for HX Expressivity}
\centering
\begin{tabular}[t]{lrrrrl}
\toprule
Experiment & n & statistic & df & p & p.signif\\
\midrule
DesertIsland & 25 & 21.2 & 2 & 0 & ****\\
SnowyMountains & 24 & 26.3 & 2 & 0 & ****\\
\bottomrule
\end{tabular}
\end{table}
\begin{table}[H]

\caption{Wilcoxon tests comparing experimental conditions for HX Expressivity}
\centering
\begin{tabular}[t]{lllrrrrl}
\toprule
Experiment & group1 & group2 & n1 & n2 & statistic & p.adj & p.adj.signif\\
\midrule
DesertIsland & Control & Static & 25 & 25 & 10.5 & 0.001 & ***\\
DesertIsland & Control & Dynamic & 25 & 25 & 19.0 & 0.001 & ***\\
DesertIsland & Static & Dynamic & 25 & 25 & 111.0 & 0.419 & ns\\
SnowyMountains & Control & Static & 24 & 24 & 10.0 & 0.000 & ***\\
SnowyMountains & Control & Dynamic & 24 & 24 & 9.5 & 0.000 & ***\\

SnowyMountains & Static & Dynamic & 24 & 24 & 65.5 & 0.085 & ns\\
\bottomrule
\end{tabular}
\end{table}
\begin{table}[H]

\caption{Friedman tests comparing experimental conditions for HX Realism}
\centering
\begin{tabular}[t]{lrrrrl}
\toprule
Experiment & n & statistic & df & p & p.signif\\
\midrule
DesertIsland & 25 & 29.1 & 2 & 0 & ****\\
SnowyMountains & 24 & 33.7 & 2 & 0 & ****\\
\bottomrule
\end{tabular}
\end{table}
\begin{table}[H]

\caption{Wilcoxon tests comparing experimental conditions for HX Realism}
\centering
\begin{tabular}[t]{lllrrrrl}
\toprule
Experiment & group1 & group2 & n1 & n2 & statistic & p.adj & p.adj.signif\\
\midrule
DesertIsland & Control & Static & 25 & 25 & 0.0 & 0.000 & ****\\
DesertIsland & Control & Dynamic & 25 & 25 & 8.0 & 0.000 & ***\\
DesertIsland & Static & Dynamic & 25 & 25 & 99.0 & 0.887 & ns\\
SnowyMountains & Control & Static & 24 & 24 & 0.0 & 0.000 & ****\\
SnowyMountains & Control & Dynamic & 24 & 24 & 6.5 & 0.000 & ***\\

SnowyMountains & Static & Dynamic & 24 & 24 & 112.0 & 0.096 & ns\\
\bottomrule
\end{tabular}
\end{table}
\begin{table}[H]

\caption{Friedman tests comparing experimental conditions for HX Involvement}
\centering
\begin{tabular}[t]{lrrrrl}
\toprule
Experiment & n & statistic & df & p & p.signif\\
\midrule
DesertIsland & 25 & 34.1 & 2 & 0 & ****\\
SnowyMountains & 24 & 27.0 & 2 & 0 & ****\\
\bottomrule
\end{tabular}
\end{table}
\begin{table}[H]

\caption{Wilcoxon tests comparing experimental conditions for HX Involvement}
\centering
\begin{tabular}[t]{lllrrrrl}
\toprule
Experiment & group1 & group2 & n1 & n2 & statistic & p.adj & p.adj.signif\\
\midrule
DesertIsland & Control & Static & 25 & 25 & 0.0 & 0.000 & ***\\
DesertIsland & Control & Dynamic & 25 & 25 & 2.5 & 0.000 & ***\\
DesertIsland & Static & Dynamic & 25 & 25 & 44.0 & 0.594 & ns\\
SnowyMountains & Control & Static & 24 & 24 & 5.5 & 0.000 & ***\\
SnowyMountains & Control & Dynamic & 24 & 24 & 19.0 & 0.000 & ***\\

SnowyMountains & Static & Dynamic & 24 & 24 & 81.0 & 0.216 & ns\\
\bottomrule
\end{tabular}
\end{table}
\begin{table}[H]

\caption{Friedman tests comparing experimental conditions for HX Harmony}
\centering
\begin{tabular}[t]{lrrrrl}
\toprule
Experiment & n & statistic & df & p & p.signif\\
\midrule
DesertIsland & 25 & 10.8 & 2 & 0.005 & **\\
SnowyMountains & 24 & 21.5 & 2 & 0.000 & ****\\
\bottomrule
\end{tabular}
\end{table}
\begin{table}[H]

\caption{Wilcoxon tests comparing experimental conditions for HX Harmony}
\centering
\begin{tabular}[t]{lllrrrrl}
\toprule
Experiment & group1 & group2 & n1 & n2 & statistic & p.adj & p.adj.signif\\
\midrule
DesertIsland & Control & Static & 25 & 25 & 35.5 & 0.005 & **\\
DesertIsland & Control & Dynamic & 25 & 25 & 25.5 & 0.005 & **\\
DesertIsland & Static & Dynamic & 25 & 25 & 83.0 & 0.642 & ns\\
SnowyMountains & Control & Static & 24 & 24 & 3.0 & 0.000 & ***\\
SnowyMountains & Control & Dynamic & 24 & 24 & 15.5 & 0.001 & ***\\

SnowyMountains & Static & Dynamic & 24 & 24 & 99.5 & 0.282 & ns\\
\bottomrule
\end{tabular}
\end{table}
\begin{table}[H]

\caption{Friedman tests comparing experimental conditions for General Presence}
\centering
\begin{tabular}[t]{lrrrrl}
\toprule
Experiment & n & statistic & df & p & p.signif\\
\midrule
DesertIsland & 25 & 20.2 & 2 & 0 & ****\\
SnowyMountains & 24 & 17.7 & 2 & 0 & ***\\
\bottomrule
\end{tabular}
\end{table}
\begin{table}[H]

\caption{Wilcoxon tests comparing experimental conditions for General Presence}
\centering
\begin{tabular}[t]{lllrrrrl}
\toprule
Experiment & group1 & group2 & n1 & n2 & statistic & p.adj & p.adj.signif\\
\midrule
DesertIsland & Control & Static & 25 & 25 & 0 & 0.001 & ***\\
DesertIsland & Control & Dynamic & 25 & 25 & 11 & 0.002 & **\\
DesertIsland & Static & Dynamic & 25 & 25 & 22 & 0.308 & ns\\
SnowyMountains & Control & Static & 24 & 24 & 0 & 0.004 & **\\
SnowyMountains & Control & Dynamic & 24 & 24 & 6 & 0.004 & **\\

SnowyMountains & Static & Dynamic & 24 & 24 & 30 & 0.824 & ns\\
\bottomrule
\end{tabular}
\end{table}
\begin{table}[H]

\caption{Friedman tests comparing experimental conditions for SPES Spatial Location}
\centering
\begin{tabular}[t]{lrrrrl}
\toprule
Experiment & n & statistic & df & p & p.signif\\
\midrule
DesertIsland & 25 & 13.3 & 2 & 0.001 & **\\
SnowyMountains & 24 & 12.4 & 2 & 0.002 & **\\
\bottomrule
\end{tabular}
\end{table}
\begin{table}[H]

\caption{Wilcoxon tests comparing experimental conditions for SPES Spatial Location}
\centering
\begin{tabular}[t]{lllrrrrl}
\toprule
Experiment & group1 & group2 & n1 & n2 & statistic & p.adj & p.adj.signif\\
\midrule
DesertIsland & Control & Static & 25 & 25 & 31.5 & 0.013 & *\\
DesertIsland & Control & Dynamic & 25 & 25 & 27.5 & 0.007 & **\\
DesertIsland & Static & Dynamic & 25 & 25 & 101.5 & 0.808 & ns\\
SnowyMountains & Control & Static & 24 & 24 & 16.5 & 0.001 & ***\\
SnowyMountains & Control & Dynamic & 24 & 24 & 38.5 & 0.027 & *\\

SnowyMountains & Static & Dynamic & 24 & 24 & 142.5 & 0.164 & ns\\
\bottomrule
\end{tabular}
\end{table}
\begin{table}[H]

\caption{Friedman tests comparing experimental conditions for SPES Possible Actions}
\centering
\begin{tabular}[t]{lrrrrl}
\toprule
Experiment & n & statistic & df & p & p.signif\\
\midrule
DesertIsland & 25 & 11.47 & 2 & 0.003 & **\\
SnowyMountains & 24 & 4.23 & 2 & 0.120 & ns\\
\bottomrule
\end{tabular}
\end{table}
\begin{table}[H]

\caption{Wilcoxon tests comparing experimental conditions for SPES Possible Actions}
\centering
\begin{tabular}[t]{lllrrrrl}
\toprule
Experiment & group1 & group2 & n1 & n2 & statistic & p.adj & p.adj.signif\\
\midrule
DesertIsland & Control & Static & 25 & 25 & 42.5 & 0.018 & *\\
DesertIsland & Control & Dynamic & 25 & 25 & 48.5 & 0.374 & ns\\
DesertIsland & Static & Dynamic & 25 & 25 & 99.5 & 0.374 & ns\\
SnowyMountains & Control & Static & 24 & 24 & 67.0 & 0.909 & ns\\
SnowyMountains & Control & Dynamic & 24 & 24 & 61.5 & 0.909 & ns\\

SnowyMountains & Static & Dynamic & 24 & 24 & 73.0 & 0.909 & ns\\
\bottomrule
\end{tabular}
\end{table}
\begin{table}[H]

\caption{Friedman tests comparing experimental conditions for Thermal Sensation}
\centering
\begin{tabular}[t]{lrrrrl}
\toprule
Experiment & n & statistic & df & p & p.signif\\
\midrule
DesertIsland & 25 & 29.4 & 2 & 0 & ****\\
SnowyMountains & 24 & 29.7 & 2 & 0 & ****\\
\bottomrule
\end{tabular}
\end{table}
\begin{table}[H]

\caption{Wilcoxon tests comparing experimental conditions for Thermal Sensation}
\centering
\begin{tabular}[t]{lllrrrrl}
\toprule
Experiment & group1 & group2 & n1 & n2 & statistic & p.adj & p.adj.signif\\
\midrule
DesertIsland & Control & Static & 25 & 25 & 5.5 & 0.000 & ***\\
DesertIsland & Control & Dynamic & 25 & 25 & 12.0 & 0.000 & ***\\
DesertIsland & Static & Dynamic & 25 & 25 & 43.0 & 0.377 & ns\\
SnowyMountains & Control & Static & 24 & 24 & 210.0 & 0.000 & ***\\
SnowyMountains & Control & Dynamic & 24 & 24 & 171.0 & 0.000 & ***\\

SnowyMountains & Static & Dynamic & 24 & 24 & 49.0 & 0.835 & ns\\
\bottomrule
\end{tabular}
\end{table}
\begin{table}[H]

\caption{Friedman tests comparing experimental conditions for Thermal Variation}
\centering
\begin{tabular}[t]{lrrrrl}
\toprule
Experiment & n & statistic & df & p & p.signif\\
\midrule
DesertIsland & 25 & 24.4 & 2 & 0 & ****\\
SnowyMountains & 24 & 26.5 & 2 & 0 & ****\\
\bottomrule
\end{tabular}
\end{table}
\begin{table}[H]

\caption{Wilcoxon tests comparing experimental conditions for Thermal Variation}
\centering
\begin{tabular}[t]{lllrrrrl}
\toprule
Experiment & group1 & group2 & n1 & n2 & statistic & p.adj & p.adj.signif\\
\midrule
DesertIsland & Control & Static & 25 & 25 & 15.0 & 0.001 & ***\\
DesertIsland & Control & Dynamic & 25 & 25 & 10.5 & 0.001 & ***\\
DesertIsland & Static & Dynamic & 25 & 25 & 18.0 & 0.178 & ns\\
SnowyMountains & Control & Static & 24 & 24 & 5.0 & 0.000 & ***\\
SnowyMountains & Control & Dynamic & 24 & 24 & 2.5 & 0.000 & ***\\

SnowyMountains & Static & Dynamic & 24 & 24 & 31.0 & 0.162 & ns\\
\bottomrule
\end{tabular}
\end{table}

\clearpage

\subsubsection{Comparisons of questionnaire answers between orders of passage}

\phantom{*}

\begin{table}[H]

\caption{Friedman tests comparing order of passage for HX Expressivity}
\centering
\begin{tabular}[t]{lrrrrl}
\toprule
Experiment & n & statistic & df & p & p.signif\\
\midrule
DesertIsland & 25 & 2.32 & 2 & 0.314 & ns\\
SnowyMountains & 24 & 3.70 & 2 & 0.157 & ns\\
\bottomrule
\end{tabular}
\end{table}
\begin{table}[H]

\caption{Wilcoxon tests comparing order of passage for HX Expressivity}
\centering
\begin{tabular}[t]{lllrrrrl}
\toprule
Experiment & group1 & group2 & n1 & n2 & statistic & p.adj & p.adj.signif\\
\midrule
DesertIsland & 1 & 2 & 25 & 25 & 113.0 & 0.669 & ns\\
DesertIsland & 1 & 3 & 25 & 25 & 88.5 & 0.669 & ns\\
DesertIsland & 2 & 3 & 25 & 25 & 148.0 & 0.772 & ns\\
SnowyMountains & 1 & 2 & 24 & 24 & 101.5 & 0.546 & ns\\
SnowyMountains & 1 & 3 & 24 & 24 & 80.0 & 0.240 & ns\\

SnowyMountains & 2 & 3 & 24 & 24 & 108.0 & 0.546 & ns\\
\bottomrule
\end{tabular}
\end{table}
\begin{table}[H]

\caption{Friedman tests comparing order of passage for HX Realism}
\centering
\begin{tabular}[t]{lrrrrl}
\toprule
Experiment & n & statistic & df & p & p.signif\\
\midrule
DesertIsland & 25 & 3.22 & 2 & 0.200 & ns\\
SnowyMountains & 24 & 2.55 & 2 & 0.279 & ns\\
\bottomrule
\end{tabular}
\end{table}
\begin{table}[H]

\caption{Wilcoxon tests comparing order of passage for HX Realism}
\centering
\begin{tabular}[t]{lllrrrrl}
\toprule
Experiment & group1 & group2 & n1 & n2 & statistic & p.adj & p.adj.signif\\
\midrule
DesertIsland & 1 & 2 & 25 & 25 & 104.0 & 0.723 & ns\\
DesertIsland & 1 & 3 & 25 & 25 & 99.0 & 0.723 & ns\\
DesertIsland & 2 & 3 & 25 & 25 & 106.5 & 0.767 & ns\\
SnowyMountains & 1 & 2 & 24 & 24 & 97.5 & 0.804 & ns\\
SnowyMountains & 1 & 3 & 24 & 24 & 71.5 & 0.645 & ns\\

SnowyMountains & 2 & 3 & 24 & 24 & 110.0 & 0.804 & ns\\
\bottomrule
\end{tabular}
\end{table}
\begin{table}[H]

\caption{Friedman tests comparing order of passage for HX Involvement}
\centering
\begin{tabular}[t]{lrrrrl}
\toprule
Experiment & n & statistic & df & p & p.signif\\
\midrule
DesertIsland & 25 & 1.927 & 2 & 0.382 & ns\\
SnowyMountains & 24 & 0.837 & 2 & 0.658 & ns\\
\bottomrule
\end{tabular}
\end{table}
\begin{table}[H]

\caption{Wilcoxon tests comparing order of passage for HX Involvement}
\centering
\begin{tabular}[t]{lllrrrrl}
\toprule
Experiment & group1 & group2 & n1 & n2 & statistic & p.adj & p.adj.signif\\
\midrule
DesertIsland & 1 & 2 & 25 & 25 & 80.0 & 0.716 & ns\\
DesertIsland & 1 & 3 & 25 & 25 & 83.0 & 0.480 & ns\\
DesertIsland & 2 & 3 & 25 & 25 & 67.0 & 0.716 & ns\\
SnowyMountains & 1 & 2 & 24 & 24 & 146.0 & 1.000 & ns\\
SnowyMountains & 1 & 3 & 24 & 24 & 106.0 & 1.000 & ns\\

SnowyMountains & 2 & 3 & 24 & 24 & 88.5 & 1.000 & ns\\
\bottomrule
\end{tabular}
\end{table}
\begin{table}[H]

\caption{Friedman tests comparing order of passage for HX Harmony}
\centering
\begin{tabular}[t]{lrrrrl}
\toprule
Experiment & n & statistic & df & p & p.signif\\
\midrule
DesertIsland & 25 & 0.621 & 2 & 0.733 & ns\\
SnowyMountains & 24 & 1.506 & 2 & 0.471 & ns\\
\bottomrule
\end{tabular}
\end{table}
\begin{table}[H]

\caption{Wilcoxon tests comparing order of passage for HX Harmony}
\centering
\begin{tabular}[t]{lllrrrrl}
\toprule
Experiment & group1 & group2 & n1 & n2 & statistic & p.adj & p.adj.signif\\
\midrule
DesertIsland & 1 & 2 & 25 & 25 & 150 & 1.000 & ns\\
DesertIsland & 1 & 3 & 25 & 25 & 102 & 1.000 & ns\\
DesertIsland & 2 & 3 & 25 & 25 & 128 & 1.000 & ns\\
SnowyMountains & 1 & 2 & 24 & 24 & 130 & 0.477 & ns\\
SnowyMountains & 1 & 3 & 24 & 24 & 118 & 0.720 & ns\\

SnowyMountains & 2 & 3 & 24 & 24 & 80 & 0.720 & ns\\
\bottomrule
\end{tabular}
\end{table}
\begin{table}[H]

\caption{Friedman tests comparing order of passage for General Presence}
\centering
\begin{tabular}[t]{lrrrrl}
\toprule
Experiment & n & statistic & df & p & p.signif\\
\midrule
DesertIsland & 25 & 0.286 & 2 & 0.867 & ns\\
SnowyMountains & 24 & 6.453 & 2 & 0.040 & *\\
\bottomrule
\end{tabular}
\end{table}
\begin{table}[H]

\caption{Wilcoxon tests comparing order of passage for General Presence}
\centering
\begin{tabular}[t]{lllrrrrl}
\toprule
Experiment & group1 & group2 & n1 & n2 & statistic & p.adj & p.adj.signif\\
\midrule
DesertIsland & 1 & 2 & 25 & 25 & 78.0 & 0.894 & ns\\
DesertIsland & 1 & 3 & 25 & 25 & 87.0 & 0.894 & ns\\
DesertIsland & 2 & 3 & 25 & 25 & 44.0 & 0.943 & ns\\
SnowyMountains & 1 & 2 & 24 & 24 & 84.0 & 0.095 & ns\\
SnowyMountains & 1 & 3 & 24 & 24 & 39.5 & 0.096 & ns\\

SnowyMountains & 2 & 3 & 24 & 24 & 66.0 & 0.379 & ns\\
\bottomrule
\end{tabular}
\end{table}
\begin{table}[H]

\caption{Friedman tests comparing order of passage for SPES Spatial Location}
\centering
\begin{tabular}[t]{lrrrrl}
\toprule
Experiment & n & statistic & df & p & p.signif\\
\midrule
DesertIsland & 25 & 1.854 & 2 & 0.396 & ns\\
SnowyMountains & 24 & 0.023 & 2 & 0.988 & ns\\
\bottomrule
\end{tabular}
\end{table}
\begin{table}[H]

\caption{Wilcoxon tests comparing order of passage for SPES Spatial Location}
\centering
\begin{tabular}[t]{lllrrrrl}
\toprule
Experiment & group1 & group2 & n1 & n2 & statistic & p.adj & p.adj.signif\\
\midrule
DesertIsland & 1 & 2 & 25 & 25 & 88 & 1 & ns\\
DesertIsland & 1 & 3 & 25 & 25 & 90 & 1 & ns\\
DesertIsland & 2 & 3 & 25 & 25 & 112 & 1 & ns\\
SnowyMountains & 1 & 2 & 24 & 24 & 114 & 1 & ns\\
SnowyMountains & 1 & 3 & 24 & 24 & 171 & 1 & ns\\

SnowyMountains & 2 & 3 & 24 & 24 & 102 & 1 & ns\\
\bottomrule
\end{tabular}
\end{table}
\begin{table}[H]

\caption{Friedman tests comparing order of passage for SPES Possible Actions}
\centering
\begin{tabular}[t]{lrrrrl}
\toprule
Experiment & n & statistic & df & p & p.signif\\
\midrule
DesertIsland & 25 & 4.886 & 2 & 0.087 & ns\\
SnowyMountains & 24 & 0.338 & 2 & 0.845 & ns\\
\bottomrule
\end{tabular}
\end{table}
\begin{table}[H]

\caption{Wilcoxon tests comparing order of passage for SPES Possible Actions}
\centering
\begin{tabular}[t]{lllrrrrl}
\toprule
Experiment & group1 & group2 & n1 & n2 & statistic & p.adj & p.adj.signif\\
\midrule
DesertIsland & 1 & 2 & 25 & 25 & 76.0 & 0.342 & ns\\
DesertIsland & 1 & 3 & 25 & 25 & 44.0 & 0.121 & ns\\
DesertIsland & 2 & 3 & 25 & 25 & 72.5 & 0.834 & ns\\
SnowyMountains & 1 & 2 & 24 & 24 & 66.0 & 1.000 & ns\\
SnowyMountains & 1 & 3 & 24 & 24 & 84.5 & 1.000 & ns\\

SnowyMountains & 2 & 3 & 24 & 24 & 80.0 & 1.000 & ns\\
\bottomrule
\end{tabular}
\end{table}
\begin{table}[H]

\caption{Friedman tests comparing order of passage for Thermal Sensation}
\centering
\begin{tabular}[t]{lrrrrl}
\toprule
Experiment & n & statistic & df & p & p.signif\\
\midrule
DesertIsland & 25 & 0.987 & 2 & 0.610 & ns\\
SnowyMountains & 24 & 0.849 & 2 & 0.654 & ns\\
\bottomrule
\end{tabular}
\end{table}
\begin{table}[H]

\caption{Wilcoxon tests comparing order of passage for Thermal Sensation}
\centering
\begin{tabular}[t]{lllrrrrl}
\toprule
Experiment & group1 & group2 & n1 & n2 & statistic & p.adj & p.adj.signif\\
\midrule
DesertIsland & 1 & 2 & 25 & 25 & 80.0 & 1.00 & ns\\
DesertIsland & 1 & 3 & 25 & 25 & 72.5 & 0.66 & ns\\
DesertIsland & 2 & 3 & 25 & 25 & 65.0 & 1.00 & ns\\
SnowyMountains & 1 & 2 & 24 & 24 & 100.0 & 1.00 & ns\\
SnowyMountains & 1 & 3 & 24 & 24 & 89.0 & 1.00 & ns\\

SnowyMountains & 2 & 3 & 24 & 24 & 71.5 & 1.00 & ns\\
\bottomrule
\end{tabular}
\end{table}
\begin{table}[H]

\caption{Friedman tests comparing order of passage for Thermal Variation}
\centering
\begin{tabular}[t]{lrrrrl}
\toprule
Experiment & n & statistic & df & p & p.signif\\
\midrule
DesertIsland & 25 & 3.46 & 2 & 0.178 & ns\\
SnowyMountains & 24 & 4.76 & 2 & 0.092 & ns\\
\bottomrule
\end{tabular}
\end{table}
\begin{table}[H]

\caption{Wilcoxon tests comparing order of passage for Thermal Variation}
\centering
\begin{tabular}[t]{lllrrrrl}
\toprule
Experiment & group1 & group2 & n1 & n2 & statistic & p.adj & p.adj.signif\\
\midrule
DesertIsland & 1 & 2 & 25 & 25 & 59.5 & 0.450 & ns\\
DesertIsland & 1 & 3 & 25 & 25 & 69.5 & 0.854 & ns\\
DesertIsland & 2 & 3 & 25 & 25 & 93.5 & 0.854 & ns\\
SnowyMountains & 1 & 2 & 24 & 24 & 68.0 & 0.544 & ns\\
SnowyMountains & 1 & 3 & 24 & 24 & 43.0 & 0.054 & ns\\

SnowyMountains & 2 & 3 & 24 & 24 & 45.5 & 0.544 & ns\\
\bottomrule
\end{tabular}
\end{table}

\clearpage

\subsubsection{Comparisons of behavioral metrics between conditions}

\phantom{*}

\begin{table}[H]

\caption{Friedman tests comparing experimental conditions for TotalLocalDistance}
\centering
\begin{tabular}[t]{lrrrrl}
\toprule
Experiment & n & statistic & df & p & p.signif\\
\midrule
DesertIsland & 25 & 1.520 & 2 & 0.468 & ns\\
SnowyMountains & 24 & 0.333 & 2 & 0.846 & ns\\
\bottomrule
\end{tabular}
\end{table}
\begin{table}[H]

\caption{Wilcoxon tests comparing experimental conditions for TotalLocalDistance}
\centering
\begin{tabular}[t]{lllrrrrl}
\toprule
Experiment & group1 & group2 & n1 & n2 & statistic & p.adj & p.adj.signif\\
\midrule
DesertIsland & Control & Static & 25 & 25 & 138 & 1 & ns\\
DesertIsland & Control & Dynamic & 25 & 25 & 167 & 1 & ns\\
DesertIsland & Static & Dynamic & 25 & 25 & 192 & 1 & ns\\
SnowyMountains & Control & Static & 24 & 24 & 173 & 1 & ns\\
SnowyMountains & Control & Dynamic & 24 & 24 & 154 & 1 & ns\\

SnowyMountains & Static & Dynamic & 24 & 24 & 148 & 1 & ns\\
\bottomrule
\end{tabular}
\end{table}
\begin{table}[H]

\caption{Friedman tests comparing experimental conditions for TotalWorldDistance}
\centering
\begin{tabular}[t]{lrrrrl}
\toprule
Experiment & n & statistic & df & p & p.signif\\
\midrule
DesertIsland & 25 & 0.32 & 2 & 0.852 & ns\\
SnowyMountains & 24 & 1.08 & 2 & 0.582 & ns\\
\bottomrule
\end{tabular}
\end{table}
\begin{table}[H]

\caption{Wilcoxon tests comparing experimental conditions for TotalWorldDistance}
\centering
\begin{tabular}[t]{lllrrrrl}
\toprule
Experiment & group1 & group2 & n1 & n2 & statistic & p.adj & p.adj.signif\\
\midrule
DesertIsland & Control & Static & 25 & 25 & 153 & 1 & ns\\
DesertIsland & Control & Dynamic & 25 & 25 & 171 & 1 & ns\\
DesertIsland & Static & Dynamic & 25 & 25 & 174 & 1 & ns\\
SnowyMountains & Control & Static & 24 & 24 & 128 & 1 & ns\\
SnowyMountains & Control & Dynamic & 24 & 24 & 120 & 1 & ns\\

SnowyMountains & Static & Dynamic & 24 & 24 & 125 & 1 & ns\\
\bottomrule
\end{tabular}
\end{table}
\begin{table}[H]

\caption{Friedman tests comparing experimental conditions for TotalLocalRotationDistance}
\centering
\begin{tabular}[t]{lrrrrl}
\toprule
Experiment & n & statistic & df & p & p.signif\\
\midrule
DesertIsland & 25 & 1.040 & 2 & 0.595 & ns\\
SnowyMountains & 24 & 0.083 & 2 & 0.959 & ns\\
\bottomrule
\end{tabular}
\end{table}
\begin{table}[H]

\caption{Wilcoxon tests comparing experimental conditions for TotalLocalRotationDistance}
\centering
\begin{tabular}[t]{lllrrrrl}
\toprule
Experiment & group1 & group2 & n1 & n2 & statistic & p.adj & p.adj.signif\\
\midrule
DesertIsland & Control & Static & 25 & 25 & 149 & 1 & ns\\
DesertIsland & Control & Dynamic & 25 & 25 & 167 & 1 & ns\\
DesertIsland & Static & Dynamic & 25 & 25 & 188 & 1 & ns\\
SnowyMountains & Control & Static & 24 & 24 & 142 & 1 & ns\\
SnowyMountains & Control & Dynamic & 24 & 24 & 135 & 1 & ns\\

SnowyMountains & Static & Dynamic & 24 & 24 & 140 & 1 & ns\\
\bottomrule
\end{tabular}
\end{table}
\begin{table}[H]

\caption{Friedman tests comparing experimental conditions for TotalWorldRotationDistance}
\centering
\begin{tabular}[t]{lrrrrl}
\toprule
Experiment & n & statistic & df & p & p.signif\\
\midrule
DesertIsland & 25 & 0.96 & 2 & 0.619 & ns\\
SnowyMountains & 24 & 0.25 & 2 & 0.882 & ns\\
\bottomrule
\end{tabular}
\end{table}
\begin{table}[H]

\caption{Wilcoxon tests comparing experimental conditions for TotalWorldRotationDistance}
\centering
\begin{tabular}[t]{lllrrrrl}
\toprule
Experiment & group1 & group2 & n1 & n2 & statistic & p.adj & p.adj.signif\\
\midrule
DesertIsland & Control & Static & 25 & 25 & 147 & 1 & ns\\
DesertIsland & Control & Dynamic & 25 & 25 & 176 & 1 & ns\\
DesertIsland & Static & Dynamic & 25 & 25 & 185 & 1 & ns\\
SnowyMountains & Control & Static & 24 & 24 & 153 & 1 & ns\\
SnowyMountains & Control & Dynamic & 24 & 24 & 153 & 1 & ns\\

SnowyMountains & Static & Dynamic & 24 & 24 & 163 & 1 & ns\\
\bottomrule
\end{tabular}
\end{table}
\begin{table}[H]

\caption{Friedman tests comparing experimental conditions for LocalRoamingEntropy}
\centering
\begin{tabular}[t]{lrrrrl}
\toprule
Experiment & n & statistic & df & p & p.signif\\
\midrule
DesertIsland & 25 & 3.440 & 2 & 0.179 & ns\\
SnowyMountains & 24 & 0.083 & 2 & 0.959 & ns\\
\bottomrule
\end{tabular}
\end{table}
\begin{table}[H]

\caption{Wilcoxon tests comparing experimental conditions for LocalRoamingEntropy}
\centering
\begin{tabular}[t]{lllrrrrl}
\toprule
Experiment & group1 & group2 & n1 & n2 & statistic & p.adj & p.adj.signif\\
\midrule
DesertIsland & Control & Static & 25 & 25 & 121 & 0.550 & ns\\
DesertIsland & Control & Dynamic & 25 & 25 & 194 & 0.550 & ns\\
DesertIsland & Static & Dynamic & 25 & 25 & 243 & 0.089 & ns\\
SnowyMountains & Control & Static & 24 & 24 & 189 & 0.831 & ns\\
SnowyMountains & Control & Dynamic & 24 & 24 & 149 & 0.989 & ns\\

SnowyMountains & Static & Dynamic & 24 & 24 & 119 & 0.831 & ns\\
\bottomrule
\end{tabular}
\end{table}
\begin{table}[H]

\caption{Friedman tests comparing experimental conditions for WorldRoamingEntropy}
\centering
\begin{tabular}[t]{lrrrrl}
\toprule
Experiment & n & statistic & df & p & p.signif\\
\midrule
DesertIsland & 25 & 0.96 & 2 & 0.619 & ns\\
SnowyMountains & 24 & 1.00 & 2 & 0.607 & ns\\
\bottomrule
\end{tabular}
\end{table}
\begin{table}[H]

\caption{Wilcoxon tests comparing experimental conditions for WorldRoamingEntropy}
\centering
\begin{tabular}[t]{lllrrrrl}
\toprule
Experiment & group1 & group2 & n1 & n2 & statistic & p.adj & p.adj.signif\\
\midrule
DesertIsland & Control & Static & 25 & 25 & 193 & 1.000 & ns\\
DesertIsland & Control & Dynamic & 25 & 25 & 174 & 1.000 & ns\\
DesertIsland & Static & Dynamic & 25 & 25 & 132 & 1.000 & ns\\
SnowyMountains & Control & Static & 24 & 24 & 160 & 0.790 & ns\\
SnowyMountains & Control & Dynamic & 24 & 24 & 117 & 0.723 & ns\\

SnowyMountains & Static & Dynamic & 24 & 24 & 108 & 0.723 & ns\\
\bottomrule
\end{tabular}
\end{table}
\begin{table}[H]

\caption{Friedman tests comparing experimental conditions for LocalMovementEntropy}
\centering
\begin{tabular}[t]{lrrrrl}
\toprule
Experiment & n & statistic & df & p & p.signif\\
\midrule
DesertIsland & 25 & 0.72 & 2 & 0.698 & ns\\
SnowyMountains & 24 & 6.08 & 2 & 0.048 & *\\
\bottomrule
\end{tabular}
\end{table}
\begin{table}[H]

\caption{Wilcoxon tests comparing experimental conditions for LocalMovementEntropy}
\centering
\begin{tabular}[t]{lllrrrrl}
\toprule
Experiment & group1 & group2 & n1 & n2 & statistic & p.adj & p.adj.signif\\
\midrule
DesertIsland & Control & Static & 25 & 25 & 147 & 1.000 & ns\\
DesertIsland & Control & Dynamic & 25 & 25 & 173 & 1.000 & ns\\
DesertIsland & Static & Dynamic & 25 & 25 & 210 & 0.630 & ns\\
SnowyMountains & Control & Static & 24 & 24 & 193 & 0.458 & ns\\
SnowyMountains & Control & Dynamic & 24 & 24 & 130 & 0.584 & ns\\

SnowyMountains & Static & Dynamic & 24 & 24 & 88 & 0.236 & ns\\
\bottomrule
\end{tabular}
\end{table}
\begin{table}[H]

\caption{Friedman tests comparing experimental conditions for WorldMovementEntropy}
\centering
\begin{tabular}[t]{lrrrrl}
\toprule
Experiment & n & statistic & df & p & p.signif\\
\midrule
DesertIsland & 25 & 0.56 & 2 & 0.756 & ns\\
SnowyMountains & 24 & 1.75 & 2 & 0.417 & ns\\
\bottomrule
\end{tabular}
\end{table}
\begin{table}[H]

\caption{Wilcoxon tests comparing experimental conditions for WorldMovementEntropy}
\centering
\begin{tabular}[t]{lllrrrrl}
\toprule
Experiment & group1 & group2 & n1 & n2 & statistic & p.adj & p.adj.signif\\
\midrule
DesertIsland & Control & Static & 25 & 25 & 174 & 1.000 & ns\\
DesertIsland & Control & Dynamic & 25 & 25 & 185 & 1.000 & ns\\
DesertIsland & Static & Dynamic & 25 & 25 & 164 & 1.000 & ns\\
SnowyMountains & Control & Static & 24 & 24 & 204 & 0.384 & ns\\
SnowyMountains & Control & Dynamic & 24 & 24 & 197 & 0.384 & ns\\

SnowyMountains & Static & Dynamic & 24 & 24 & 148 & 0.966 & ns\\
\bottomrule
\end{tabular}
\end{table}
\begin{table}[H]

\caption{Friedman tests comparing experimental conditions for NumberOfTeleportations}
\centering
\begin{tabular}[t]{lrrrrl}
\toprule
Experiment & n & statistic & df & p & p.signif\\
\midrule
DesertIsland & 25 & 1.670 & 2 & 0.434 & ns\\
SnowyMountains & 24 & 0.452 & 2 & 0.798 & ns\\
\bottomrule
\end{tabular}
\end{table}
\begin{table}[H]

\caption{Wilcoxon tests comparing experimental conditions for NumberOfTeleportations}
\centering
\begin{tabular}[t]{lllrrrrl}
\toprule
Experiment & group1 & group2 & n1 & n2 & statistic & p.adj & p.adj.signif\\
\midrule
DesertIsland & Control & Static & 25 & 25 & 132 & 1 & ns\\
DesertIsland & Control & Dynamic & 25 & 25 & 164 & 1 & ns\\
DesertIsland & Static & Dynamic & 25 & 25 & 140 & 1 & ns\\
SnowyMountains & Control & Static & 24 & 24 & 154 & 1 & ns\\
SnowyMountains & Control & Dynamic & 24 & 24 & 154 & 1 & ns\\

SnowyMountains & Static & Dynamic & 24 & 24 & 162 & 1 & ns\\
\bottomrule
\end{tabular}
\end{table}

\clearpage

\subsubsection{Comparisons of behavioral metrics between order of passage}

\phantom{*}

\begin{table}[H]

\caption{Friedman tests comparing order of passage for TotalLocalDistance}
\centering
\begin{tabular}[t]{lrrrrl}
\toprule
Experiment & n & statistic & df & p & p.signif\\
\midrule
DesertIsland & 25 & 28.1 & 2 & 0 & ****\\
SnowyMountains & 24 & 22.3 & 2 & 0 & ****\\
\bottomrule
\end{tabular}
\end{table}
\begin{table}[H]

\caption{Wilcoxon tests comparing order of passage for TotalLocalDistance}
\centering
\begin{tabular}[t]{lllrrrrl}
\toprule
Experiment & group1 & group2 & n1 & n2 & statistic & p.adj & p.adj.signif\\
\midrule
DesertIsland & 1 & 2 & 25 & 25 & 301 & 0.000 & ****\\
DesertIsland & 1 & 3 & 25 & 25 & 320 & 0.000 & ****\\
DesertIsland & 2 & 3 & 25 & 25 & 250 & 0.017 & *\\
SnowyMountains & 1 & 2 & 24 & 24 & 266 & 0.001 & ***\\
SnowyMountains & 1 & 3 & 24 & 24 & 281 & 0.000 & ***\\

SnowyMountains & 2 & 3 & 24 & 24 & 226 & 0.029 & *\\
\bottomrule
\end{tabular}
\end{table}
\begin{table}[H]

\caption{Friedman tests comparing order of passage for TotalWorldDistance}
\centering
\begin{tabular}[t]{lrrrrl}
\toprule
Experiment & n & statistic & df & p & p.signif\\
\midrule
DesertIsland & 25 & 17.4 & 2 & 0.000 & ***\\
SnowyMountains & 24 & 12.3 & 2 & 0.002 & **\\
\bottomrule
\end{tabular}
\end{table}
\begin{table}[H]

\caption{Wilcoxon tests comparing order of passage for TotalWorldDistance}
\centering
%
\end{table}
\begin{table}[H]

\caption{Friedman tests comparing order of passage for TotalLocalRotationDistance}
\centering
\begin{tabular}[t]{lrrrrl}
\toprule
Experiment & n & statistic & df & p & p.signif\\
\midrule
DesertIsland & 25 & 31 & 2 & 0 & ****\\
SnowyMountains & 24 & 31 & 2 & 0 & ****\\
\bottomrule
\end{tabular}
\end{table}
\begin{table}[H]

\caption{Wilcoxon tests comparing order of passage for TotalLocalRotationDistance}
\centering
%
\end{table}
\begin{table}[H]

\caption{Friedman tests comparing order of passage for TotalWorldRotationDistance}
\centering
%
\end{table}
\begin{table}[H]

\caption{Wilcoxon tests comparing order of passage for TotalWorldRotationDistance}
\centering
%
\end{table}
\begin{table}[H]

\caption{Friedman tests comparing order of passage for LocalRoamingEntropy}
\centering
%
\end{table}
\begin{table}[H]

\caption{Wilcoxon tests comparing order of passage for LocalRoamingEntropy}
\centering
%
\end{table}
\begin{table}[H]

\caption{Friedman tests comparing order of passage for WorldRoamingEntropy}
\centering
%
\end{table}
\begin{table}[H]

\caption{Wilcoxon tests comparing order of passage for WorldRoamingEntropy}
\centering
%
\end{table}
\begin{table}[H]

\caption{Friedman tests comparing order of passage for LocalMovementEntropy}
\centering
%
\end{table}
\begin{table}[H]

\caption{Wilcoxon tests comparing order of passage for LocalMovementEntropy}
\centering
%
\end{table}
\begin{table}[H]

\caption{Friedman tests comparing order of passage for WorldMovementEntropy}
\centering
%
\end{table}
\begin{table}[H]

\caption{Wilcoxon tests comparing order of passage for WorldMovementEntropy}
\centering
%
\end{table}
\begin{table}[H]

\caption{Friedman tests comparing order of passage for NumberOfTeleportations}
\centering
%
\end{table}
\begin{table}[H]

\caption{Wilcoxon tests comparing order of passage for NumberOfTeleportations}
\centering
%
\end{table}

\clearpage

\subsubsection{Correlations}

\phantom{*}

\begin{table}[H]

\caption{Spearman correlations between participant individual characteristics and post-immersion questionnaires for Desert Island}
\centering
%
\end{table}

\begin{table}[H]

\caption{Spearman correlations between participant individual characteristics and post-immersion questionnaires for Snowy Mountains}
\centering
%
\end{table}

\subsection{Discussion - Additional Figures}

\begin{figure}[H]
    \centering
    \subfloat[]{
        \includegraphics[width=0.8\linewidth]{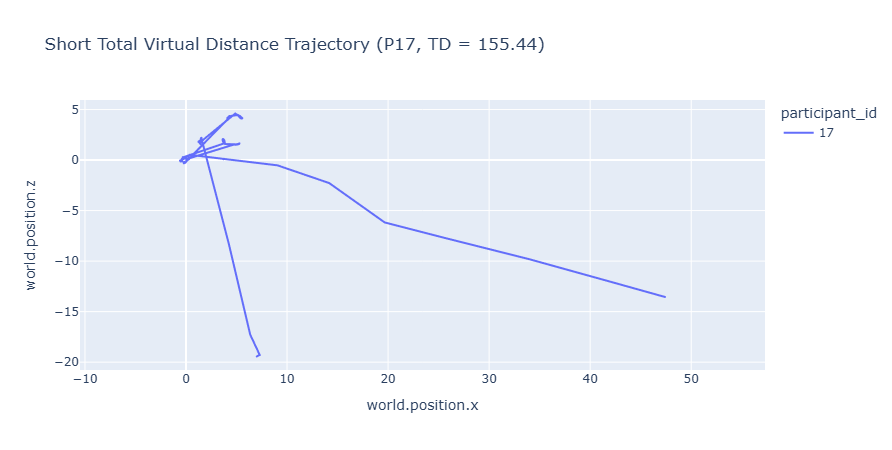}
        \label{short_distance_DI}
    }
    \\
    \subfloat[]{
        \includegraphics[width=0.8\linewidth]{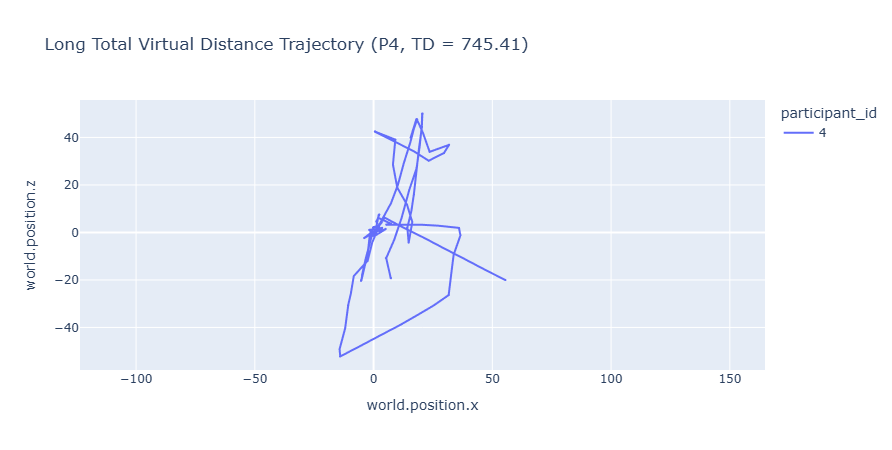}
        \label{long_distance_DI}
    }
    \caption{Visual comparison of a short and long virtual trajectory in Desert Island.}
    \label{fig:thermals}
\end{figure}

\begin{figure}[H]
    \centering
    \subfloat[]{
        \includegraphics[width=0.8\linewidth]{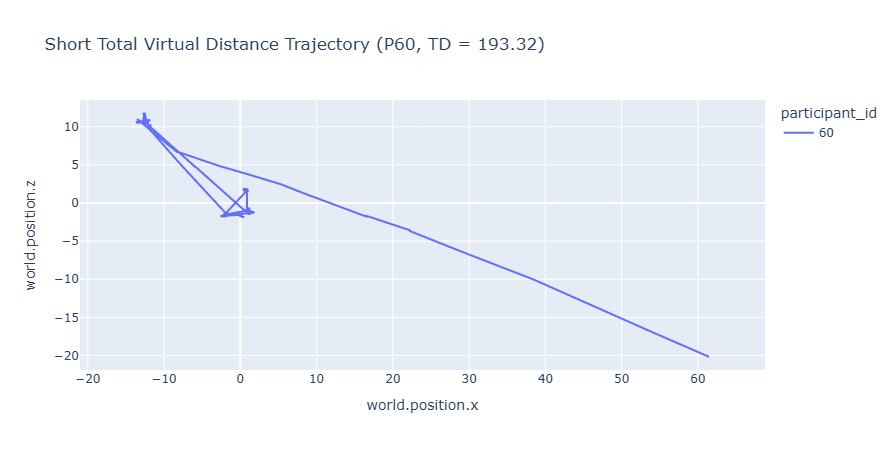}
        \label{short_distance_SM}
    }
    \\
    \subfloat[]{
        \includegraphics[width=0.8\linewidth]{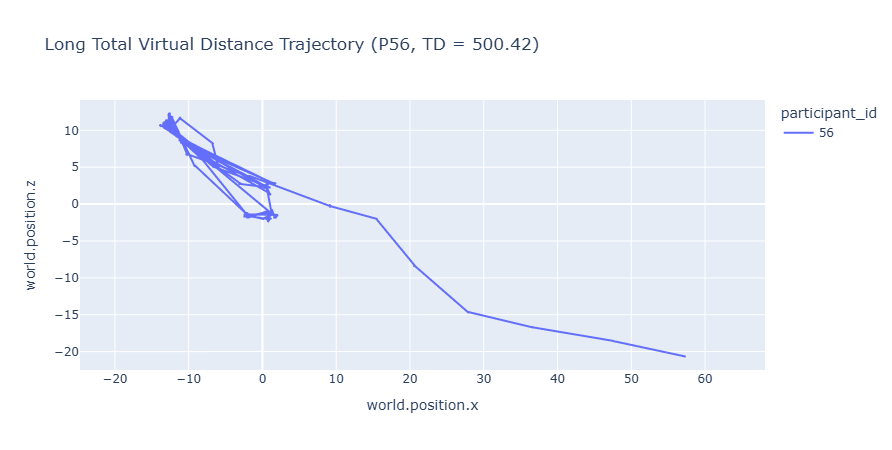}
        \label{long_distance_SM}
    }
    \caption{Visual comparison of a short and long virtual trajectory in Snowy Mountains.}
    \label{fig:thermals}
\end{figure}

\bibliographystyle{IEEEtran}
\bibliography{bibliography}